%% file: main.tex
\documentclass[preprint,3p]{elsarticle}

\usepackage{lineno}

\input{utilities/commands}

\input{utilities/packages}
\input{utilities/cpscolors}

\input{utilities/tudacolors}

\journal{Computer Methods in Applied Mechanics and Engineering}

\begin{document}

\begin{frontmatter}


\title{Immersed boundary-conformal isogeometric methods for magnetostatics}
\author[TUDa]{\texorpdfstring{Yusuf T. Elbadry\corref{cor1}}{Yusuf T. Elbadry}}
\cortext[cor1]{Corresponding author. Email: elbadry@cps.tu-darmstadt.de}
\author[EPFL]{Giuliano Guarino}
\author[EPFL]{Pablo Antolín}
\author[TUDa]{Oliver Weeger}

\affiliation[TUDa]{organization={Cyber-Physical Simulation, Department of Mechanical Engineering and Graduate School Computational Engineering, Technical University of Darmstadt},
            addressline={Dolivostr.\ 15}, 
            city={Darmstadt},
            postcode={64293}, 
            state={Hessen},
            country={Germany}}
\affiliation[EPFL]{organization={Institute of Mathematics, École Polytechnique Fédérale de Lausanne},
            addressline={Bâtiment MA, Station 8}, 
            city={Lausanne},
            postcode={CH-1015}, 
            country={Switzerland}}
            
\begin{abstract}
Isogeometric analysis was proposed to bridge the gap between computer-aided design and numerical discretization. However, standard multi-patch isogeometric analysis mandates conformal discretizations across patch interfaces, posing challenges for multi-material domain problems. In the context of electric machines, this requirement becomes evident in a large number of patches needed to represent machines consisting of several domains and materials.
In this work, we adopt, extend, and evaluate three non-conformal discretization strategies for magnetostatic problems: a fully immersed approach, the union with non-conformal patches, and the union with conformal layers. 
In all three methods, boundary-conformal high-order quadrature rules are employed for integration over trimmed boundary and interface elements.
In the two union approaches, material regions are, as far as possible, represented by independent non-conformal spline patches that are embedded within a background patch and coupled weakly through Nitsche’s method. In the latter framework, critical interfaces are additionally surrounded by conformal layers that enable the strong imposition of boundary conditions and improved resolution of interface behavior. 
The proposed approaches are assessed through several magnetostatic benchmark problems and an industrial application. 
The numerical results show that the union methods achieve highly accurate solutions, while the fully immersed approach struggles with discontinuities in field gradients across material interfaces.
Nevertheless, these methods significantly reduce the geometric preprocessing effort compared to conventional, conformal multi-patch analysis and require substantially fewer patches.
Overall, this demonstrates that our immersed boundary-conformal isogeometric framework possesses great potential for efficient simulation of complex electromagnetic devices.
\end{abstract}

\begin{graphicalabstract}
\includegraphics[width = \linewidth]{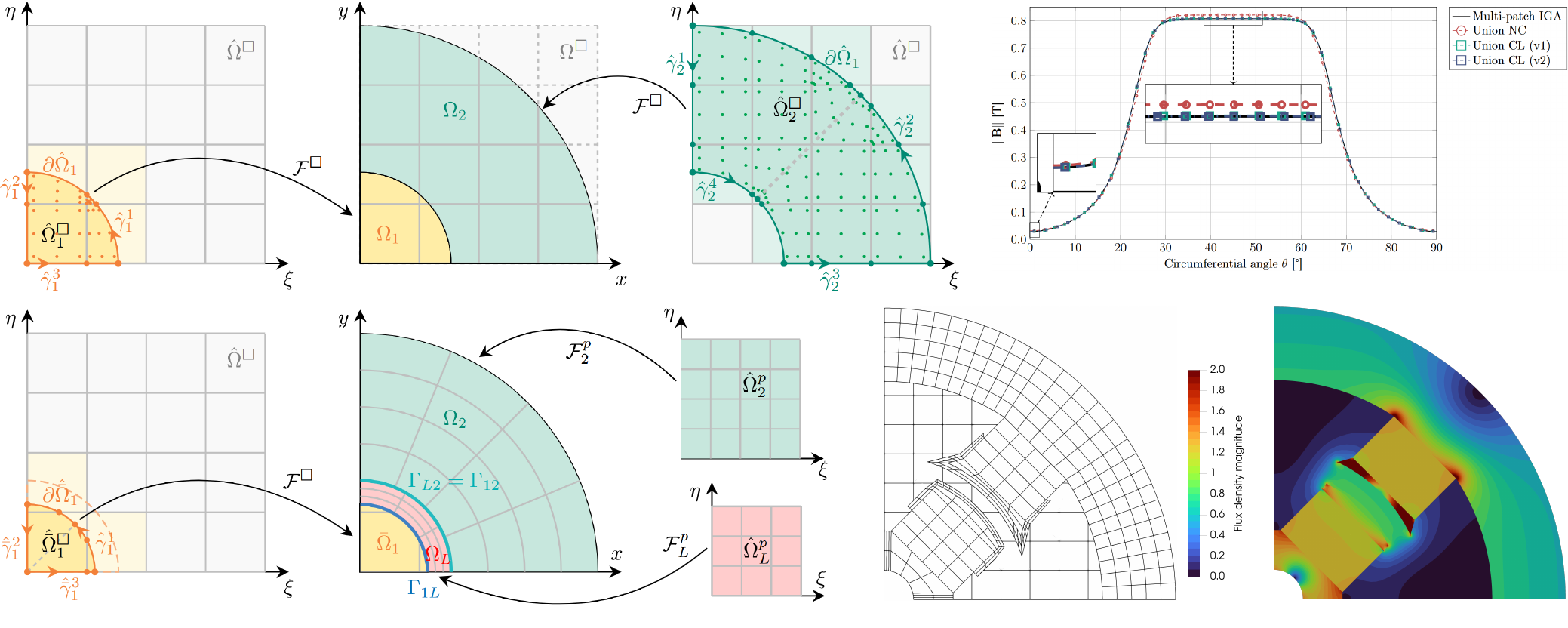}
\end{graphicalabstract}

\begin{highlights}
\item Accurate integration over trimmed elements via boundary-conformal quadrature method
\item Reduced geometry preprocessing effort through minimization of the required patch count
\item Mitigated need for Lagrange multiplier-type methods for coupling
\item Accurate solution for complex geometries without imposing conformality conditions
\item Demonstrated efficacy of the proposed method for magnetostatic applications
\end{highlights}


\end{frontmatter}


\input{sections/introduction}
\input{sections/magnetostatics}

\input{sections/iga}
\input{sections/ibcm}
\input{sections/results}

\input{sections/conclusions}

\section*{Acknowledgments}
Y.T.E.\ and O.W.\ acknowledge financial support by funds from LOEWE -- the Hessian State Offensive for the Development of Scientific and Economic Excellence (Landes-Offensive zur Entwicklung Wissenschaftlich-ökonomischer Exzellenz, Förderlinie 3: KMU-Verbundvorhaben), HA project no.\ 1450/23-04, and of the Graduate School Computational Engineering at the Technical University of Darmstadt, Germany.

G.G.\ and P.A.\ acknowledge the financial support of the Swiss National Science Foundation through the project \emph{FLAS$_h$} with no.\ 200021\_214987.

The authors would like to thank Mr.~Daniel Salazar for his assistance in performing the simulation of the horseshoe magnet with the fully immersed method. 

\section*{CRediT authorship contribution statement}
Yusuf T. Elbadry: Writing – original draft, Visualization, Validation, Software, Investigation, Conceptualization;
Giuliano Guarino: Writing – review \& editing, Visualization, Validation, Software, Conceptualization;
Pablo Antolín: Writing – review \& editing, Software, Methodology, Conceptualization;
Oliver Weeger: writing—review \& editing, Visualization, Methodology, Supervision, Resources, Funding acquisition, Conceptualization.

\section*{Data Availability}

None.

\section*{Disclosure Statement}
The authors declare that they have no known competing financial interests or personal relationships that could have appeared to influence the work reported in this paper.

\section*{ Declaration of generative AI and AI-assisted technologies in the writing process}
The authors disclose the use of ChatGPT, a large language model developed by OpenAI \cite{chatgpt2025}, as a language editing tool to enhance the clarity and readability of the manuscript. The authors are solely responsible for the content of this work.


\bibliographystyle{elsarticle-num} 
\bibliography{references}

\end{document}

%% file: utilities/commands.tex
\newcommand{\bxi}{\boldsymbol{\xi}}

\newcommand{\rmd}{\mathrm{d}}

\newcommand{\tbb}{\textbf{b}}

\newcommand{\tbn}{\textbf{n}}

\newcommand{\tbu}{\textbf{u}}
\newcommand{\tbx}{\textbf{x}}

\newcommand{\tbA}{\textbf{A}}
\newcommand{\tbB}{\textbf{B}}

\newcommand{\tbH}{\textbf{H}}
\newcommand{\tbK}{\textbf{K}}

\newcommand{\tbJ}{\textbf{J}}

\newcommand{\bbR}{\mathbb{R}}


%% file: utilities/packages.tex
\usepackage{graphicx}%
\usepackage[dvipsnames]{xcolor}
\usepackage{amssymb}
\usepackage{amsmath}
\usepackage{amsthm}

\usepackage{tikz}
\usetikzlibrary{patterns}
\usetikzlibrary{external}
\usetikzlibrary{decorations.markings}
\usetikzlibrary{arrows.meta,intersections}
\usepackage{pgfplots}
\usepackage{pgfplotstable}
\usetikzlibrary{spy}
\pgfplotsset{compat=newest}
\usepgfplotslibrary{groupplots}
\usepgfplotslibrary{fillbetween}
\usepgfplotslibrary{colormaps}
\usepgfplotslibrary{patchplots}
\usetikzlibrary{shapes,arrows,matrix,positioning}
\usepackage{tikzscale}
\usepackage[font=small]{caption}
\usepackage{subcaption}
\usepackage{wrapfig}

\usepackage{hyperref}
\usepackage[capitalise]{cleveref}

\usepackage{lipsum}

%% file: utilities/cpscolors.tex
\definecolor{CPSgreen}{RGB}{22,164,138}
\definecolor{CPSlightblue}{RGB}{104,143,198}
\definecolor{CPSdarkblue}{RGB}{67,83,132}
\definecolor{CPSgrey}{RGB}{204, 204, 204}
\definecolor{CPSorange}{RGB}{246,163,21}
\definecolor{CPSred}{RGB}{194,76,76}

%% file: utilities/tudacolors.tex
\definecolor{TUDa-0d}{cmyk/RGB/HTML}{0,0,0,.8/83,83,83/535353}
\definecolor{TUDa-0c}{cmyk/RGB/HTML}{0,0,0,.6/137,137,137/898989}
\definecolor{TUDa-0b}{cmyk/RGB/HTML}{0,0,0,.4/181,181,181/B5B5B5}
\definecolor{TUDa-0a}{cmyk/RGB/HTML}{0,0,0,.2/220,220,220/DCDCDC}

\definecolor{TUDa-1a}{cmyk/RGB/HTML}{.7,.4,0,0/93,133,195/5D85C3}
\definecolor{TUDa-2a}{cmyk/RGB/HTML}{0.8,.2,0,0/0,156,218/009CDA}
\definecolor{TUDa-3a}{cmyk/RGB/HTML}{0.7,0,.5,0/80,182,149/50B695}
\definecolor{TUDa-4a}{cmyk/RGB/HTML}{.4,0,.8,0/175,204,80/AFCC50}
\definecolor{TUDa-5a}{cmyk/RGB/HTML}{.2,0,.8,0/221,223,72/DDDF48}
\definecolor{TUDa-6a}{cmyk/RGB/HTML}{0,.1,.7,0/255,224,92/FFE05C}
\definecolor{TUDa-7a}{cmyk/RGB/HTML}{0,.3,.8,0/248,186,60/F8BA3C}
\definecolor{TUDa-8a}{cmyk/RGB/HTML}{0,.6,.8,0 /238,122,52/EE7A34}
\definecolor{TUDa-9a}{cmyk/RGB/HTML}{0,.8,.7,0/233,80,62/E9503E}
\definecolor{TUDa-10a}{cmyk/RGB/HTML}{.2,.9,0,0/201,48,142/C9308E}
\definecolor{TUDa-11a}{cmyk/RGB/HTML}{.6,.8,0,0/128,69,151/804597}
\definecolor{TUDa-1b}{cmyk/RGB/HTML}{1,.6,0,0/0,90,169/005AA9}
\definecolor{TUDa-2b}{cmyk/RGB/HTML}{1,.3,0,0/0,131,204/0083CC}
\definecolor{TUDa-3b}{cmyk/RGB/HTML}{1,0,.6,0/0,157,129/009D81}
\definecolor{TUDa-4b}{cmyk/RGB/HTML}{.5,0,1,0/153,192,0/99C000}
\definecolor{TUDa-5b}{cmyk/RGB/HTML}{.3,0,1,0/201,212,0/C9D400}
\definecolor{TUDa-6b}{cmyk/RGB/HTML}{0,.2,1,0/253,202,0/FDCA00}
\definecolor{TUDa-7b}{cmyk/RGB/HTML}{0,.4,1,0/245,163,0/F5A300}
\definecolor{TUDa-8b}{cmyk/RGB/HTML}{0,.7,1,0/236,101,0/EC6500}
\definecolor{TUDa-9b}{cmyk/RGB/HTML}{0,1,.9,0/230,0,26/E6001A}
\definecolor{TUDa-10b}{cmyk/RGB/HTML}{.4,1,0,0/166,0,132/A60084}
\definecolor{TUDa-11b}{cmyk/RGB/HTML}{.7,1,0,0/114,16,133/721085}
\definecolor{TUDa-1c}{cmyk/RGB/HTML}{1,.7,.2,0/0,78,138/004E8A}
\definecolor{TUDa-2c}{cmyk/RGB/HTML}{1,.5,.2,0/0,104,157/00689D}
\definecolor{TUDa-3c}{cmyk/RGB/HTML}{1,.2,.6,0/0,136,119/008877}
\definecolor{TUDa-4c}{cmyk/RGB/HTML}{.6,.1,1,0/127,171,22/7FAB16}
\definecolor{TUDa-5c}{cmyk/RGB/HTML}{.4,.1,1,0/177,189,0/B1BD00}
\definecolor{TUDa-6c}{cmyk/RGB/HTML}{.2,.3,1,0/215,172,0/D7AC00}
\definecolor{TUDa-7c}{cmyk/RGB/HTML}{.2,.5,1,0/210,135,0/D28700}
\definecolor{TUDa-8c}{cmyk/RGB/HTML}{.2,.8,1,0/204,76,3/CC4C03}
\definecolor{TUDa-9c}{cmyk/RGB/HTML}{.3,1,.9,0/185,15,34/B90F22}
\definecolor{TUDa-10c}{cmyk/RGB/HTML}{.5,1,.3,0/149,17,105/951169}
\definecolor{TUDa-11c}{cmyk/RGB/HTML}{.8,1,.2,0/97,28,115/611C73}
\definecolor{TUDa-1d}{cmyk/RGB/HTML}{1,.9,.3,0/36,53,114/243572}
\definecolor{TUDa-2d}{cmyk/RGB/HTML}{1,.7,.4,0/0,78,115/004E73}
\definecolor{TUDa-3d}{cmyk/RGB/HTML}{1,.4,.7,0/0,113,94/00715E}
\definecolor{TUDa-4d}{cmyk/RGB/HTML}{.7,.3,1,0/106,139,55/6A8B22}
\definecolor{TUDa-5d}{cmyk/RGB/HTML}{.5,.2,1,0/153,166,4/99A604}
\definecolor{TUDa-6d}{cmyk/RGB/HTML}{.4,.4,1,0/174,142,0/AE8E00}
\definecolor{TUDa-7d}{cmyk/RGB/HTML}{.3,.6,1,0/190,111,0/BE6F00}
\definecolor{TUDa-8d}{cmyk/RGB/HTML}{.4,.8,1,0/169,73,19/A94913}
\definecolor{TUDa-9d}{cmyk/RGB/HTML}{.5,1,.9,0/156,28,38/961C26}
\definecolor{TUDa-10d}{cmyk/RGB/HTML}{.7,1,.5,0/115,32,84/732054}
\definecolor{TUDa-11d}{cmyk/RGB/HTML}{.9,1,.3,0/76,34,106/4C226A}

%% file: sections/introduction.tex
\section{Introduction}
\label{sec:intro}

Numerical simulation has become an essential tool in engineering and applied sciences, reducing overall design time, and enabling detailed analysis across a wide range of applications, such as solid mechanics \cite{baiges2017,codony2019,alzatecobo2025, guarino2023}, fluid dynamics \cite{munthe2000,bazilevs2007, hamada2023}, and electric machines \cite{merkel2021a,kapidani2022,bontinck2018b}. However, geometry preprocessing still constitutes a significant portion of the overall simulation process. The finite element method (FEM) represents one of the cornerstones of numerical simulation and is widely used across a broad range of engineering applications. Nonetheless, it suffers from limitations related to the approximation of geometric representations, particularly in the presence of curved boundaries and complex features. To address this issue, isogeometric analysis (IGA) was introduced to establish a tighter integration between computer-aided design (CAD) and finite element simulations \cite{hughes2005, cottrell2009}, by employing Non-Uniform Rational B-Splines (NURBS) for the simultaneous representation of both geometry and solution fields \cite{piegl1997a}.
Despite these advances, geometric preprocessing remains a time-consuming step, especially for complex domains. Generation of body-fitted finite element meshes typically requires substantial manual intervention. 
In the context of IGA, conformal multi-patch discretizations must be employed \cite{pauley2015isogeometric}. 
This limits the degree of automation achievable in the overall simulation workflow, also in view of upcoming agent-based and large language model-empowered computer-aided engineering \cite{guoLLM-CAE2026}. 

To mitigate the meshing effort, immersed boundary methods (IBM) provide an alternative approach to body-fitted, conformal meshes. A large number of methods adopt this concept, including the level-set method \cite{peskin1972,osher1988}, marker-and-cell methods \citep{harlow1965, griebel1998numerical}, volume-of-fluid methods \citep{hirt1981,pilliod2004}, cut-cell methods \citep{1988nmfd.proc..375L}, fictitious domain and penalty methods \citep{ramiere2007, bishop2003}, the eXtended finite element method \cite{sukumar2001, gerstenberger2008, becker2011, schmidt2023}, CutFEM \citep{burman2015}, and the finite cell method \citep{parvizian2007}. 
However, the main challenge with such approaches lies in accurately resolving the geometry, particularly the interfaces between domains, either with adaptive discretizations or just the integration scheme, which are both also challenging. Moreover, Dirichlet boundary conditions must typically be imposed weakly while maintaining stability of the formulation.

In the context of magnetostatics and electric machine analysis, simulations typically involve multiple computational domains composed of different materials with strong material property contrasts, e.g., of the magnetic permeability of air vs.\ iron. Consequently, the meshing process becomes increasingly challenging, as material interfaces must be explicitly identified and accurately resolved. When multi-patch discretizations are employed, this results in a large number of patches, as the patches must be topologically square and the spline discretizations conformal across interfaces \cite{pauley2015isogeometric,wiesheu2025,komann2024a}. Constructing these patches, enforcing conformity conditions, and assembling them into a multi-patch mesh constitutes a tedious and time-consuming process, which may require several hours or even days of manual labor to complete for geometries with complex features. 
Alternatively, to loosen at least the conformity requirement, Nitsche's method \cite{nitsche1971} can be used to couple multi-patches in a weak manner \cite{nguyen2014}. However, this can be applied to domains topologically equivalent to a square, a condition that is not met in complex geometries. Hence, such a technique eliminates the conformity condition, but large number of patches will be needed to represent a complex geometry. Furthermore, when rotational motion is considered, as in rotor–stator configurations, the coupling of interfaces between rotating and stationary subdomains becomes particularly challenging, since maintaining mesh conformity typically necessitates repeated re-meshing of the rotating regions \cite{razek2000a,bontinck2018b}.

Multiple methods have addressed the coupling of rotating subdomains effectively by weakly coupling the interface between the rotor-stator subdomains. Mortar-type methods \cite{bernardi1989nna, bernardi1994new} were applied to Maxwell’s equations, including non-overlapping mortar formulations \cite{belgacem1999, razek2000a, benbelgacem2001}. For overlapping subdomains, mortar-based techniques have been applied to coupled magneto-mechanical problems \cite{rapetti2010a}. A local discontinuous Galerkin method was proposed to treat the air-gap region in electric machine simulations \cite{alotto2002a}. In addition, interface treatment based on Lagrange multipliers has been investigated in \cite{lange2010a, lange2012a, boehmer2013a}, adopting non-conformal mapping approaches in \cite{boy2018a, desenfans2024a}. Beyond Lagrange multiplier-based techniques, an implicit boundary finite element method was proposed in \cite{zhang2010a} for magnetostatic analysis; in this method, material interfaces are handled without explicitly conforming meshes. In the context of IGA for electric machines, harmonic rotor-stator coupling techniques \cite{degersem2004} have been combined with mortar methods \cite{bontinck2018b}. In these approaches, the rotor and stator are constructed independently using conforming multi-patch discretizations that are subsequently coupled weakly through harmonic basis functions. However, while these methods effectively address the interface coupling between rotating and stationary domains, they generally still require mesh conformity within each subdomain (e.g., within the rotor and stator regions). This limits their applicability to complex geometries and reduces the overall level of automation.

To overcome challenges related to geometric pre-processing, patch conformity, and the accurate resolution of interfaces between geometric patches, the concept of the immersed boundary conformal method (IBCM) was introduced \cite{wei2021,guarino2024a}. The method was applied to linear elasticity \cite{wei2021}, contact problems \cite{lapina2024a}, and linear shell theory \cite{guarino2024a}. In this concept, multiple non-conforming patches are coupled weakly by means of Nitsche’s method, thereby avoiding the need for fully matching discretizations at interfaces. For integration over trimmed elements and interfaces, a high-order quadrature rule is employed. The approach can be further extended by surrounding the non-conformal patches with conformal boundary layers, which enables the strong imposition of Dirichlet boundary conditions, and smooths the transition between different material domains.

In the present work, we transfer three different concepts to two-dimensional magnetostatics, namely: \emph{fully immersed} isogeometric analysis \cite{elbadry2025} , the \emph{union with non-conformal patches} (``Union NC''), and the \emph{union with conformal layers} (``Union CL'') \cite{wei2021,guarino2024a}.
We aim to establish a computational framework, in which complex geometries involving multiple materials can be represented using a minimal number of patches, without enforcing discretization conformity between them. In this way, strict multi-patch and conformal meshing requirements are circumvented while maintaining solution accuracy. The performance of the proposed approaches is assessed through a series of benchmark problems that feature, in contrast to previous works on heat diffusion or linear elasticity \cite{wei2021,guarino2024a}, multiple material subdomains with challenges such as discontinuous material coefficients, abrupt variations in current density across subdomain interfaces, and a remanence source term that involves the gradient of the test functions.

The remainder of this manuscript is organized as follows: \cref{sec:magnetostatic} introduces the governing equations of magnetostatics as foundational model, the bilinear form, and the usage of Nitsche's method for weakly imposition of essential boundary conditions and multi-domain coupling. A brief overview of isogeometric finite element analysis is presented in \cref{sec:iga}, in addition to the discretized weak form. \cref{sec:ibcm} then describes the three different numerical techniques, namely: \emph{fully immersed}, \emph{union with non-conformal patches}, and \emph{union with conformal layers}. Moreover, a note is given on the topic of stability and conditioning. In \cref{sec:results}, the proposed techniques are numerically evaluated through a series of benchmark examples, including configurations with impressed currents, material discontinuities, and a complex, industry-relevant multi-material geometry. Finally, \cref{sec:conclusion} concludes the paper and outlines directions for future research.

%% file: sections/magnetostatics.tex
\section{Fundamentals of magnetostatics}
\label{sec:magnetostatic}

\subsection{Basics of magnetostatics}

The mathematical modeling of electric machines is governed by Maxwell’s equations. However, in the case of low-frequency rotating machines, particularly when displacement currents and induced eddy currents can be neglected, the full set of Maxwell’s equations can be reduced to the magnetostatic formulation \cite{vanoost2014a,buffa2000}.

We consider a physical domain $\Omega \subset \mathbb{R}^3$, on which the magnetostatic formulation is given by:
\begin{subequations}
\label{eq:magnetostatics}
\begin{align}
    \nabla \times \tbH &= \tbJ,\\
    \nabla \cdot \tbB &= 0,
\end{align} 
\end{subequations}
where $\tbH$ denotes the magnetic field strength, $\tbB$ is the magnetic flux density, and $\tbJ$ represents the source current density.
In this work, we assume all materials to be isotropic and linear. Thus, the field strength $\tbH$ and the flux density $\tbB$ are linked through the constitutive equation:
\begin{equation}
\label{eq:constitutiveBH}
\tbB = \mu \tbH,
\end{equation}
where $\mu$ is the material's magnetic permeability. It is independent of the field strength and given as $\mu = \mu_r \mu_0$, with $\mu_r$ denoting the relative permeability of the material and $\mu_0 = 4\pi \cdot 10^{-7}\,\mathrm{H/m}$ being the permeability of air. 

For materials exhibiting remanent behavior, such as permanent magnets,  the constitutive relation reads \cite{griffiths2013}:
\begin{equation}
    \label{eq:constlawBHBr}
    \tbB = \mu \tbH + \tbB_r,
\end{equation}
where $\tbB_r$ denotes the permanent magnet's remanence.

To explicitly satisfy the divergence-free property of the magnetic flux density $\tbB$, we introduce the vector potential function $\tbA$ such that:
\begin{equation}
\label{eq: vectorpotential}
    \tbB = \nabla \times \tbA .
\end{equation}
Plugging \cref{eq: vectorpotential} and the constitutive relation \cref{eq:constlawBHBr} into \cref{eq:magnetostatics} yields:
\begin{equation}
\label{eq:curlcurlA}
    \nabla \times \left( \nu \left( \nabla \times \tbA - \tbB_{r} \right) \right) = \tbJ,
\end{equation}
where $\nu = \mu^{-1}$ is the magnetic reluctivity.


As commonly done for modeling rotating electric machines with uniform cross-sections, we restrict our analysis to two-dimensional geometries $\Omega\subset\mathbb{R}^2$ \cite{salon1995finite,vanoost2014a}.
Accordingly, for the two-dimensional formulation, only the out-of-plane, \( z \)-components of the magnetic vector potential \( A_z \) and the current density \( J_z \) are considered. Under these assumptions, the strong form of the magnetostatic problem reads:
\begin{equation}
    \label{eq:strongform2Dmagneto}
    -\nabla \cdot \left( \nu \left( \nabla A_z - \tbB_{r}^{\perp} \right) \right) = J_z \quad \text{in } \Omega, 
\end{equation}
where the in-plane remanence $\tbB_{r}^{\perp}$ is parameterized as $\tbB_{r}^{\perp} = B_r\,(-\sin{\theta_r}, \cos{\theta_r})^\top$ with $B_r$ being the remanence magnitude and $\theta_r$  the remanence direction.

To complete the boundary value problem, the following boundary conditions are imposed:
\begin{subequations}
\label{eq:BCsstrongform}
\begin{align}
A_z &= g \quad \text{on } \Gamma_{\mathrm D} \subset \partial\Omega, 
 \label{eq:BCD} \\
\mathbf n \cdot \big( \nabla A_z\big) &= 0 \quad \text{on } \Gamma_{\mathrm N} \subset \partial\Omega, 
\label{eq:BCN}
\end{align}
\end{subequations}
where $\mathbf n$ is the outward unit normal to the boundary and
$\Gamma_{\mathrm D} \cup \Gamma_{\mathrm N} = \partial\Omega$ with
$\Gamma_{\mathrm D} \cap \Gamma_{\mathrm N} = \emptyset$.

\subsection{Weak form of 2D magnetostatics}

Applying the standard procedure of multiplying the strong form of the partial differential equation \eqref{eq:strongform2Dmagneto} with a test function $v$, integrating over the domain $\Omega$, performing integration by parts, applying the divergence theorem, and exploiting the boundary conditions given in \cref{eq:BCsstrongform}, one obtains the weak form of the 2D magnetostatic problem:
Find $A_z \in {H}^{1}_g$ such that
\begin{equation}
\label{eq:finalweakform}
    \int_{\Omega} \nu\, \nabla v \cdot \nabla A_z \ \rmd \Omega = \int_{\Omega} v J_z  \, \rmd\Omega + \int_{\Omega} \nu \nabla v \cdot \tbB^{\perp}_r\ \rmd\Omega
    \quad \forall \, v \in {H}^{1}_0.
\end{equation}
Here, the boundary integral resulting from the integration by parts vanishes due to the imposition of the homogeneous Neumann condition on $\Gamma_{\mathrm N}$, which physically represents a boundary where the magnetic flux is directed purely in the normal direction. 
Furthermore, the trial and test function spaces are defined as:
\begin{subequations}
    \label{eq:trialspace}
\begin{align}    
    {H}^{1}_g = \{u \in H^{1}(\Omega) : u = g\quad \text{on } \, \Gamma_{\mathrm D}\}, \label{eq:trialspaceU}\\
    {H}^{1}_0 = \{v \in H^{1}(\Omega) : v = 0\quad \text{on } \, \Gamma_{\mathrm D}\}. \label{eq:trialspaceV}
\end{align}
\end{subequations}

Using from now on $u=A_z$, the weak form in \cref{eq:finalweakform} can be written in a more concise, abstract mathematical form as: Find $u \in {H}^{1}_g$ such that
\begin{equation}
    \label{eq:bilinearform_1}
    a(u,v) = l(v)\quad \forall\, v\in {H}^{1}_0,
\end{equation}
with the symmetric positive definite bilinear form:
\begin{equation} \label{eq:bilinearform_a}
    a(u,v) = \int_{\Omega} \nu\, \nabla v \cdot \nabla u \ \rmd \Omega,
\end{equation}
and the linear form:
\begin{equation} \label{eq:bilinearform_l}
    l(v) = \int_{\Omega} v J_z  \, \rmd\Omega + \int_{\Omega} \nu \nabla v \cdot \tbB^{\perp}_r\ \rmd\Omega.
\end{equation}
Note here that, as linear elliptic partial differential equations, the bilinear forms of heat diffusion, linear elasticity, and magnetostatics are similar, but magnetostatics poses the additional challenges that (i) usually subdomains with different material properties must be modeled, resulting in discontinuously varying reluctivity $\nu$, source current density $J_z$, and remanence $\tbB_r^{\perp}$ (which are often non-zero) and (ii) the remanence term in the linear form includes the gradient of the test function $\nabla v$.

\subsection{Weak imposition of Dirichlet boundary conditions}
\label{subsec:DBC_nitsche_trim}

In immersed boundary method settings, coincidence of the physical domain boundaries with the computational domain boundaries is not guaranteed. In such cases, strong imposition of the Dirichlet boundary conditions is not possible. There are different methods for weak imposition of the Dirichlet boundary conditions, such as the penalty method \cite{babuska1973}, Nitsche's method \cite{nitsche1971}, and Lagrange multiplier methods \cite{babuska1973a}. In this work, in case that strong imposition of Dirichlet boundary conditions $u=g\;\text{on}\;\Gamma_{\text{D}}$ is not possible, we apply Nitsche's method for preserving the consistency and symmetry of the weak form. 

Consequently, the bilinear form and linear form given in \crefrange{eq:bilinearform_1}{eq:bilinearform_l} are amended to incorporate the additional penalty, consistency, and symmetry terms: 
Find $u \in {H}^{1}$ such that
\begin{equation}
    \label{eq:coaxial_bilinear}
    a(u,v) + a_g(u,v)= l(v) + l_g(v)\quad \forall\, v \in {H}^{1},
\end{equation}
where
\begin{equation}
    \label{eq:coaxial_lhs_nitsche}
    \begin{aligned}
            a_g(u,v) =
            - \int_{\Gamma_{\text{D}}}\nu  (\nabla u \cdot \tbn)v\, \rmd\Gamma - \int_{\Gamma_{\text{D}}}\nu  (\nabla v\cdot \tbn) u\,\rmd \Gamma + \beta \int_{\Gamma_{\text{D}}}\nu\,  v u\, \rmd \Gamma,
    \end{aligned}
\end{equation}
and 
\begin{equation}
    \label{eq:coaxial_rhs_nitsche}
    l_g(v) = 
    - \int_{\Gamma_{\text{D}}}\nu  (\nabla v\cdot \tbn) g\,\rmd \Gamma + \beta \int_{\Gamma_{\text{D}}}\nu  v g\, \rmd \Gamma .
\end{equation}
Here, 
$\beta>0$ represents the stabilization parameter, which needs to be selected sufficiently large to impose the prescribed Dirichlet boundary condition.

\subsection{Weak coupling of non-conformal domains}
\label{subsec:nitsche}

For coupling non-conformal domains along their interface, e.g., in the setting of non-conformal multi-patch discretizations, we also employ Nitsche's method \cite{nitsche1971}, leveraging its consistent and symmetric formulation to weakly enforce interface continuity conditions.

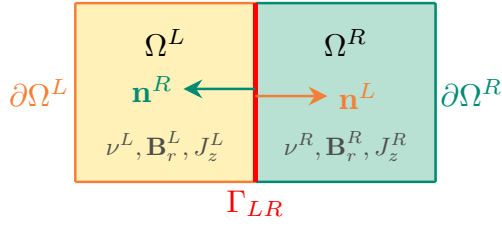
\begin{figure}[tb!]
    \centering
    \resizebox{!}{30mm}{\input{figures/nitsche_coupling}} 
    \caption{Coupling of two domains $\Omega^L$ and $\Omega^R$ across their common interface $\Gamma_{LR}$ via Nitsche's method}
    \label{fig:nitsche_coupling}
\end{figure}

For illustrating the concept in the two-dimensional magnetostatic setting, we consider the coupling of two disjunct domains $\Omega^L$ and $\Omega^R$, $\Omega^{L} \cap \Omega^{R} = \emptyset$,  across their interface $\Gamma_{LR}= \partial \Omega^{L} \cap \partial \Omega^{R}$, see \cref{fig:nitsche_coupling}.
Though discontinuities in material parameters, magnetic remanences or source current densities can arise at the interfaces, the normal component of the magnetic flux density $\tbB$ and the tangential component of the magnetic field strength $\tbH$ should be continuous \cite[Ch.~I5]{jackson2012classical}:
\begin{subequations}
    \label{eq:interfaceconditions}
    \begin{align}
        [[\tbB]] \cdot \tbn &= 0 
        \quad\Rightarrow& 
        \tbn \times [[A_z]] &= 0 &\text{on}\quad &\Gamma_{LR},\\
        \tbn \times [[\tbH]] &= 0 
        \quad\Rightarrow& 
        \tbn \times [[\nu \nabla \times A_z]] &= 0  &\text{on}\quad &\Gamma_{LR},
    \end{align}
\end{subequations}
with interface jumps on $\Gamma_{LR}$ defined as $[[\bullet]] = \bullet^R - \bullet^L$.
The outward unit normal to the interface $\Gamma_{LR}$ is defined as $\tbn = \tbn^{R}=-\tbn^L$, where $\tbn^{L}$ and $\tbn^{R}$ are the outward unit normals to $\partial \Omega^{L}$ and $\partial\Omega^{R}$, respectively.
In terms of the magnetic scalar potential $u=A_z$, these conditions translate as:
\begin{subequations}
    \label{eq:strongmagnetowithjumps}
    \begin{align}
            [[u]] = u^{L} - u^{R} = 0  \quad &\text{on}\quad \Gamma_{LR},
            \label{eq:jump_u}\\
           [[\nu\nabla u]]\cdot\tbn =
           \nu^{L} \nabla u^{L} \cdot \tbn + \nu^{R} \nabla u^{R} \cdot \tbn = 0 \quad &\text{on}\quad \Gamma_{LR} 
           \label{eq:jump_H},
    \end{align}
\end{subequations}
where $u^{R} = u|_{\Omega^{R}}$ and $u^{L} = u|_{\Omega^{L}}$ denote the potentials in their corresponding domains, and $\nu^L$ and $\nu^R$ are the magnetic reluctivities in $\Omega^L$ and $\Omega^R$, respectively.

To weakly enforce these two conditions using Nitsche's method, the bilinear form in \cref{eq:bilinearform_1} is adapted: Find $u \in {V}^{g}$ such that:
\begin{equation}
    \label{eq:bilinearnitschecouple}
    a(u,v) + a_n(u,v) = l(v), \quad \forall\, v \in {V}^{0}.
\end{equation}
Here, the global trial and test function space are defined as:
\begin{equation}
    \label{eq:modapproxspace}
    {V}^{\alpha} = \left\{v\in {H}^{1}(\Omega^L): v|_{\Gamma_D \cap \partial \Omega^{L}} = \alpha \right\} \oplus \left\{v\in {H}^{1}(\Omega^R): v|_{\Gamma_D\cap \partial \Omega^{R}} = \alpha  \right\},
\end{equation}
where $\alpha:\Gamma_D\to \bbR$ denotes the prescribed Dirichlet boundary condition on the joint Dirichlet boundary of $\partial\Omega^L$ and $\partial\Omega^R$, such that it is $\alpha = g$ for ${V}^{g}$ and $\alpha = 0$ for ${V}^{0}$, respectively.
More importantly, the additional bilinear form to weakly enforce \cref{eq:strongmagnetowithjumps} by Nitsche's method is given as:
\begin{equation}
\label{eq:bilinearnitscheexpand}
    \begin{aligned}
a_n(u,v)
=&  -\int_{\Gamma_{LR}} \big\{\nu \nabla u \cdot \mathbf{n} \big\}_{\gamma_2}\, [[v]] \, d\Gamma 
-\gamma_1 \int_{\Gamma_{LR}} \big\{\nu \nabla v \cdot \mathbf{n} \big\}_{\gamma_2}\, [[u]] \, d\Gamma\\
&
+\beta\,(h_R^{-1}+h_{L}) \int_{\Gamma_{LR}} \nu_{\text{max}}\, [[u]]\, [[v]] \, d\Gamma ,
\end{aligned}
\end{equation}
with:

\begin{equation}
\label{eq:avg_jump_defs}
\big\{\nu \nabla u \cdot \mathbf{n} \big\}_{\gamma_2}
:= \gamma_2 \big(\nu^R \nabla u^R \cdot \mathbf{n}\big)
+ (1-\gamma_2)\big(\nu^{L} \nabla u^{L} \cdot \mathbf{n}\big).
\end{equation}
This latter expression represents a weighted flux jump on the interface $\Gamma_{LR}$, compare \cref{eq:jump_H}.
The parameter $\gamma_2 \in [0,1]$ controls the weighting of the average operator and determines toward which domain the average is biased. 
In the present formulation, if we assumed that $\Omega^R$ is discretized in an untrimmed, boundary-conformal fashion, i.e., as an individual patch, whereas domain $\Omega^L$ is immersed and thus trimmed, then the choice $\gamma_2 = 1$ leads to a one-sided flux evaluation, thereby eliminating the symmetric averaging procedure. Consequently, the penalty term depends solely on the mesh size $h_R$ of the untrimmed ``master'' domain $\Omega^R$, rather than on an averaged mesh parameter that also considers the mesh size $h_L$ of the trimmed domain $\Omega^L$. 
In our formulation, it will be essential that the selected master domain remains untrimmed in order to preserve stability. More elaboration on the trimming process and the union approach will be discussed in \cref{subsec:trimsingle,subsec:unionofnonconf}.

In \cref{eq:bilinearnitscheexpand}, the parameter $\gamma_1 \in \left\{-1,0,1\right\}$ controls the symmetry of the formulation. Here, we set $\gamma_1 = 1$ to maintain symmetry of the bilinear form $a_n$. Furthermore, the stabilization parameter $\beta > 0$ in the last term of \cref{eq:bilinearnitscheexpand} is required to ensure coercivity. It must be chosen sufficiently large to enforce the  interface condition in a stable and consistent manner. 
Here, we select $\beta \in [10,10^4]$. Additionally, the reluctivity used in the stabilization term is chosen as the maximum value of the reluctivity of the two domains in order to ensure robustness of the formulation, $\nu_{\text{max}}=\max\{\nu^L,\nu^R\}$.

%% file: figures/nitsche_coupling.tex
\begin{tikzpicture}
    \fill [Goldenrod!40] (0, 0) rectangle (0.5,0.5);

    \fill [SeaGreen!40] (0.5, 0) rectangle (1,0.5);

\draw[Orange, very thin] (0, 0) -- (0.5,0);
\draw[PineGreen, very thin] (0.5, 0) -- (1,0);
\draw[PineGreen, very thin] (1, 0) -- (1,0.5);
\draw[PineGreen, very thin] (1, 0.5) -- (0.5,0.5);
\draw[Orange, very thin] (0.5, 0.5) -- (0,0.5);
\draw[Orange, very thin] (0, 0.5) -- (0,0);



\draw[red, thin] (.5,0) -- (.5, 0.5);


\node[scale = 0.25] at (0.5,-0.06) {\color{red} $\Gamma_{LR}$};

\node[scale = 0.25] at (0.25,0.39) {\color{black} $\Omega^{L}$};
\node[scale = 0.2] at (0.25,0.10) {\color{black!70} $\nu^{L}, \tbB_{r}^{L}, J_{z}^{L}$};

\node[scale = 0.25] at (0.75,0.39) {\color{black} $\Omega^{R}$};
\node[scale = 0.2] at (0.75,0.10) {\color{black!70} $\nu^{R}, \tbB_{r}^{R}, J_{z}^{R}$};

\node[scale = 0.25] at (-0.1,0.25) {\color{Orange} $\partial \Omega^{L}$};
\node[scale = 0.25] at (1.1,0.25) {\color{PineGreen} $\partial \Omega^{R}$};



\draw[Orange, ->, line width=0.2pt, 
      >={Stealth[length=1.5pt, width=1.5pt, inset=0.5pt]}] 
      (0.5,0.24) -- (0.7,0.24)
      node[right, scale=0.25] at (0.7,0.24) {\color{Orange} $\tbn^{L}$}; 

\draw[PineGreen, ->, line width=0.2pt, 
      >={Stealth[length=1.5pt, width=1.5pt, inset=0.5pt]}] 
      (0.5,0.26) -- (0.3,0.26)
      node[left, scale=0.25] at (0.3,0.26) {\color{PineGreen} $\tbn^{R}$}; 
\end{tikzpicture}

%% file: sections/iga.tex
\section{Isogeometric analysis}
\label{sec:iga}

This section introduces the basic concepts of isogeometric finite element analysis and fixes the notation used throughout the manuscript. Further details can be found in \cite{piegl1997a, hughes2005, cottrell2009}. First, we introduce B-splines, followed by an isogeometric finite element discretization of the magnetostatic weak form  \cref{eq:finalweakform}.

\subsection{B-splines}

Let $\{\xi_1,\ldots,\xi_{n+p+1}\}$ be a knot vector defining the parametric domain $\Omega_{\xi} = [\xi_1,\xi_{n+p+1}]$. The Cox–de Boor recursion formula is used to define the $n$ B-spline basis functions $N_{i}^{p} : \Omega_{\xi} \rightarrow \mathbb{R}$ of degree $p \ge 0$ as:
\begin{equation}
\label{eq:bspline}
\begin{aligned}
N_{i}^{0}(\xi) &=
  \begin{cases}
    1, & \text{if } \xi_i \le \xi < \xi_{i+1},\\
    0, & \text{otherwise},
  \end{cases} \\[1ex]
N_{i}^{p}(\xi) &=
  \frac{\xi - \xi_i}{\xi_{i+p} - \xi_i}\, N_{i}^{p-1}(\xi)
  + \frac{\xi_{i+p+1} - \xi}{\xi_{i+p+1} - \xi_{i+1}}\, N_{i+1}^{p-1}(\xi).
\end{aligned}
\end{equation}
Here, it is assumed that any quotient of the form $0/0$ is taken to be zero. The B-spline basis functions possess several important properties, including non-negativity, linear independence, and partition of unity. Moreover, their regularity can be controlled by the knot multiplicity; in particular, they are $(p-1)$-times continuously differentiable at interior knots of multiplicity one. To enforce the interpolation property at the boundary, open knot vectors are employed, for which the first and last knots are repeated $(p+1)$-times.

Bivariate B-splines basis functions $N_i(\bxi)$ with $\bxi=(\xi, \eta)$ can be constructed by the tensor product of two univariate B-spline basis functions as
\begin{equation}
\label{eq:multivariateBspline}
\begin{aligned}    
N_i(\xi, \eta) = N_{j,p}(\xi) \cdot N_{k,q}(\eta)\quad 
\text{for}\quad i =(j-1)l+k \quad\text{with}\quad\left\{\begin{array}{l}
j=1,\ldots,l,\\ k=1,\ldots,m
\end{array}\right. .
\end{aligned}
\end{equation}
Here, $N_{j,p}:\Omega_\xi\to\mathbb{R}$ and $N_{k,q}:\Omega_\eta\to\mathbb{R}$ are the $l$ and $m$ univariate B-spline basis functions of degree $p$ and $q$, respectively, as defined in \cref{eq:bspline}.
The $n=l\cdot m $ bivariate B-splines basis functions $N_i(\bxi):\hat\Omega\to\mathbb{R}$ are then defined on the two-dimensional parameter domain $\hat\Omega=\Omega_\xi\times\Omega_\eta$. For simplicity of the notation, $i=1,\ldots,n$ is used as a combined index and the dependence of $N_i$ on the degrees $p$ and $q$ is dropped. In the following, for the sake of clarity, we always set $p=q$. %
\subsection{Isogeometric discretization}

A two-dimensional B-spline surface $\Omega \ \subset \bbR^2$ is represented through the mapping $\mathcal{F} : \hat{\Omega} \to \Omega$ as a linear combination of $n$ bivariate B-splines basis functions $N_i(\boldsymbol{\xi})$ with control points $\tbx_i \in \bbR^2$: 

\begin{equation}
\label{eq:dispfield}
    \mathcal{F}(\boldsymbol{\xi}) = \tbx^{h}(\boldsymbol{\xi}) = \sum_{i=1}^{n} N_{i}(\boldsymbol{\xi})\, \tbx_{i}.
\end{equation}

The B-splines surface parametrization is not limited to describe the geometry in terms of the position vector $\tbx^h\in\Omega$, but also to discretize and approximate the field variable and test functions:
\begin{equation}\label{eq:vecpotentfield}
        A_z \approx u^{h}(\boldsymbol{\xi}) = \sum_{i=1}^{n} N_{i}(\boldsymbol{\xi})\, {u}_{i}, \qquad
v \approx v^{h}(\boldsymbol{\xi}) = \sum_{i=1}^{n} N_{i}(\boldsymbol{\xi})\, {v}_{i}.
\end{equation}
Here, $u_i,v_i\, \in \bbR$ are the $n$ control points of the discretized scalar potential field $u^h : \hat{\Omega} \to \bbR$ and the scalar test function $v^h : \hat{\Omega} \to \bbR$, respectively.

Under the isoparametric assumption, the field variables on the physical domain $\Omega$ are obtained by composing the discrete field with the inverse of the geometric mapping \cref{eq:dispfield}, e.g., $u^h(\tbx)=u^h(\bxi)\circ\mathcal{F}(\bxi)^{-1}$,
such that $u^h,v^h\in H^1(\Omega)$. 
Hence, the gradients with respect to the physical domain coordinates needs to be computed using the chain rule, e.g.,
\begin{equation}\label{eq:trafoJacobian}
    \nabla_{\tbx} u^h(\bxi) = \nabla_{\bxi} u^h(\bxi) \cdot (\nabla_{\bxi} \mathcal{F}(\bxi))^{-1},
\end{equation}
with

\begin{equation}\label{eq:parametricgradients}
    \nabla_{\bxi} u^h(\bxi) = \sum_{i=1}^{n} \nabla_{\bxi} N_{i}(\boldsymbol{\xi})\, {u}_{i}, \qquad
    \tbJ(\bxi) =\nabla_{\bxi} \mathcal{F}(\bxi) = \sum_{i=1}^{n} \nabla_{\bxi} N_{i}(\boldsymbol{\xi})\, \tbx_{i},
\end{equation}
where $\tbJ(\bxi)$ here refers to the Jacobian of the geometric parameterization (not to be confused with the source current density, which is also denoted as $\tbJ$ in \cref{eq:magnetostatics}, but reduced to the use of $J_z$ for 2D magnetostatics, see \cref{eq:strongform2Dmagneto,eq:finalweakform}).

\subsection{Isogeometric finite elements}

Now, the discretization introduced in \cref{eq:vecpotentfield} is substituted into the magnetostatic weak form given in \cref{eq:finalweakform}, such that the bilinear and linear form can be written as:
\begin{equation}\label{eq:LHSweak1}
\begin{aligned}
    a(u^h,v^h) 
    &= \int_{\Omega} \nu\, \nabla_{\tbx} v^h \cdot \nabla_{\tbx} u^h \, \mathrm{d}\Omega 
    = \int_{\hat{\Omega}} \nu \left(\nabla_{\bxi} v^h \cdot \tbJ^{-1}\right) \cdot \left(\nabla_{\bxi} u^h \cdot \tbJ^{-1} \right) \, \det \tbJ ~\rmd \hat{\Omega} \\
    &= \sum_{i=1}^n v_i \sum_{j=1}^n u_j \int_{\hat{\Omega}} \nu \left(\nabla_{\bxi} N_{i}\cdot \tbJ^{-1}\right)\cdot \,\left(\nabla_{\bxi} N_{j}\cdot \tbJ^{-1}\right) \, \det \tbJ~ \rmd \hat{\Omega} ,
\end{aligned}
\end{equation}
\begin{equation} \label{eq:RHSweak1}
    \begin{aligned}
        l(v^h) 
        &= \int_{\Omega} v^h\, J_z \, \mathrm{d}\Omega + \int_{\Omega} \nu\, \nabla v^h \cdot \tbB^{\perp}_r\, \rmd\Omega 
        = \int_{\hat{\Omega}} v^h\, J_z \,  \det \tbJ~ \rmd \hat{\Omega} + \int_{\hat{\Omega}} \nu \left(\nabla_{\bxi} v^h \cdot \tbJ^{-1} \right) \cdot \,\tbB^{\perp}_r \,\det \tbJ~ \rmd \hat{\Omega}\\
        &=  \sum_{i=1}^n v_i \int_{\hat{\Omega}} \left (N_i \, J_z  + \nu\left( \nabla_{\bxi} N_{i} \cdot \tbJ^{-1}\right) \cdot \tbB^{\perp}_r \right) \,  \det \tbJ ~\rmd \hat{\Omega}.
    \end{aligned}
\end{equation}
Here, the integration over the physical domain $\Omega$ and the evaluation of all shape functions and fields is pulled back to the parametric domain $\hat\Omega$, as $\Omega=\mathcal{F}(\hat\Omega)$.

Since $a(u^h,v^h)=l(v^h)$ must hold for all $v^h$, the $v_i$ can be chosen arbitrarily, which results in the linear system of equations $\tbK\tbu=\tbb$, where $\tbu=(u_1,\ldots,u_n)^\top\in\mathbb{R}^n$ is the sought vector of the control points of the scalar magnetic potential and the coefficients of the system matrix $\tbK$ and the RHS vector $\tbb$ are given as:
\begin{align}
    \label{eq:systemmatrix1}
    K_{ij} &= \int_{\hat{\Omega}} \nu \left(\nabla_{\bxi} N_{i}\cdot \tbJ^{-1}\right)\cdot \,\left(\nabla_{\bxi} N_{j}\cdot \tbJ^{-1}\right) \, \det \tbJ~ \rmd \hat{\Omega},\\
    \label{eq:rhsvector1}
    b_i &= \int_{\hat{\Omega}} \left (N_i \, J_z  + \nu\left( \nabla_{\bxi} N_{i} \cdot \tbJ^{-1}\right) \cdot \tbB^{\perp}_r \right) \,  \det \tbJ ~\rmd \hat{\Omega}.
\end{align}

Thanks to the piecewise polynomial definition and limited support of the B-spline shape functions $N_i$, for the numerical evaluation of the integrals arising from the discretized weak form, the parametric domain $\hat{\Omega}$ is partitioned into $n_e$ disjoint elements $\hat{\Omega}^e \subset \hat{\Omega}$ such that $\bigcup_{e=1}^{n_e} \hat{\Omega}^e = \hat{\Omega}$. These elements are defined by the knot spans of the non-repeated knot vectors $\{\xi_1,\dots,\xi_{l+p+1}\} \times \{\eta_1,\dots,\eta_{m+q+1}\}$, for example $\hat{\Omega}^e = [\xi_j,\xi_{j+1}] \times [\eta_k,\eta_{k+1}]$.

The corresponding elements in the physical domain are given by $\Omega^e = \mathcal{F}(\hat{\Omega}^e)$. 
Analogously to the isoparametric finite element method, a tensor-product Gauss-Legendre quadrature rule with $n_{qp} = (p+1)^2$ quadrature points is employed on each element \cite{hughes2000finite,hughes2005}. 
Ultimately, the numerical integration of \cref{eq:systemmatrix1,eq:rhsvector1} leads to:
\begin{align}
    \label{eq:systemmatrix}
    K_{ij} &\approx \sum_{e=1}^{n_e} \sum_{r=1}^{n_{qp}} w_r^e\, \nu(\bxi_{r}^{e}) \left(\nabla_{\bxi} N_{i}(\bxi_{r}^{e})\cdot \tbJ^{-1}(\bxi_{r}^{e})\right)\cdot \,\left(\nabla_{\bxi} N_{j}(\bxi_{r}^{e})\cdot \tbJ^{-1}(\bxi_{r}^{e})\right) \, \det \tbJ(\bxi_{r}^{e}),\\
    \label{eq:rhsvector}
    b_i &\approx \sum_{e=1}^{n_e} \sum_{r=1}^{n_{qp}} w_r^e \left (N_i(\bxi_{r}^{e}) \, J_z(\bxi_{r}^{e})  + \nu\left( \nabla_{\bxi} N_{i}(\bxi_{r}^{e}) \cdot \tbJ^{-1}(\bxi_{r}^{e})\right) \cdot \tbB^{\perp}_r(\bxi_{r}^{e}) \right) \det \tbJ(\bxi_{r}^{e}).
\end{align}
where $w_r^e$ are the weights and $\bxi_r^e$ are the points of a standard, $n_{qp}$-point Gauss-Legendre quadrature rule on the parameter element $\hat\Omega^e$. 
Note that in practice, the element-wise contributions to the coefficients $K_{ij}$ and $b_i$ are only evaluated on elements within the support of $N_i$ and $N_j$, as otherwise they are zero anyway.

Importantly, the Dirichlet boundary conditions contained in the definition of the trial and test functions spaces, see \cref{eq:trialspace}, still must be enforced on the linear system $\tbK\tbu=\tbb$.
In conformal IGA and FEM, this is typically done by elimination, resulting in a smaller, positive definite system to be solved.
Similarly, also anti-periodic Dirichlet boundary conditions can be applied.
If a weak enforcement of Dirichlet boundary conditions or of multi-domain coupling conditions using Nitsche's method is necessary, see \cref{subsec:DBC_nitsche_trim,subsec:nitsche}, the additional terms appearing in the formulations of the bilinear forms and linear forms must be discretized accordingly, which is not discussed in detail here. 

%% file: sections/ibcm.tex
\section{Immersed boundary conformal methods}
\label{sec:ibcm}

In conformal isogeometric finite element discretizations, the physical domain $\Omega$ is represented by a B-spline surface through the mapping defined in \cref{eq:dispfield}. However, for such a mapping $\mathcal{F}:\hat{\Omega} \to \Omega$ to exist, $\Omega$ must be topologically equivalent to a square. For most engineering applications with complex geometries and multi-material domains, this requirement is not satisfied. Within the framework of isogeometric analysis, decomposing the domain into a multi-patch geometry offers an alternative approach \cite{pauley2015isogeometric}. However, as noted previously, this process is generally cumbersome and requires significant manual effort, especially when conformity of the patches is to be maintained.

In this work, we adopt a hierarchy of three immersed isogeometric concepts to overcome this limitation: 
\begin{enumerate}
    \item \emph{Fully immersed}, see \cref{subsec:trimsingle} and \cite{elbadry2025}: Spline-based boundary representations (B-rep) of the material subdomains of the physical domain are constructed and subsequently immersed into a single background patch, the extended domain, over which the problem solution is discretized. 

    A boundary-conformal quadrature strategy is used to accurately resolve the integration over the different regions that are trimmed from the background patch through the boundary curves of the subdomains.
    \item \emph{Union with non-conformal patches} (``Union NC''), see \cref{subsec:unionofnonconf} and \cite{antolin2021,johansson2019}: To minimize the preprocessing effort, as far as desired and possible, each topologically square material subdomain of the physical domain is represented by an individual patch, which all overlay the extended domain. Thus, the overall domain consists of the union of the non-conformal patches with the trimmed regions on background patch. Nitsche's method is used to couple non-conformal  patches of the subdomains and the boundary-conformal quadrature strategy is used on the trimmed background patch.
    \item \emph{Union with conformal layers} (``Union CL''), see \cref{subsec:ibcmexpl} and \cite{guarino2024a,wei2021}: In addition to the \emph{union of non-conformal patches}, selected interfaces between individual patches of the subdomains and the background patch are wrapped with a conformal, overlaying patch that smoothens the transition between adjacent regions and is further trimmed from the background patch.
\end{enumerate}

In the following, these three approaches are presented in more detail.
Generally, we assume that the overall problem domain $\Omega$ is composed of several subdomains $\Omega_i$ such that $\Omega=\bigcup_i\Omega_i$ and $\Omega_i\cap\Omega_j=\emptyset$.
For the parameterization, the background patch $\Omega^\square$ is defined such that it contains the whole domain, $\Omega\subseteq\Omega^\square$.
Each subdomain $\Omega_i$ is parameterized either by immersion into the background patch $\Omega^\square$ and trimming of the parameter domain $\hat\Omega^\square$, where the ``trimmed region''  corresponding to the subdomain $\Omega_i$ is then denoted as $\hat\Omega^\square_i$, or by an individual patch with parameter domain $\hat\Omega_i^p$.
The subdomain itself is then either obtained by the mapping of the background patch, $\Omega_i=\mathcal{F}^\square(\hat\Omega_i^\square)$ or by the mapping of its individual patch, $\Omega_i=\mathcal{F}^p_i(\hat\Omega_i^p)$, respectively.
Without loss of generality, in the following we assume a physical domain consisting of two disjunct material subdomains, $\Omega=\Omega_1\cup\Omega_2$.

\subsection{Fully immersed}
\label{subsec:trimsingle}

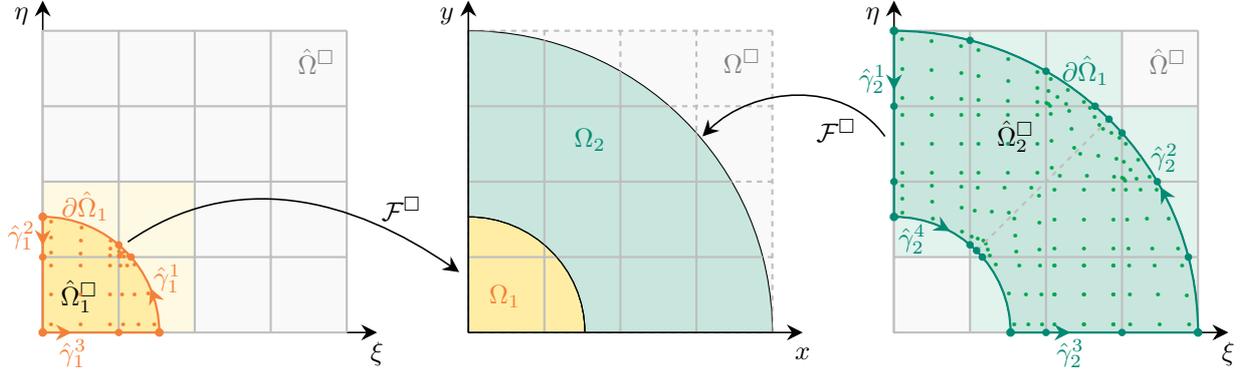
\begin{figure*}[t!]
    \centering
    \resizebox{!}{48mm}{\input{figures/trim_new_explain}} 
    \caption{
    Illustration of the \emph{fully immersed} concept. The physical domain $\Omega=\Omega_1\cup\Omega_2$ is immersed into an extended domain, the background patch $\Omega^{\square}$, here consisting of $4\times 4$ elements in its parameter domain $\hat\Omega^\square$. 
    B-reps of the subdomains $\Omega_{i}$ consisting of loops of boundary curves $\partial\Omega_i=\bigcup_k \gamma^k_{i}$ are used to determine the immersed trimmed regions $\hat{\Omega}_i^\square$. 
    To determine the boundary-conformal quadrature points, the trimmed elements of each region $\hat{\Omega}_i^\square$ are re-parameterized 
    \label{fig:singpatchtrimm}}
\end{figure*}

First, we adopt the immersed IGA concept with boundary conformal quadrature, see \cite{elbadry2025}. 
We immerse the physical domain $\Omega$ into a background domain $\mathcal{F}^{\square} :\hat{\Omega}^{\square} \to \Omega^{\square}$ such that $\Omega \subseteq \Omega^{\square}$. 
Each subdomain $\Omega_1$ and $\Omega_2$ is then represented in the parameter domain of the background patch by a region $\hat\Omega^{\square}_1$ and $\hat\Omega^{\square}_2$. Each region is identified in the parametric domain by the closed loop representing its B-rep, 
$\partial \hat{\Omega}_i = \bigcup_k \hat{\gamma}_{i}^{k}$, see \cite{antolin2019}, as illustrated in \cref{fig:singpatchtrimm}. 
Here, each curve $\hat{\gamma}_{i}^{k}={\mathcal{F}^{\square}}^{-1} ({\gamma}_{i}^{k})$ represents the pull-backward of the boundary curves of the physical subdomain $\partial\Omega_i$.
While this transformation may appear cumbersome, in practice it can be trivially realized by choosing $\mathcal{F}$ as the identity, i.e., $\hat\Omega^\square$ is identical to $\Omega^\square$, or as an affine transformation or scaling, which simply translates to an affine transformation or scaling of the control points of $\gamma_i^k$. 
For each region constructed through trimming the background patch $\hat\Omega^{\square}$ with the trimming loop $\partial \hat\Omega_i$, we accordingly restrict our mapping into that active region, hence the associated spline space over that region is defined as:
\begin{equation}
    \label{eq:restrictbspline}
    \mathcal{N}|_{\Omega_{i}^{\square}} = \mathrm{span}\Big\{N_{j}\circ \mathcal{F^{\square}}^{-1}(\tbx): j\in \mathcal{I},\, \mathrm{supp}(N_{j}) \cap \mathcal{F^{\square}}^{-1}\big(\Omega_{i}^{\square}\big)\neq \emptyset \Big\},
\end{equation}
where $\mathrm{supp}(N_{j})\subseteq\hat\Omega^\square$ is the support portion of the shape function $N_{j}(\bxi)$. This basis function can have high continuity across material interfaces, which can cause cross-talk between different subdomains, as discontinuities cannot be correctly represented. This will be further discussed and evaluated numerically in \cref{sec:results}.

To achieve accurate numerical integration over the physical domain $\Omega$, we perform the integration over each region $\hat{\Omega}^{\square}_i$ through the following procedure. Based on the representation of $\partial \hat{\Omega}_i$, the intersections between the B-rep and the vertical and horizontal knot lines of the parametric domain, i.e., the elements, are computed, see \cref{fig:singpatchtrimm} left and right.
This operation, referred to as \emph{slicing}, must be performed with sufficiently high precision, since subsequent operations, including the accuracy of the ensuing PDE analysis, are governed by this precision\footnote{For computing those intersections we use the \texttt{irit} library \citep{irit} that allows to compute them up to the desired precision. Note that most CAD kernels limit the maximum precision achievable in such operations. E.g., the open source kernel \texttt{Open CASCADE} \citep{occt} limits the maximum precision of such operations to $10^{-9}$, what would preclude the level of accuracy reached in the convergence analyses included in \cref{sec:results}.}. 
As a result of the slicing procedure, the elements of the background patch, $\hat{\Omega}^{\square} = \bigcup_e \hat\Omega^e$, are categorized into three distinct groups with respect to each region, as illustrated by different colors in \cref{fig:singpatchtrimm}, e.g. with respect to the region $\hat{\Omega}^{\square}_1$: \emph{active} where $| \hat\Omega^e \cap \hat{\Omega}^{\square}_1 | = | \hat\Omega^e |$, \emph{trimmed} are those with $0 < | \hat\Omega^e \cap \hat{\Omega}^{\square}_1 |  < |\hat\Omega^e|$ with an active portion $\hat\Omega^e \cap \hat{\Omega}^{\square}_1$, and \emph{non-active} elements with $| \hat\Omega^e \cap \hat{\Omega}^{\square}_1 | = 0$, where $| \bullet |$ represents the area.
The same procedure is subsequently applied to the subdomain $\Omega_2$, see \cref{fig:singpatchtrimm}. Its boundary $\partial \Omega_2$ is represented by a spline-based B-rep, and the slicing operation is performed accordingly. This results in a corresponding classification of elements into \emph{active}, \emph{trimmed}, and \emph{non-active} categories with respect to $\hat{\Omega}^{\square}_2$.

On active non-trimmed elements, the standard Gauss-Legendre quadrature rule is used to integrate the discretized weak form, c.f.\ \cref{eq:systemmatrix,eq:rhsvector}. For the active portions of the trimmed elements, that are represented again through B-reps, we adopt the boundary conformal quadrature method \citep{wei2021}. 
The technique depends on a re-parameterization of the trimmed elements and consists of three stages:
\begin{enumerate}
    \item \emph{Approximation.} For every trimmed element, the B-rep segment of the boundary of the physical domain is approximated by a Bézier curve with a polynomial degree $p$ of the same order as the spline discretization in \cref{eq:dispfield}. This allows to keep consistency under control and ensure optimal convergence for elliptic problems, as discussed in \citep{antolin2019}..

    \item \emph{Decomposition.} Every trimmed element is partitioned into active and non-active regions, distinguished by the varying colors in \cref{fig:singpatchtrimm}. When an integration cell does not conform to a standard topological triangle or quadrilateral, such as the two pentagonal active regions along the diagonal in \cref{fig:singpatchtrimm} (right), it is further subdivided along the trimming curve into triangular or quadrilateral sub-cells. Comprehensive details of this procedure are available in \citep{wei2021}.
    \item \emph{Re-parametrization.} In the final step, the integration cells are re-parameterized as B-spline surfaces of degree $p$ and quadrature points are then distributed within these cells according to a standard Gauss-Legendre rule with $n_{qp}=(p+1)^2$ points, see \cref{fig:singpatchtrimm}. 
\end{enumerate}
In this way, the boundaries $\partial\Omega_1, \partial\Omega_2$ and the integration over $\Omega_1,\Omega_2$ are resolved numerically exact up to the approximation errors that stems from the re-parameterization of the trimmed cell boundaries and the quadrature order employed.

It is important to note that the decomposition and re-parameterization of trimmed elements are performed solely for the purposes of quadrature and do not alter the discretization or the number of degrees of freedom.

When employing the \emph{fully immersed} technique, imposing Dirichlet boundary conditions $u = g$ on $\Gamma_{\text{D}} \subset \partial \Omega$, can be tricky if $\partial \Omega$ does not coincide with $\partial \Omega^\square$. In such case, Dirichlet boundary conditions cannot be imposed strongly. In our work, we impose Dirichlet boundary conditions via Nitsche's method \cite{nitsche1971}, see \cref{subsec:DBC_nitsche_trim}.

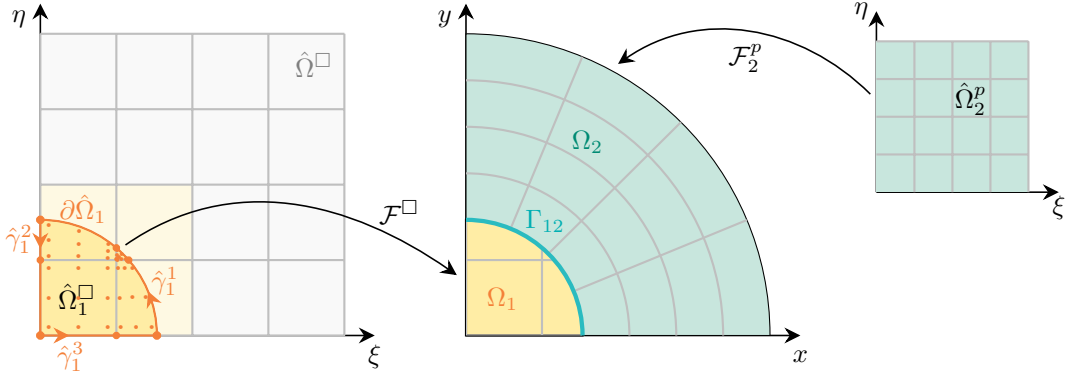
\begin{figure*}[t!]
    \centering
    \resizebox{!}{48mm}{\input{figures/union_new_explain}} 
    \caption{
    Illustration of the \emph{union with non-conformal patches} concept. 
    As in the \emph{fully immersed} method, the subdomain $\Omega_1$ is immersed into the background patch $\Omega^\square$.
    However, here $\Omega_2$ is parameterized by an individual patch with the mapping $\mathcal{F}_2^p$. 
    Thus, continuity at the interface $\Gamma_{12}=\partial\Omega_1\cap\partial\Omega_2$ is weakly enforced using Nitsche's method
    \label{fig:union}
    }
\end{figure*}

\subsection{Union with non-conformal patches}
\label{subsec:unionofnonconf}

To improve the accuracy of the discretization at interfaces between subdomains, but nonetheless avoid the effort of generating a conformal multi-patch representation with only topologically square patches, we adopt the concept of overlapping, non-conformal patches, see also \cite{antolin2021,wei2021,wei2023,larson2024,larson2024a}.

In \cref{fig:union}, we illustrate this method again for the physical domain consisting of two material subdomains, $\Omega=\Omega_1\cup\Omega_2$.
As in the \emph{fully immersed} method, the (not topologically square) subdomain $\Omega_1$ is immersed into the background patch $\Omega^\square$ as a trimmed region $\hat\Omega_1^\square$.  The trimming operation and the mapping back to the physical domain as $\Omega_1=\mathcal{F}^\square(\hat\Omega^\square_1)$ are conducted in accordance with \cref{subsec:trimsingle}. 
However, now the topologically square subdomain $\Omega_2$ is parameterized by its own individual patch with the spline mapping $\mathcal{F}_{2}^p : \hat{\Omega}_{2}^p \rightarrow \Omega_{2}$.
Thus, the active part of the background patch $\Omega^\square$ corresponding to $\Omega_{1}$ is combined with the overlaying patch of $\Omega_{2}$ to construct the overall computational domain as $\Omega = \Omega_{1} \cup \Omega_{2}$, see \cref{fig:union}. 

The fundamental concept of the \emph{union with non-conformal patches} consists of weakly coupling the active parts of the background patch and the overlapping patches across their interfaces.
Here, this means the continuity conditions of \cref{eq:strongmagnetowithjumps} have to be enforced on the interface $\Gamma_{12}=\partial\Omega_1\cap\partial\Omega_2$.
As described in \cref{subsec:nitsche}, this is realized by means of Nitsche’s method \cite{nitsche1971}.
Importantly, along the interface $\Gamma_{12}$, the discretization of the overlaying patch $\Omega_2$ is conformal with the interface, whereas $\Omega_{1}$ is a trimmed region of the background patch $\Omega^{\square}$. Thus, the weighting introduced in 
Nitsche's method is always put on the untrimmed, overlaying patches as master domains, i.e., here $\Omega_2$, see \cref{eq:avg_jump_defs}.
A comprehensive discussion on the construction of interface quadrature schemes used for the evaluation of interface integrals can be found in \cite{wei2021,antolin2021}.

Note further that while the interface conditions between the background patch and overlaying patches must be enforced weakly, the \emph{union with non-conformal patches} provides the opportunity to strongly enforce Dirichlet boundary conditions on the outer boundaries of the overlaying patches, e.g., here on $\Gamma_\text{D}\cap\partial\Omega_2$.

\subsection{Union with conformal layers}
\label{subsec:ibcmexpl}

In the \emph{union with non-conformal patches}, spurious effects may still occur at interfaces with strong material discontinuities, since the discretization of immersed subdomains are no conformal with the interface. To further improve the accuracy in such cases, we adopt the \emph{union with conformal layers} approach, which is called ``immersed boundary conformal method'' in \cite{wei2021,wei2023,guarino2024a}.

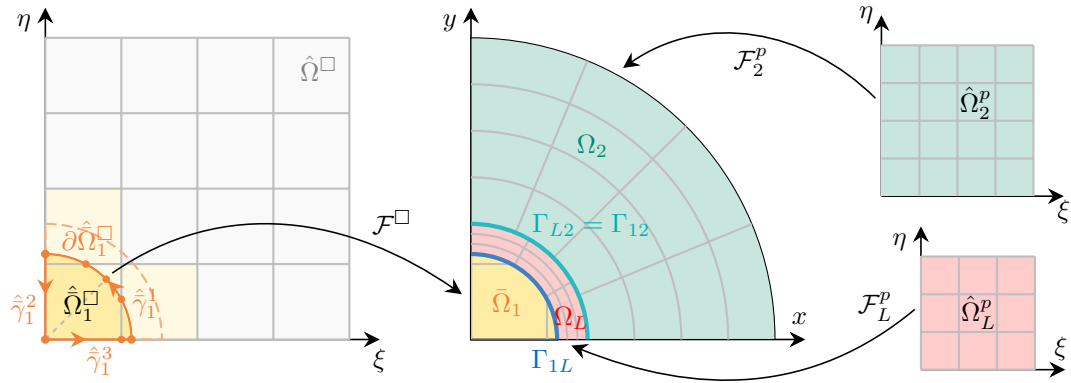
\begin{figure}[t!]
    \centering
    \resizebox{!}{48mm}{\input{figures/unionConf_new_explain}} 
    \caption{
    Illustration of the \emph{union with conformal layers} concept. 
    As in the \emph{union with non-conformal patches} method, the subdomain $\Omega_2$ is parameterized by an individual patch with the mapping $\mathcal{F}_2^p$.
    Beyond, here the subdomain $\Omega_1$ is partitioned into two parts: Along the interface with $\Omega_2$, the new subdomain $\Omega_L$ forms a geometrically conformal layer, which is offset from the boundary $\Gamma_{12}$ and also parameterized by an individual patch with the mapping $\mathcal{F}_L^p$. The cropped subdomain $\bar\Omega_1=\Omega_1\backslash \Omega_L$ remains immersed in the background domain as region $\hat{\bar\Omega}_1^\square$, which is defined by a modified set of boundary curves $\partial\hat{\bar\Omega}_1=\bigcup_k \hat{\bar\gamma}^k_{1}$.
    Continuity at the two interfaces $\Gamma_{1L}= \partial\bar\Omega_1\cap\partial\Omega_L$ and $\Gamma_{L2}= \partial\Omega_L\cap\partial\Omega_2$ is again weakly enforced using Nitsche's method  
    \label{fig:bconflayer}
    }
\end{figure}

As in the \emph{union with non-conformal patches} method, the subdomain $\Omega_2$ is parameterized by an individual patch with parameter domain $\hat\Omega_2^p$ and mapping $\mathcal{F}_2^p$.
However, as illustrated in \cref{fig:bconflayer}, the subdomain $\Omega_1$ is now partitioned into two parts: $\Omega_1=\bar\Omega_1\cup\Omega_L$. Along the interface with $\Omega_2$, the new subdomain $\Omega_L$ forms a geometrically conformal layer, which is generated by extruding $\Gamma_{12}$ into $\Omega_1$, yielding a new interface curve $\Gamma_ {1L}$.
This subdomain is also parameterized by an individual patch with the mapping $\mathcal{F}_{L}^p : \hat{\Omega}_{L}^p \rightarrow \Omega_{L}$ and maintains the same material parameters as $\Omega_1$. The cropped subdomain $\bar\Omega_1=\Omega_1\backslash \Omega_L$ remains immersed into the background domain as region $\hat{\bar\Omega}_1^\square$. It is defined by a modified set of boundary curves $\partial\hat{\bar\Omega}_1=\bigcup_k \hat{\bar\gamma}^k_{1}$, which is determined from the original boundary $\partial\Omega_1$ through the extrusion operation that establishes $\Omega_L$.

Here, continuity at the two interfaces $\Gamma_{1L}= \partial\bar\Omega_1\cap\partial\Omega_L$ and $\Gamma_{L2}= \partial\Omega_L\cap\partial\Omega_2$ is again weakly enforced using Nitsche's method, as described in \cref{subsec:nitsche}.
For $\Gamma_{1L}$, $\Omega_L$ must be the master domain, as it is discretized conformal with the interface, whereas $\bar\Omega_1$ is trimmed.
For $\Gamma_{L2}$, either patch could be the master, as both $\Omega_L$ and $\Omega_2$ are geometrically conformal.
We emphasize, however,  that conformity is understood purely in a geometric sense: the boundary layer $\Omega_{L}$ inherits the geometric shape of the interface $\Gamma_{12}$, but no constraint is imposed on the discrete function spaces. In particular, the discretization of $\Omega_{L}$ is not required to be conforming with that of $\Omega_{2}$, i.e., the degrees and number of elements along the interface direction may vary in each patch. 
Thus, a weak enforcement of the interface conditions is also applied on $\Gamma_{L2}$.

%


\subsection{Preconditioning and stability}
\label{subsec:precond}

The application of immersed methods in FEM/IGA is often associated with stability issues and ill-conditioning \cite{deprenter2023}. Although these issues are closely related, their treatment generally requires different techniques. 

Both issues are directly related to the small cut element problem, where elements are trimmed, leaving only a tiny active part. In the context of stability, this becomes evident when Dirichlet boundary conditions need to be imposed on a boundary that trims the background domain, resulting in tiny active parts. In such configurations, the weak imposition of Dirichlet boundary conditions via, e.g., Nitsche's method leads to an ill-posed numerical formulation, requiring a penalty parameter that tends to infinity to maintain coercivity of the formulation.
In the context of immersed and trimmed IGA, several unstable configurations were investigated, including cases where trimming produces high aspect ratio rectangular trimmed elements \cite{buffa2020}. In such cases, instabilities were observed to be associated with the flux term. In the framework of unfitted FEM, similar stability issues have been reported, and the reader is referred to \cite{deprenter2018b}. Additional discussions on stability aspects in immersed and trimmed spline methods can be found in \citep{hollig2001, elfverson2019, jonsson2019, marussig2017, marussig2018b}. However, in the present work, no additional stabilization technique is required in the union methods, as the weighted flux term in 
Nitsche's method is evaluated exclusively from the untrimmed domain, see \cref{eq:avg_jump_defs} in \cref{subsec:nitsche}, as discussed in \citep{antolin2021}.

Furthermore, ill-conditioning in immersed methods arises due to the small cut element problem, where some basis functions lose a significant portion of their support, leading to near-linear dependencies in the discrete system. Detailed discussions on this phenomenon can be found in \citep{lang2014, lehrenfeld2017, deprenter2017, deprenter2020}. In the present work, diagonal scaling in the form of Jacobi preconditioning is applied to improve the conditioning of the system matrix. Furthermore, a sparse direct solver is used in \texttt{MATLAB}, and no conditioning issues are observed in the numerical investigation.

%% file: figures/trim_new_explain.tex
\begin{tikzpicture}
\path[use as bounding box] (-1.3,-0.1) rectangle (2.2,1.1);
\def\R{1}    
\def\step{0.25} 


\fill[gray!5] (0,0) rectangle (1,1);
\draw[step=0.25, gray!50, dash pattern=on 0.5pt off 0.5pt, very thin]  (0,0) grid (1,1);
\node[scale = 0.25] at (0.9,0.9) {\color{gray} ${\Omega}^\square$};

\draw[fill = Goldenrod!50, draw=black, ultra thin] (0,0) -- (0:1/3+0.05) arc (0:90:1/3+0.05) -- cycle;

\draw[fill = SeaGreen!30, draw=black, ultra thin] (0:1/3+0.05) arc (0:90:1/3+0.05) -- (0,1) -- (90:1) arc (90:0:1) -- cycle;

\begin{scope}
    \clip (0,0) -- (0:\R) arc (0:90:\R) -- cycle;
    
    \draw[step=\step, gray!50, very thin] (0,0) grid (\R,\R);
\end{scope}

\draw[black, ultra thin] (0:1/3+0.05) arc (0:90:1/3+0.05);

\node[scale = 0.25] at (0.12,0.12) {\color{Orange} $\Omega_{1}$};
\node[scale = 0.25] at (0.4,0.64) {\color{PineGreen} $\Omega_{2}$};

\draw[->, line width=0.15pt, >={Stealth[length=1.5pt,width=1.5pt,inset=0.5pt]}] (0,0) -- (1.1,0) node[scale=0.25] at (1.1,-0.07) {$x$};

\draw[->, line width=0.15pt, >={Stealth[length=1.5pt,width=1.5pt,inset=0.5pt]}] (0,0) -- (0,1.1) node[scale=0.25] at (-0.07,1.05) {$y$};


\begin{scope}[xshift=-14mm]

\draw[->, line width=0.15pt, >={Stealth[length=1.5pt,width=1.5pt,inset=0.5pt]}] (0,0) -- (1.1,0) node[scale=0.25] at (1.1,-0.07) {$\xi$};

\draw[->, line width=0.15pt, >={Stealth[length=1.5pt,width=1.5pt,inset=0.5pt]}] (0,0) -- (0,1.1) node[scale=0.25] at (-0.07,1.05) {$\eta$};

\fill[gray!5] (0,0) rectangle (1,1);

\fill[Goldenrod!15] (0,0) rectangle (0.5,0.5);

\fill[Goldenrod!50] (0,0) -- (0:1/3+0.05) arc (0:90:1/3+0.05) -- cycle;



\draw[step=0.25, gray!50, very thin] (0,0) grid (1,1);

\begin{scope}[Orange,very thin,decoration={
    markings,
    mark=at position 0.25 with {\arrow{Stealth[length=1.5pt,width=1.5pt]}}}
    ] 
    \draw[postaction={decorate}] (0:1/3+0.05)  arc (0:90:1/3+0.05);
    \draw[postaction={decorate}]  (90:1/3+0.05) --  (0,0);
    \draw[postaction={decorate}] (0:0)  -- (0:1/3+0.05);
\end{scope} 

\node[scale = 0.25] at (0.12,0.12) {$\hat{\Omega}_{1}^\square$};
\node[scale = 0.25] at (0.9,0.9) {\color{gray} $\hat{\Omega}^\square$};

\node[scale = 0.25] at (0.41,0.18) {\textcolor{Orange}{$\hat{\gamma}^1_1$}};
\node[scale = 0.25] at (-0.07,0.32) {\textcolor{Orange}{$\hat{\gamma}^2_1$}};
\node[scale = 0.25] at (0.1,-0.07) {\textcolor{Orange}{$\hat{\gamma}^3_1$}};
\node[scale = 0.25] at (0.14,0.42) {\textcolor{Orange}{$\partial \hat{\Omega}_1$}};

\filldraw[Orange] (0:1/3+0.05) circle (0.20pt) node{};
\filldraw[Orange] (90:1/3+0.05) circle (0.20pt) node{};
\filldraw[Orange] (0,0) circle (0.20pt) node{};

\pgfplotstableread{data/ibcm_new_qp_inner.txt}\loadedtable

\pgfplotstablegetrowsof{\loadedtable}
\pgfmathtruncatemacro\loadedtablerowcount{\pgfplotsretval-1} 

\foreach \i in {0,...,\loadedtablerowcount}{
    \pgfplotstablegetelem{\i}{[index]0}\of{\loadedtable}
    \pgfmathsetmacro{\x}{\pgfplotsretval}
    \pgfplotstablegetelem{\i}{[index]1}\of{\loadedtable}
    \pgfmathsetmacro{\y}{\pgfplotsretval}
    \draw (\x,\y) node[scale=0.6]{\color{Orange} \tiny $\cdot$};
}
%

\filldraw[Orange] (0,0.383333) circle (0.15pt) node{};
\filldraw[Orange] (0,0.25) circle (0.15pt) node{};
\filldraw[Orange] (0.290498,0.25) circle (0.15pt) node{};
\filldraw[Orange] (0.25,0.290498) circle (0.15pt) node{};
\filldraw[Orange] (0.383333,0) circle (0.15pt) node{};

\filldraw[Orange] (0.0,0) circle (0.15pt) node{};
\filldraw[Orange] (0.25,0) circle (0.15pt) node{};
    
\end{scope}

\draw[->, line width=0.15pt, 
      >={Stealth[length=1.5pt, width=1.5pt, inset=0.5pt]}] 
      (-1.12,0.3) to[out=35, in=145] (-0.03,0.2)
      node[scale=0.25] at (-0.22,0.41) {$\mathcal{F}^{\square}$};


\begin{scope}[xshift=14mm]

\draw[->, line width=0.15pt, >={Stealth[length=1.5pt,width=1.5pt,inset=0.5pt]}] (0,0) -- (1.1,0) node[scale=0.25] at (1.1,-0.07) {$\xi$};

\draw[->, line width=0.15pt, >={Stealth[length=1.5pt,width=1.5pt,inset=0.5pt]}] (0,0) -- (0,1.1) node[scale=0.25] at (-0.07,1.05) {$\eta$};

\fill[gray!5] (0,0) rectangle (1,1);

\fill[SeaGreen!15] (0,0.25) rectangle (0.75,1);
\fill[SeaGreen!15] (0.25,0) rectangle (1,0.75);


\draw[fill = SeaGreen!30, draw=black, ultra thin] (0:1/3+0.05) arc (0:90:1/3+0.05) -- (0,1) -- (90:1) arc (90:0:1) -- cycle;

\draw[fill = SeaGreen!30, draw=black, ultra thin] (0:1/3+0.05) arc (0:90:1/3+0.05) -- (0,1) -- (90:1) arc (90:0:1) -- cycle;

\draw[step=0.25, gray!50, very thin] (0,0) grid (1,1);

\begin{scope}[PineGreen,very thin,decoration={
    markings,
    mark=at position 0.32 with {\arrow{Stealth[length=1.5pt,width=1.5pt]}}}
    ] 
    \draw[postaction={decorate}] (90:1/3+0.05)  arc (90:0:1/3+0.05);
    \draw[postaction={decorate}]  (0,1) -- (90:1/3+0.05) ;
    \draw[postaction={decorate}] (0:1)  arc (0:90:1);
    \draw[postaction={decorate}] (0:1/3+0.05)  -- (0:1);
\end{scope}

\node[scale = 0.25] at (-0.07,0.84) {\textcolor{PineGreen}{$\hat{\gamma}^1_2$}};
\node[scale = 0.25] at (0.89,0.58) {\textcolor{PineGreen}{$\hat{\gamma}^2_2$}};
\node[scale = 0.25] at (0.58,-0.07) {\textcolor{PineGreen}{$\hat{\gamma}^3_2$}};
\node[scale = 0.25] at (0.06,0.31) {\textcolor{PineGreen}{$\hat{\gamma}^4_2$}};
\node[scale = 0.25] at (0.63,0.87) {\textcolor{PineGreen}{$\partial \hat{\Omega}_1$}};
\filldraw[PineGreen] (1,0) circle (0.20pt) node{};
\filldraw[PineGreen] (0,1) circle (0.20pt) node{};
\filldraw[PineGreen] (0:1/3+0.05) circle (0.20pt) node{};
\filldraw[PineGreen] (90:1/3+0.05) circle (0.20pt) node{};

\node[scale = 0.25] at (0.4,0.65) {$\hat{\Omega}_{2}^\square$};
\node[scale = 0.25] at (0.9,0.9) {\color{gray} $\hat{\Omega}^\square$};

\filldraw[PineGreen] (0,1) circle (0.15pt) node{};
\filldraw[PineGreen] (0.25,0.968246) circle (0.15pt) node{};
\filldraw[PineGreen] (0.50,0.866025) circle (0.15pt) node{};
\filldraw[PineGreen] (0.661438,0.75) circle (0.15pt) node{};
\filldraw[PineGreen] (0.75,0.661438) circle (0.15pt) node{};
\filldraw[PineGreen] (0.866025,0.50) circle (0.15pt) node{};
\filldraw[PineGreen] (0.968246,0.25) circle (0.15pt) node{};
\filldraw[PineGreen] (1,0) circle (0.15pt) node{};

\filldraw[PineGreen] (0,0.75) circle (0.15pt) node{};
\filldraw[PineGreen] (0,0.50) circle (0.15pt) node{};
\filldraw[PineGreen] (0,0.383333) circle (0.15pt) node{};

\filldraw[PineGreen] (0.290498,0.25) circle (0.15pt) node{};
\filldraw[PineGreen] (0.25,0.290498) circle (0.15pt) node{};
\filldraw[PineGreen] (0.383333,0) circle (0.15pt) node{};

\filldraw[PineGreen] (0.5,0) circle (0.15pt) node{};
\filldraw[PineGreen] (0.75,0) circle (0.15pt) node{};

\pgfplotstableread{data/ibcm_new_qp.txt}\loadedtable

\pgfplotstablegetrowsof{\loadedtable}
\pgfmathtruncatemacro\loadedtablerowcount{\pgfplotsretval-1} 

\foreach \i in {0,...,\loadedtablerowcount}{
    \pgfplotstablegetelem{\i}{[index]0}\of{\loadedtable}
    \pgfmathsetmacro{\x}{\pgfplotsretval}
    \pgfplotstablegetelem{\i}{[index]1}\of{\loadedtable}
    \pgfmathsetmacro{\y}{\pgfplotsretval}
    \draw (\x,\y) node[scale=0.6]{\color{Green} \tiny $\cdot$};
}
\draw[gray!50, dash pattern=on 0.5pt off 0.5pt, very thin] 
(0.707106,0.707106) -- (0.271529,0.271529);
\filldraw[PineGreen] (0.707106,0.707106) circle (0.15pt) node{};
\filldraw[PineGreen] (0.271529,0.271529) circle (0.15pt) node{};

\end{scope}

\draw[->, line width=0.15pt, 
      >={Stealth[length=1.5pt, width=1.5pt, inset=0.5pt]}] 
      (1.37,0.65) to[out=125, in=45] (0.77,0.65)
      node[scale=0.25] at (1.21,0.67) {$\mathcal{F}^{\square}$};

\end{tikzpicture}

%% file: figures/union_new_explain.tex
\begin{tikzpicture}

\def\R{1}        
\def\step{0.25}  
\path[use as bounding box] (-1.3,-0.1) rectangle (1.65,1.1);


\begin{scope}[xshift=-14mm]
\path[use as bounding box] (-0.12,-0.12) rectangle (1.05,1.05);
\def\R{1}    
\def\step{0.5} 

\draw[->, line width=0.15pt, >={Stealth[length=1.5pt,width=1.5pt,inset=0.5pt]}] (0,0) -- (1.1,0) node[scale=0.25] at (1.1,-0.07) {$\xi$};

\draw[->, line width=0.15pt, >={Stealth[length=1.5pt,width=1.5pt,inset=0.5pt]}] (0,0) -- (0,1.1) node[scale=0.25] at (-0.07,1.05) {$\eta$};

\fill[gray!5] (0,0) rectangle (1,1);

\fill[Goldenrod!15] (0,0) rectangle (0.5,0.5);

\draw[fill = Goldenrod!50, draw=black, ultra thin] (0,0) -- (0:1/3+0.05) arc (0:90:1/3+0.05) -- cycle;

\draw[step=0.25, gray!50, very thin] (0,0) grid (1,1);

\begin{scope}[Orange,very thin,decoration={
    markings,
    mark=at position 0.25 with {\arrow{Stealth[length=1.5pt,width=1.5pt]}}}
    ] 
    \draw[postaction={decorate}] (0:1/3+0.05)  arc (0:90:1/3+0.05);
    \draw[postaction={decorate}]  (90:1/3+0.05) --  (0,0);
    \draw[postaction={decorate}] (0:0)  -- (0:1/3+0.05);
\end{scope}

\node[scale = 0.25] at (0.12,0.12) {$\hat{\Omega}_{1}^\square$};
\node[scale = 0.25] at (0.9,0.9) {\color{gray} $\hat{\Omega}^\square$};

\node[scale = 0.25] at (0.41,0.18) {\textcolor{Orange}{$\hat{\gamma}^1_1$}};
\node[scale = 0.25] at (-0.07,0.32) {\textcolor{Orange}{$\hat{\gamma}^2_1$}};
\node[scale = 0.25] at (0.1,-0.07) {\textcolor{Orange}{$\hat{\gamma}^3_1$}};
\node[scale = 0.25] at (0.14,0.42) {\textcolor{Orange}{$\partial \hat{\Omega}_1$}};

\filldraw[Orange] (0:1/3+0.05) circle (0.20pt) node{};
\filldraw[Orange] (90:1/3+0.05) circle (0.20pt) node{};
\filldraw[Orange] (0,0) circle (0.20pt) node{};

\pgfplotstableread{data/ibcm_new_qp_inner.txt}\loadedtable

\pgfplotstablegetrowsof{\loadedtable}
\pgfmathtruncatemacro\loadedtablerowcount{\pgfplotsretval-1} 

\foreach \i in {0,...,\loadedtablerowcount}{
    \pgfplotstablegetelem{\i}{[index]0}\of{\loadedtable}
    \pgfmathsetmacro{\x}{\pgfplotsretval}
    \pgfplotstablegetelem{\i}{[index]1}\of{\loadedtable}
    \pgfmathsetmacro{\y}{\pgfplotsretval}
    \draw (\x,\y) node[scale=0.6]{\color{Orange} \tiny $\cdot$};
}
%

\filldraw[Orange] (0,0.383333) circle (0.15pt) node{};
\filldraw[Orange] (0,0.25) circle (0.15pt) node{};
\filldraw[Orange] (0.290498,0.25) circle (0.15pt) node{};
\filldraw[Orange] (0.25,0.290498) circle (0.15pt) node{};
\filldraw[Orange] (0.383333,0) circle (0.15pt) node{};

\filldraw[Orange] (0.0,0) circle (0.15pt) node{};
\filldraw[Orange] (0.25,0) circle (0.15pt) node{};

\end{scope}

\draw[->, line width=0.15pt, 
      >={Stealth[length=1.5pt, width=1.5pt, inset=0.5pt]}] 
      (-1.12,0.3) to[out=35, in=145] (-0.03,0.2)
      node[scale=0.25] at (-0.22,0.41) {$\mathcal{F}^{\square}$};



\draw[->, line width=0.15pt, >={Stealth[length=1.5pt,width=1.5pt,inset=0.5pt]}] (0,0) -- (1.1,0) node[scale=0.25] at (1.1,-0.07) {$x$};

\draw[->, line width=0.15pt, >={Stealth[length=1.5pt,width=1.5pt,inset=0.5pt]}] (0,0) -- (0,1.1) node[scale=0.25] at (-0.07,1.05) {$y$};

\draw[fill = Goldenrod!50, draw=black, very thin] (0,0) -- (0:1/3+0.05) arc (0:90:1/3+0.05) -- cycle;

\begin{scope}
    \clip (0,0) -- (0:\R) arc (0:90:\R) -- cycle;
    
    \draw[step=\step, gray!50, very thin] (0,0) grid (\R,\R);
\end{scope}

\draw[fill = SeaGreen!30, draw=black, ultra thin] (0:1/3+0.05) arc (0:90:1/3+0.05) -- (0,1) -- (90:1) arc (90:0:1) -- cycle;  
  
\def\n{4}
\foreach \i in {1,...,3} {
    \pgfmathsetmacro{\r}{1/3+0.05 + (\i/\n)*(1-1/3-0.05)}
    \draw[very thin, gray!50] (0:\r) arc (0:90:\r);
    \draw[very thin, gray!50] (\i*90/\n:1/3+0.05) -- (\i*90/\n:1);
}

\node[scale = 0.25] at (0.12,0.12) {\color{Orange} $\Omega_{1}$};
\node[scale = 0.25] at (0.4,0.64) {\color{PineGreen} $\Omega_{2}$};

\draw[draw=TealBlue!80, thin] (0:1/3+0.05) arc (0:90:1/3+0.05) 
node[scale=0.25] at (0.26,0.38) {\color{TealBlue} $\Gamma_{12}$};


\begin{scope}[xshift=1mm,yshift=1mm]

\draw[->, line width=0.15pt, >={Stealth[length=1.5pt,width=1.5pt,inset=0.5pt]}] (1.25,0.375) -- (1.85,0.375) node[scale=0.25] at (1.85,0.325) {$\xi$};

\draw[->, line width=0.15pt, >={Stealth[length=1.5pt,width=1.5pt,inset=0.5pt]}] (1.25,0.375) -- (1.25,0.975) node[scale=0.25] at (1.2,0.975) {$\eta$};

\fill[SeaGreen!30] (1.25,0.375) rectangle (1.75,0.875);
\draw[step=0.125, gray!50, very thin] (1.25,0.375) grid (1.75,0.875);

\node[scale=0.25] at (1.56,0.685) {$\hat{\Omega}_2^p$};

\end{scope}

\draw[->, line width=0.15pt, 
      >={Stealth[length=1.5pt, width=1.5pt, inset=0.5pt]}] 
      (1.33,0.8) to[out=135, in=35] (0.5,0.9)
      node[scale=0.25] at (0.92,0.92) {$\mathcal{F}_2^{p}$};

\end{tikzpicture}

%% file: figures/unionConf_new_explain.tex
\begin{tikzpicture}

\def\R{1}        
\def\step{0.25}  
\path[use as bounding box] (-1.3,-0.1) rectangle (1.65,1.1);


\begin{scope}[xshift=-14mm]
\path[use as bounding box] (-0.12,-0.12) rectangle (1.05,1.05);
\def\R{1}    
\def\step{0.5} 

\draw[->, line width=0.15pt, >={Stealth[length=1.5pt,width=1.5pt,inset=0.5pt]}] (0,0) -- (1.1,0) node[scale=0.25] at (1.1,-0.07) {$\xi$};

\draw[->, line width=0.15pt, >={Stealth[length=1.5pt,width=1.5pt,inset=0.5pt]}] (0,0) -- (0,1.1) node[scale=0.25] at (-0.07,1.05) {$\eta$};

\fill[gray!5] (0,0) rectangle (1,1);

\fill[Goldenrod!15] (0,0) rectangle (0.5,0.25);
\fill[Goldenrod!15] (0,0) rectangle (0.25,0.5);

\draw[fill = Goldenrod!50, draw=black, ultra thin] (0,0) -- (0:1/3-0.05) arc (0:90:1/3-0.05) -- cycle;

\draw[step=0.25, gray!50, very thin] (0,0) grid (1,1);

\draw[Orange!60,very thin, dash pattern=on 1pt off 0.5pt] (0:0)  -- (0:1/3+0.05)  arc (0:90:1/3+0.05) --  (0,0);

\begin{scope}[Orange,very thin,decoration={
    markings,
    mark=at position 0.48 with {\arrow{Stealth[length=1.5pt,width=1.5pt]}}}
    ] 
    \draw[postaction={decorate}] (0:1/3-0.05)  arc (0:90:1/3-0.05);
    \draw[postaction={decorate}]  (90:1/3-0.05) --  (0,0);
    \draw[postaction={decorate}] (0:0)  -- (0:1/3-0.05);
\end{scope}

\node[scale = 0.25] at (0.9,0.9) {\color{gray} $\hat{\Omega}^\square$};

\node[scale = 0.25] at (0.33,0.12) {\textcolor{Orange}{$\hat{\bar\gamma}^1_1$}};
\node[scale = 0.25] at (-0.07,0.10) {\textcolor{Orange}{$\hat{\bar\gamma}^2_1$}};
\node[scale = 0.25] at (0.18,-0.07) {\textcolor{Orange}{$\hat{\bar\gamma}^3_1$}};
\node[scale = 0.25] at (0.14,0.335) {\textcolor{Orange}{$\partial \hat{\bar{\Omega}}_1^\square$}};

\filldraw[Orange] (0:1/3-0.05) circle (0.20pt) node{};
\filldraw[Orange] (90:1/3-0.05) circle (0.20pt) node{};
\filldraw[Orange] (0,0) circle (0.20pt) node{};

\pgfplotstableread{data/ibcm_new_qp_inner.txt}\loadedtable


%

\draw[gray!50, dash pattern=on 0.5pt off 0.5pt, very thin] 
(0,0) -- (0.20,0.20);

\filldraw[Orange] (0.25,0.133333) circle (0.15pt) node{};
\filldraw[Orange] (0.20,0.20) circle (0.15pt) node{};
\filldraw[Orange] (0.13333,0.25) circle (0.15pt) node{};
\filldraw[Orange] (0.0,0) circle (0.15pt) node{};
\filldraw[Orange] (0.25,0) circle (0.15pt) node{};

\node[scale = 0.25] at (0.12,0.12) {$\hat{\bar{\Omega}}_{1}^\square$};

\end{scope}

\draw[->, line width=0.15pt, 
      >={Stealth[length=1.5pt, width=1.5pt, inset=0.5pt]}] 
      (-1.18,0.22) to[out=35, in=145] (-0.03,0.15)
      node[scale=0.25] at (-0.26,0.39) {$\mathcal{F}^{\square}$};



\draw[->, line width=0.15pt, >={Stealth[length=1.5pt,width=1.5pt,inset=0.5pt]}] (0,0) -- (1.1,0) node[scale=0.25] at (1.08,0.07) {$x$};

\draw[->, line width=0.15pt, >={Stealth[length=1.5pt,width=1.5pt,inset=0.5pt]}] (0,0) -- (0,1.1) node[scale=0.25] at (-0.07,1.05) {$y$};

\draw[fill = Goldenrod!50, draw=black, very thin] (0,0) -- (0:1/3-0.05) arc (0:90:1/3-0.05) -- cycle;

\begin{scope}
    \clip (0,0) -- (0:\R) arc (0:90:\R) -- cycle;
    
    \draw[step=\step, gray!50, very thin] (0,0) grid (\R,\R);
\end{scope}

\draw[fill = red!20, draw=black, ultra thin] (0:1/3-0.05) arc (0:90:1/3-0.05) -- (0,1/3+0.05) -- (90:1/3+0.05) arc (90:0:1/3+0.05) -- cycle;  
  
\def\n{3}
\foreach \i in {1,...,2} {
    \pgfmathsetmacro{\r}{1/3-0.05 + (\i/\n)*(2*0.05)}
    \draw[very thin, gray!50] (0:\r) arc (0:90:\r);
    \draw[very thin, gray!50] (\i*90/\n:1/3-0.05) -- (\i*90/\n:1/3+0.05);
}

\draw[fill = SeaGreen!30, draw=black, ultra thin] (0:1/3+0.05) arc (0:90:1/3+0.05) -- (0,1) -- (90:1) arc (90:0:1) -- cycle;  
  
\def\n{4}
\foreach \i in {1,...,3} {
    \pgfmathsetmacro{\r}{1/3+0.05 + (\i/\n)*(1-1/3-0.05)}
    \draw[very thin, gray!50] (0:\r) arc (0:90:\r);
    \draw[very thin, gray!50] (\i*90/\n:1/3+0.05) -- (\i*90/\n:1);
}

\draw[draw=TealBlue!80, thin] (0:1/3+0.05) arc (0:90:1/3+0.05) 
node[scale=0.25, right] at (0.17,0.38) {\color{TealBlue} $\Gamma_{L2}=\Gamma_{12}$};

\draw[draw=NavyBlue!80, thin] (0:1/3-0.05) arc (0:90:1/3-0.05) 
node[scale=0.25] at (0.27,-0.07) {\color{NavyBlue} $\Gamma_{1L}$};

\node[scale = 0.25] at (0.33,0.07) {\color{red}$\Omega_L$};
\node[scale = 0.25] at (0.12,0.12) {\color{Orange} $\bar\Omega_{1}$};
\node[scale = 0.25] at (0.4,0.64) {\color{PineGreen} $\Omega_{2}$};


\begin{scope}[xshift=1mm,yshift=1mm]

\draw[->, line width=0.15pt, >={Stealth[length=1.5pt,width=1.5pt,inset=0.5pt]}] (1.25,0.375) -- (1.85,0.375) node[scale=0.25] at (1.85,0.325) {$\xi$};

\draw[->, line width=0.15pt, >={Stealth[length=1.5pt,width=1.5pt,inset=0.5pt]}] (1.25,0.375) -- (1.25,0.975) node[scale=0.25] at (1.2,0.975) {$\eta$};

\fill[SeaGreen!30] (1.25,0.375) rectangle (1.75,0.875);
\draw[step=0.125, gray!50, very thin] (1.25,0.375) grid (1.75,0.875);

\node[scale=0.25] at (1.56,0.685) {$\hat{\Omega}_2^p$};

\end{scope}

\draw[->, line width=0.15pt, 
      >={Stealth[length=1.5pt, width=1.5pt, inset=0.5pt]}] 
      (1.33,0.8) to[out=135, in=35] (0.5,0.9)
      node[scale=0.25] at (0.92,0.92) {$\mathcal{F}_2^{p}$};


\begin{scope}[xshift=14.8mm,yshift=-1mm]
    
\draw[->, line width=0.15pt, >={Stealth[length=1.5pt,width=1.5pt,inset=0.5pt]}] (0,0) -- (0.475,0) node[scale=0.25] at (0.46,0.07) {$\xi$};

\draw[->, line width=0.15pt, >={Stealth[length=1.5pt,width=1.5pt,inset=0.5pt]}] (0,0) -- (0,0.475) node[scale=0.25] at (-0.07,0.425) {$\eta$};

\fill[red!20] (0,0) rectangle (0.375,0.375);
\draw[step=0.125, gray!50, very thin] (0,0) grid (0.375,0.375);

\node[scale=0.25] at (0.19,0.19) {$\hat{\Omega}_L^p$};

\end{scope}

\draw[->, line width=0.15pt, 
      >={Stealth[length=1.5pt, width=1.5pt, inset=0.5pt]}] 
      (1.46,0.1) to[out=-140, in=-22] (0.33,-0.02)
      node[scale=0.25] at (1.33,0.11) {$\mathcal{F}_L^{p}$};
    
\end{tikzpicture}

%% file: sections/results.tex
\section{Numerical results}
\label{sec:results}

In the following, we apply immersed IGA with the three aforementioned concepts to two-dimensional magnetostatic problems to demonstrate their efficiency and accuracy. The current work is implemented within the framework of the open-source \texttt{MATLAB} code \texttt{geoPDEs} \cite{vazquez2016,geopdes} and an in-house library based on \texttt{irit} \citep{irit} and \texttt{Open CASCADE} \citep{occt} for obtaining the boundary conformal quadrature points.

Throughout this manuscript, the ``$k$-version'' of IGA is adopted as the default refinement strategy, i.e., $k$-refinement is used by increasing the smoothness at internal knots along with the degree \cite{hughes2005}. Any deviation from this approach, such as the use of the FEM $p$-refinement, will be explicitly stated where applicable.

In the following, we evaluate the accuracy of the obtained numerical solutions $u^h$ against reference solutions $u^{ref}$, which could be either analytical or highly-refined multipatch IGA solutions, using the $L^2$-norm and the $H^1_s$-seminorm of the error, defined as follows:
\begin{equation}
    \label{eq:errors}
    \begin{aligned}
        \left\| u^h - u^{ref} \right\|_{L^2}^{2} &= \int_{\Omega} \left\| u^h - u^{ref} \right\|_{2}^{2}\, \rmd \Omega,\\
        \left\| u^h - u^{ref} \right\|_{H_{s}^{1}}^{2} &= \int_{\Omega} \left\| \nabla u^h - \nabla u^{ref} \right\|_{2}^{2}\,\rmd \Omega.
    \end{aligned}
\end{equation}

\subsection{Coaxial cable}

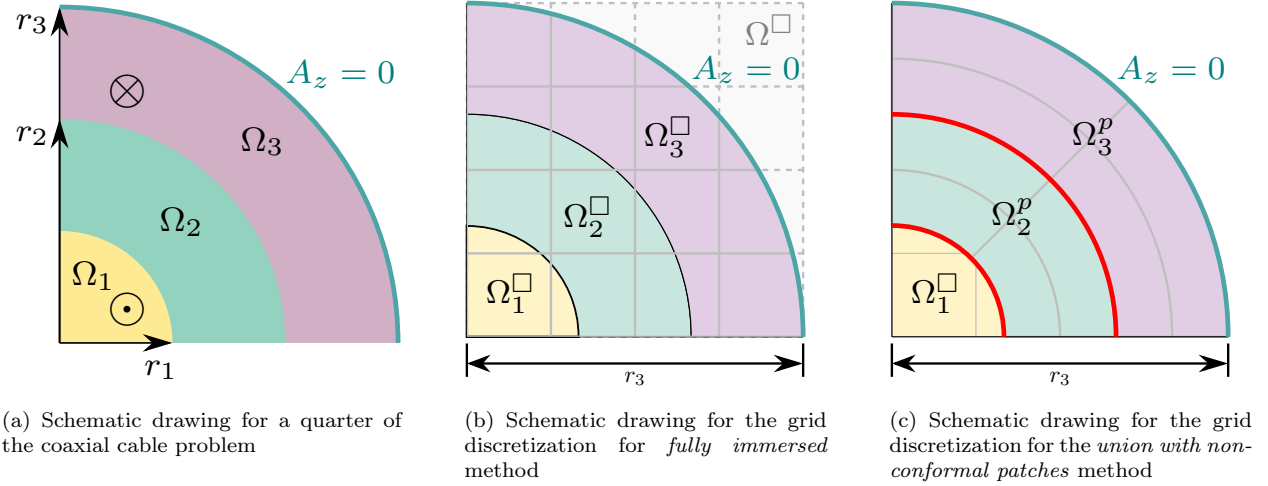
\begin{figure*}[bt!]
    \centering
    \begin{subfigure}[t]{0.32\linewidth}
        \centering
        \resizebox{!}{52mm}{\input{figures/coaxial_schem_quarter}} 
        \caption{Schematic drawing for a quarter of the coaxial cable problem\label{fig:coaxialschem:a}}
    \end{subfigure}
    \hfill
    \begin{subfigure}[t]{0.29\linewidth}
        \centering
        \resizebox{!}{52mm}{\input{figures/coaxial_trim_a}} 
        \caption{Schematic drawing for the grid discretization for \emph{fully immersed} method \label{fig:coaxialschem:b}}
    \end{subfigure}
    \hfill
    \begin{subfigure}[t]{0.29\linewidth}
        \centering
        \resizebox{!}{52mm}{\input{figures/coaxial_union_schem_a}}
        \caption{Schematic drawing for the grid discretization for the \emph{union with non-conformal patches} method \label{fig:coaxialschem:c}}
    \end{subfigure}
    \caption{Illustration of the coaxial cable problem setup 
    }
\end{figure*}

First, we consider a coaxial cable as benchmark problem, see \cref{fig:coaxialschem:a} and \cite{zhang2010a, griffiths2013, houssein2024} for similar studies. The geometry consists of three concentric cylindrical regions: an inner conducting core (gold), a surrounding insulating layer (green), and an outer conductor (orchid). 
Due to the double symmetry of the problem, we only solve quarter of the cable geometry.
The core has radius $r_1$, the insulator occupies the annular region $r_1 < r < r_2$, and the outer conductor extends for $r_2<r < r_3$. 
Here, the radii are specified as $r_1=\tfrac{1}{3},\, r_2=\tfrac{2}{3},\, r_3=1$.
An axial current flows in the core in the positive $z$-direction (out of plane) with magnitude $I_{1} = 1000$. The outer conductor carries an equal current in the opposite direction, $I_{3} = -I_{1}$. We assume a unified permeability for the three materials, equal to the one of air $\mu = \mu_0 = 4\pi \times 10^{-7}\, \mathrm{H/m}$. It is important to note that the remanence term in \cref{eq:coaxial_bilinear} is omitted, since all materials are assumed to be non-magnetic, i.e., $\mathbf{B}_r = 0$.  

A Dirichlet boundary condition is imposed on the outer radius of the cable $r=r_3$ with $A_z = 0$. No additional boundary conditions are needed at the left and bottom symmetry boundaries. There, the magnetic field is perpendicular to the surface, which implies the natural Neumann condition $\nabla A_z\cdot \tbn = 0$. This condition is inherently satisfied in the weak formulation, see \cref{eq:finalweakform,eq:bilinearform_1}. 

The analytical solution for the magnetic flux density is formulated in terms of the radial coordinate $r=\sqrt{x^2+y^2}$, see \cite{zhang2010a,griffiths2013}, as follows:
\begin{equation}
\label{eq:coaxial_exact}
    B_{\theta}(r) = \begin{cases}
    \frac{\mu_{0} I}{2 \pi r_{1}^{2}} r, & 0 \le r < r_1, \\[1.5ex] %
    \frac{\mu_{0} I}{2 \pi r}, & r_1 \le r < r_2, \\[1.5ex] %
    \frac{\mu_{0} I (r_{3}^{2}-r^{2})}{2 \pi r (r_{3}^{2}-r_{2}^{2})}, & r_2 \le r < r_3, \\[1.5ex] %
    0, & r \ge r_3.
\end{cases}
\end{equation}

The scalar potential $A_z=:u^{ref}$ can be obtained by integrating \cref{eq:coaxial_exact} with integration constants determined by continuity conditions.
\begin{equation}
    \label{eq:coaxial_exact_az}
    A_{z}(r) = - \int_0^r B_{\theta}(\rho) \, \rmd \rho.
\end{equation}

First, we solve the coaxial problem using the \emph{fully immersed} approach, see \cref{subsec:trimsingle}. The physical domain $\Omega$ is embedded into an extended square domain $\Omega^{\square} = [0,\,r_3]^2$. Similar to the example shown in \cref{fig:singpatchtrimm}, the present physical domain consists of multiple subregions. Accordingly, a separate trimming loop is defined for each subdomain $\Omega_i$. Each trimming loop is applied to the extended domain $\Omega^{\square}$, resulting in the identification of the corresponding active and trimmed elements associated with each subregion, in addition to the corresponding integration points for trimmed elements and, at each quadrature point on the boundary, the corresponding outward normal vector. 
Finally, we impose weakly the Dirichlet boundary condition $A_z = 0$ on the outer radius $r= r_3$  via Nitsche's method as this domain boundary does not coincide with the boundary $\partial \Omega^\square$. 
Furthermore, the integration domain is restricted exclusively to the physical domain $\Omega$, while all degrees of freedom associated solely with the fictitious domain $\Omega^\square \backslash \Omega$ (the gray region shown in \cref{fig:coaxialschem:b}) are excluded from the system. By only integrating over the active part of the trimmed elements, there will be some basis functions with very small support, which may result in an ill-conditioned system matrix. To alleviate this issue, a diagonal preconditioner is employed.

\begin{figure*}[t!]
    \centering
    \begin{subfigure}[b]{0.41\linewidth}
        \centering
	    \resizebox{!}{60mm}{\input{figures/coaxial_H1snorm_href}}
	    \caption{$H_{s}^{1}$-seminorm error over $h$}
        \label{fig:coaxial_error_h1s_helem}
    \end{subfigure}
    \begin{subfigure}[b]{0.58\linewidth}
        \centering
	    \resizebox{!}{60mm}{\input{figures/coaxial_L2norm_href}}
	    \caption{$L^2$-norm error over $h$}
        \label{fig:coaxial_error_l2_helem}
    \end{subfigure}
\vspace*{0.1ex}

    \begin{subfigure}[b]{0.41\linewidth}
        \centering
	    \resizebox{!}{60mm}{\input{figures/coaxial_H1snorm_sqrtdofs}}
	    \caption{$H_{s}^{1}$-seminorm error over $\sqrt{\text{DOFs}}$}
        \label{fig:coaxial_error_h1s_ndofs}
    \end{subfigure}
    \begin{subfigure}[b]{0.58\linewidth}
        \centering
	    \resizebox{!}{60mm}{\input{figures/coaxial_L2norm_sqrtdofs}}
	    \caption{$L^2$-norm error over $\sqrt{\text{DOFs}}$}
        \label{fig:coaxial_error_l2_ndofs}
    \end{subfigure}
    \caption{Comparison of convergence behaviors of the \emph{fully immersed} and \emph{union with non-conformal patches} methods over the element size $h$ and the average number of degrees of freedom $\sqrt{\text{DOFs}}$ for the coaxial cable problem 
    } \label{fig:coaxial_error_ndofs}
\end{figure*}
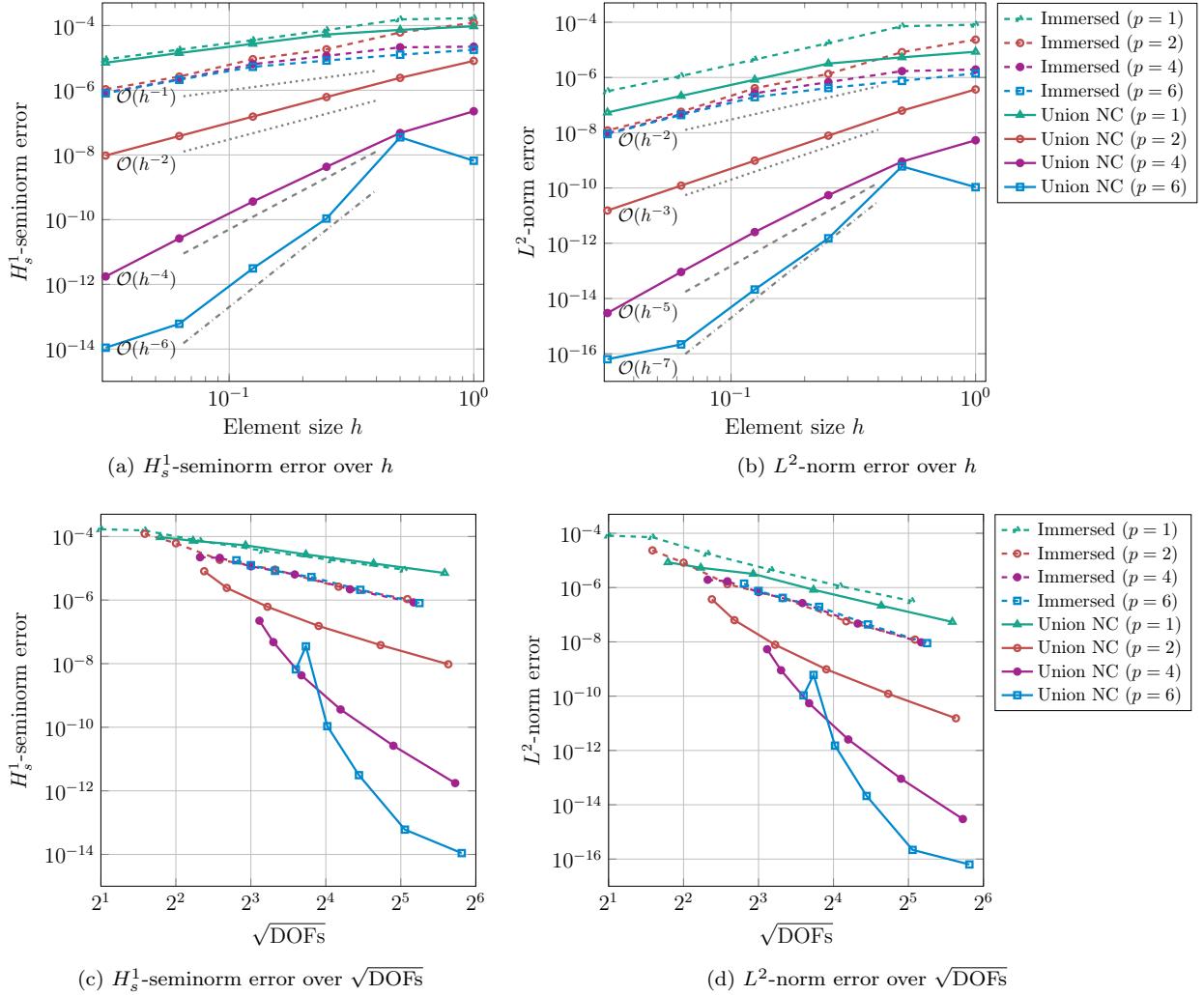

Second, we solve the problem using \emph{union with non-conformal patches} approach, following \cref{subsec:unionofnonconf}. In \cref{fig:coaxialschem:c}, the physical domain $\Omega$ is defined as the union of the overlapping patches, $\Omega = \Omega_{1}^\square \cup \Omega^p_2 \cup \Omega^p_3$, where $\Omega_{1}^{\square}$ denotes the active portion of the trimmed patch $\Omega_1$. Here, we impose the Dirichlet boundary condition $A_z = 0$ strongly via elimination.

For both methods, the \emph{fully immersed} and \emph{union with non-conformal patches}, we set the stabilization parameter in Nitsche's method to be $\beta\nu_{\text{air}} p / h$, where $\beta \in \{100,1000\}$ and $h$ denotes the smallest element size in the domain. 

The $h$-refinement analysis at different polynomial degrees in \cref{fig:coaxial_error_l2_helem,fig:coaxial_error_h1s_helem} shows optimal convergence rates for both the $L^2$-norm error and the $H_{s}^{1}$-seminorm error for the \emph{Union NC} method.
However, for the \emph{fully immersed} case, the $p$-convergence behavior is pre-asymptotic and hence the convergence rates for $p>1$ are not optimal. This can be attributed to the different inter-patch continuity properties. In the union formulation, only $C^0$ continuity is enforced weakly across the patch interfaces. By contrast, in the \emph{fully immersed} approach, both the solution and its gradients (i.e., the flux density) are represented by basis functions exhibiting $C^{p-1}$ continuity across the regions' interfaces, as we are solving on a single patch. This explains the similarity in accuracy for the case $p=1$, as the gradients are represented by non-smooth basis functions. This can be seen clearly in \cref{fig:coaxial_error_h1s_ndofs,fig:coaxial_error_l2_ndofs}, where the errors are nearly identical in terms of the $H^{1}_{s}$-seminorm and remain very close in the $L^2$-norm for the case $p=1$. Overall, the union approach provides higher accuracy for higher polynomial orders while maintaining a comparable number of degrees of freedom.

\begin{figure*}[t!]
    \centering
    \begin{subfigure}[b]{0.47\linewidth}
    \centering
    \includegraphics[width = 0.99\linewidth]{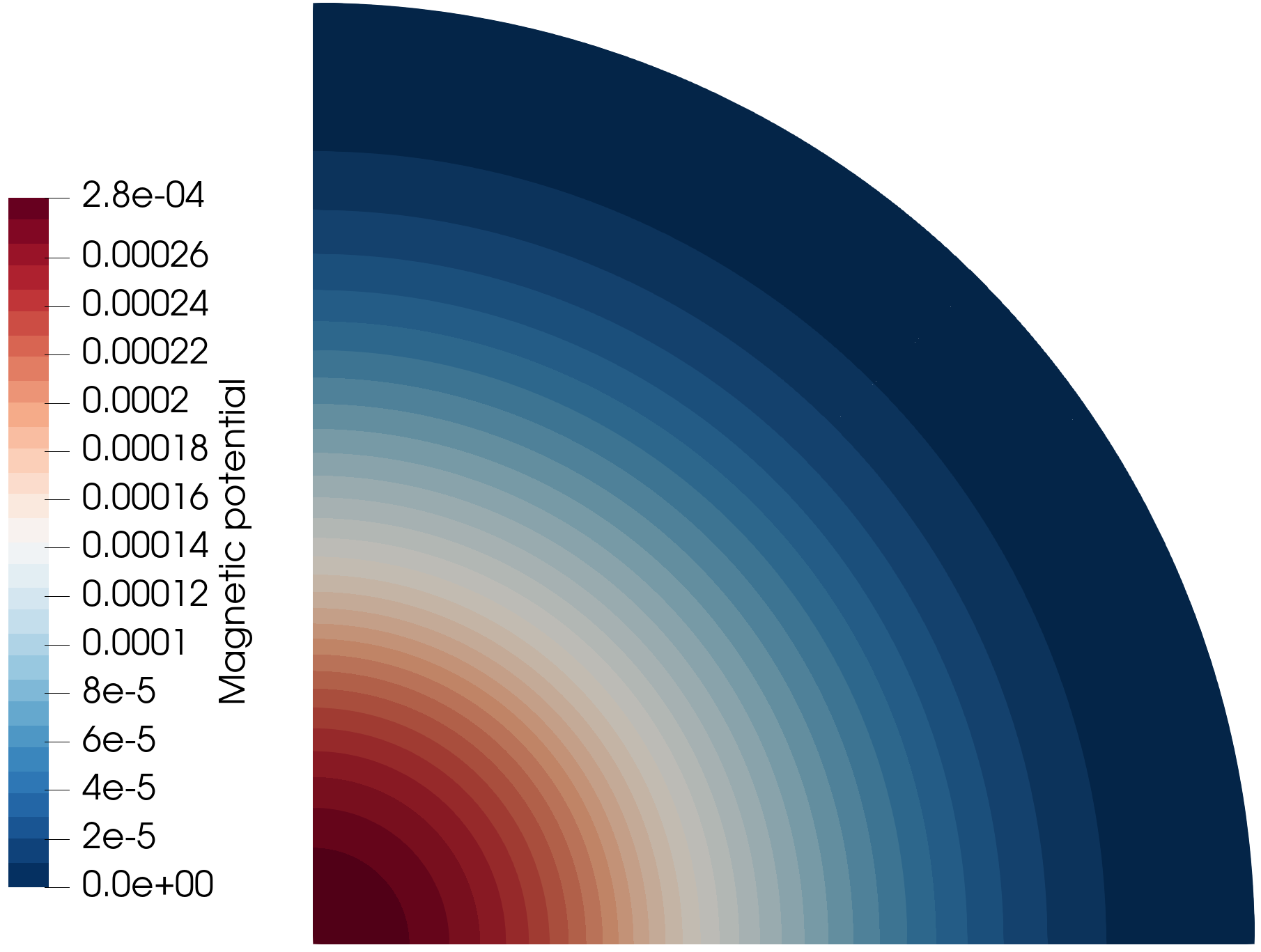}
    \caption{Contours of scalar potential $A_z$ (from \emph{union})}
    \label{fig:coaxial_contours_scalar}
    \end{subfigure}
    \hfill
    \begin{subfigure}[b]{0.47\linewidth}
    \centering
    \includegraphics[width = 0.99\linewidth]{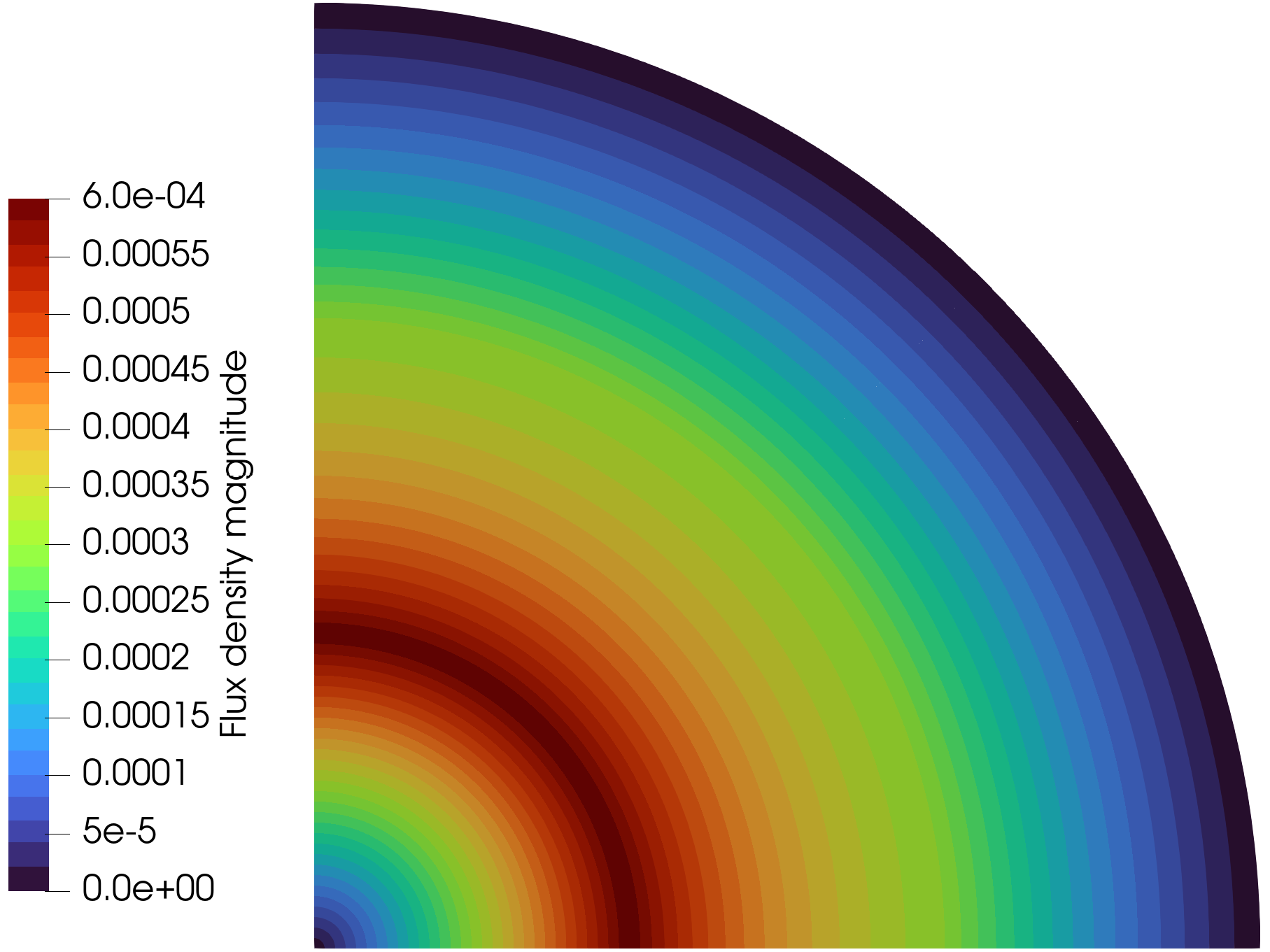}
    \caption{Contours of magnitude of flux density $B_\theta$ (from \emph{union})}
    \label{fig:coaxial_contours_flux}
    \end{subfigure}\\ 
    
    \vspace*{1ex}
    
    \begin{subfigure}[b]{0.49\linewidth}
    \centering
    \resizebox{0.99\linewidth}{!}{\input{figures/coaxial_exact_sol_comp}}
    \caption{Comparison of scalar potential $A_z$ along $x=y$}
    \label{fig:coaxial_comp_exact_union_sol}
    \end{subfigure}
    \hfill
    \begin{subfigure}[b]{0.49\linewidth}
    \centering
    \resizebox{0.99\linewidth}{!}{\input{figures/coaxial_exact_grad_comp}}
    \caption{Comparison of magnitude of flux density $B_\theta$ along $x=y$}
    \label{fig:coaxial_comp_exact_union_grad}
    \end{subfigure}
    \caption{Comparison of the scalar potential $A_z$ and the magnitude of the flux density $B_\theta$ between the analytical solution and the numerical results obtained using the \emph{fully immersed} and \emph{union with non-conformal patches} methods with $p=2$ and $32\times32$ elements for each patch} \label{fig:coaxial_exact_sol}
\end{figure*}

Additionally, \cref{fig:coaxial_contours_scalar,fig:coaxial_contours_flux} shows the contours of the scalar potential $A_z$ and the magnitude of the flux density $B_\theta$ for a discretization with degree $p=2$ and $32\times32$ elements for each patch for the union method.
To further assess the performance, we compare the exact solution along the line $y=x$ with the corresponding numerical results obtained using both methods. \Cref{fig:coaxial_exact_sol} demonstrates excellent agreement between the analytical and numerical solutions already for this coarse, low-degree discretization for the \emph{fully immersed} and \emph{union of non-conformal patches} methods. However, as the insets show, the kinks of $B_\theta$ at the interfaces are less well resolved by the \emph{fully immersed} approach, which may explain the lack of higher-order convergence.

\subsection{Horseshoe magnet}
\label{subsec:horseshoe_magnet}

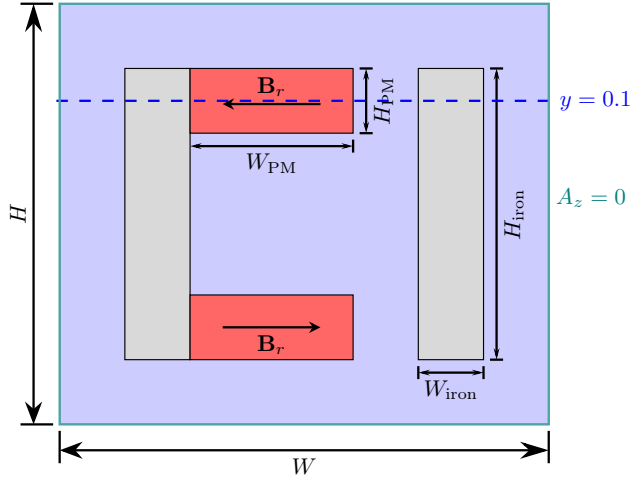
\begin{figure}[t!]
    \centering
    \resizebox{0.54\linewidth}{!}{\input{figures/StriaghtHSM_schem_a}}
    \caption{Schematic drawing of the horseshoe magnet problem. The red rectangular regions denote permanent magnets with different magnetization directions $\tbB_r$. The gray rectangular regions correspond to soft iron rods. The surrounding blue domain represents the air region} \label{fig:horseshoe_schem}
\end{figure}

To further assess the performance of the three proposed methods on a more complex problem, we consider a straight horseshoe magnet. The problem consists of two soft iron rods, one of which is connected to two permanent magnets (PMs) having opposite remanence directions, as illustrated in \cref{fig:horseshoe_schem}. The computational domain has a width of $W = 0.15$ mm and a height of $H = 0.13$ mm. 
Each soft iron rod has a width of $W_{\text{iron}} = 0.02$ mm and a height of $H_{\text{iron}} = 0.09$ mm. The permanent magnets each have dimensions $W_{\text{PM}} = 0.05$ mm and $H_{\text{PM}} = 0.02$ mm. The magnitude of the remanence of each magnet is prescribed as $B_r = 1.4$. The relative permeability of the permanent magnets is $\mu_{r}^{\text{PM}} = 1.05$, such that their permeability is given by $\mu_{\text{PM}} = \mu_{r}^{\text{PM}} \mu_0$, where $\mu_0$ denotes the vacuum permeability. The soft iron rods are modeled with a relative permeability of $\mu_{r}^{\text{iron}} = 4000$, while the surrounding air region is assigned the vacuum permeability, i.e., $\mu^{\text{air}} = \mu_0$. A Dirichlet boundary condition is defined on the outer boundary of the computational domain with $A_z = 0$. 

\begin{figure}[t!]
\centering
    \begin{subfigure}[t]{0.48\linewidth}
        \centering
        \resizebox{0.85\linewidth}{!}{\input{figures/StriaghtHSM_schem_mpiga}}
        \caption{\emph{Standard multi-patch IGA}: The domain is represented by 30 conformal patches \label{fig:straight_HSM_mpiga}}
    \end{subfigure}\hfill
    \begin{subfigure}[t]{0.48\linewidth}
        \centering
        \resizebox{0.85\linewidth}{!}{\input{figures/StriaghtHSM_schem_b}}
        \caption{\emph{Fully immersed}: The subdomains of the magnets and iron rods are immersed in the air background grid\label{fig:straight_HSM_trim}}
    \end{subfigure}\\[3mm]
    \begin{subfigure}[t]{0.48\linewidth}
        \centering
        \resizebox{0.85\linewidth}{!}{\input{figures/StriaghtHSM_schem_c}}
        \caption{\emph{Union NC}: Each magnet and iron component is defined by its own patch, and then union with the remaining active part of the air patch\label{fig:straight_HSM_union}}
    \end{subfigure} \hfill
    \begin{subfigure}[t]{0.48\linewidth}
        \centering
        \resizebox{0.85\linewidth}{!}{\input{figures/StriaghtHSM_schem_d}}
        \caption{\emph{Union CL}: Two boundary layers are wrapping the magnet and iron rod patches  \label{fig:straight_HSM_ibcm}}
    \end{subfigure}
    \caption{Schematic drawings for the discretizations of the horseshoe magnet problem with standard multi-patch IGA and the three proposed immersed IGA methods}
    \label{fig:StriaghtHSM_schem}
\end{figure}
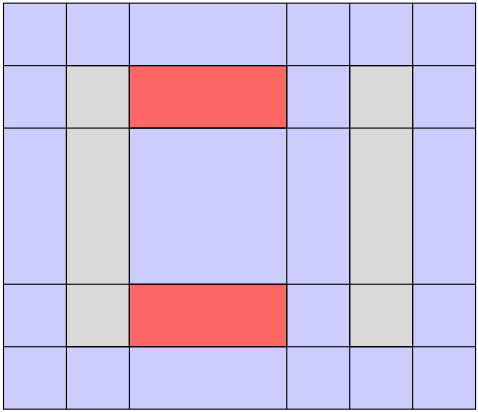
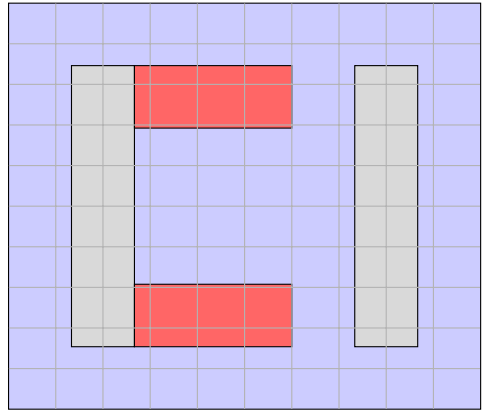
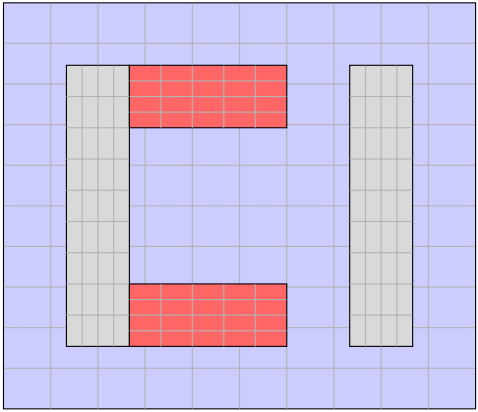
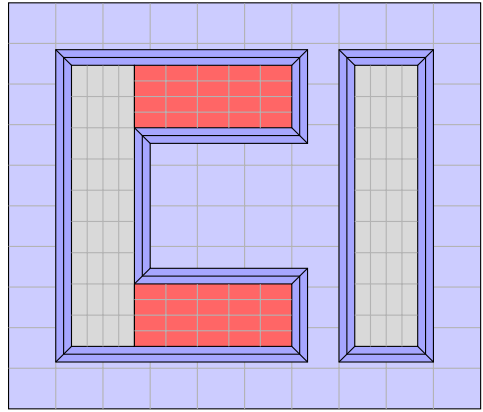

Here, we compare all three immersed IGA methods, see \cref{fig:StriaghtHSM_schem} for illustrations of the meshes.
To have a reference solution $u^{ref}$, we solve the horseshoe magnet problem with conformal multi-patch IGA using 30 conformal patches with each patch being discretized with $20\times20$ elements and polynomial degree $p=5$, with a total of 17,545 DOFs, see \cref{fig:straight_HSM_mpiga}.
To compare the numerical solution of the \emph{fully immersed} approach with the conformal multi-patch solution, we evaluate the error norms, see \cref{eq:errors}, as well as the magnitude of the flux density over the horizontal line $y=0.1$, which is indicated in blue in \cref{fig:horseshoe_schem}.

First, we start with the \emph{fully immersed} approach, where trimming loops are defined for each soft iron rod and permanent magnet component. We follow the same procedure as in \cref{subsec:trimsingle}, where every active and trimmed element corresponding to a specific region is classified. 

The system matrix and RHS vector, see \cref{eq:systemmatrix,eq:rhsvector}, are then obtained by integrating over the corresponding elements for all the different regions/components (magnets, iron rods, air). 

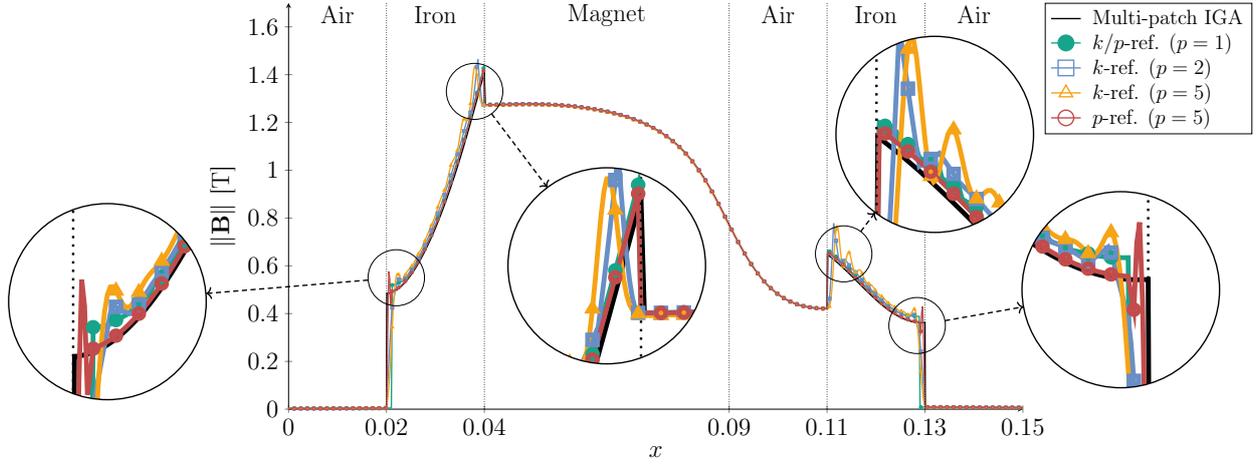
\begin{figure}[t!]
    \centering
    \resizebox{\linewidth}{!}{\input{figures/y01-h127_trim}}
    \caption{Evaluation of the magnetic flux density magnitude along the line $y = 0.1$ for different polynomial orders $p = {1,2,5}$ for both $k$- and $p$-refinement strategies with a background grid of $128\times128$ knot spans for the \emph{fully immersed} method}\label{fig:bmag_y01}
\end{figure}

The results shown in \cref{fig:bmag_y01} demonstrate that, for both $k$- and $p$-refinement strategies, with a background grid of $128\times128$ knot spans, the magnetic flux density magnitude is largely consistent with the reference solution within each material subdomain and sufficiently far from the material interfaces. 
However, for the cases employing $k$-refinement with highly smooth basis functions with $C^{p-1}$-continuity across knot spans, a significant discrepancy compared to the reference solution arises in the vicinity of the material interfaces. In particular, the lowest polynomial order $p = 1$ exhibits the smallest deviation from the reference solution relative to the higher orders $p = \{2,5\}$. This behavior can be explained by the low continuity of the basis functions at $p=1$, where the gradients are discontinuous across element boundaries. Since the magnetic flux density is obtained as the spatial derivative of the magnetic potential, this discontinuity enables a more accurate representation of the magnetic flux density at material interfaces. In contrast, higher polynomial orders combined with high inter-element continuity enforce smoothness of both the solution and its derivatives, i.e., also the magnetic flux density, across knot spans. As a consequence, the magnetic flux density is approximated by overly smooth basis functions across interfaces where discontinuities in material properties occur. This leads to spurious oscillations characteristic of Gibbs-type phenomena, as observed in \cref{fig:bmag_y01}. These spurious oscillations can be mitigated by employing $p$-refinement with $C^0$-continuity, where the continuity across knot spans is reduced to $C^0$. 

The lower inter-element continuity allows for a more flexible representation of gradient variations at material interfaces, thereby reducing non-physical oscillatory behavior. Nevertheless, if a material interface is located within the interior of an element rather than coinciding with element boundaries, spurious oscillations may still arise, as observed in \cref{fig:bmag_y01} in the magnifiers at $x=0.02$ and $x=0.13$. In such cases, the approximation space remains smooth within the element, preventing an accurate representation of the discontinuity in the material coefficients and consequently leading to localized oscillatory artifacts.
Thus, it is evident that the \emph{fully immersed} strategy is not suitable for accurately handling multiple materials within the same patch.

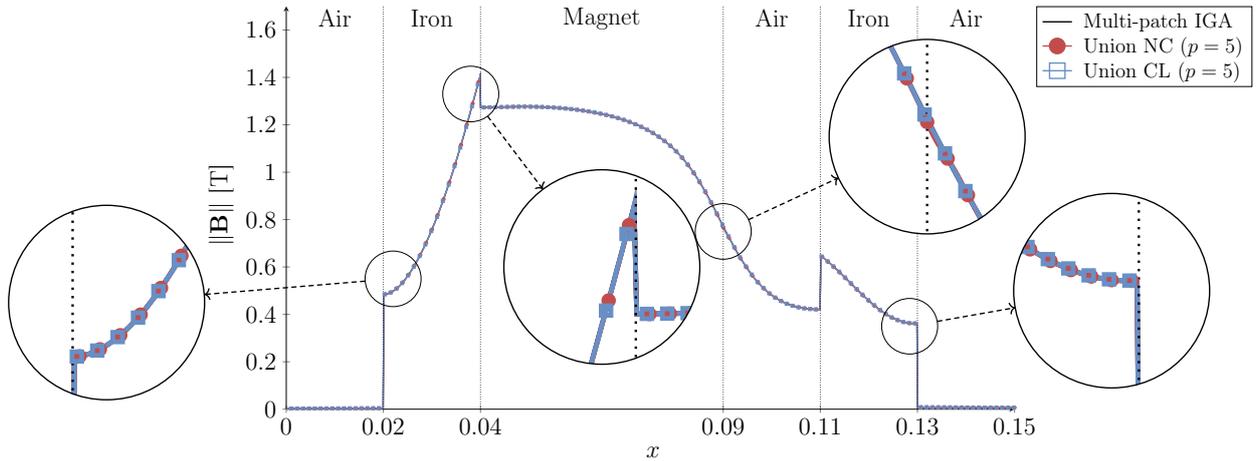
\begin{figure}[t!]
    \centering
    \resizebox{\linewidth}{!}{\input{figures/y01_union}}
    \caption{Evaluation of the magnetic flux density magnitude along the line $y = 0.1$ for polynomial order $p = 5$ for the \emph{union with non-conformal patches} approach and the \emph{union with conformal layers} method with two conformal layers, both with element size $h=0.01$}\label{fig:bmag_y01_union_IBCM}
\end{figure}

To mitigate the fictitious oscillations over the material interfaces, we employ the \emph{union with non-conformal patches} approach and the \emph{union with conformal layers} approach. In the former approach, we couple four non-conformal patches alongside the trimmed background patch, following the procedures in \cref{subsec:unionofnonconf}, as illustrated in \cref{fig:straight_HSM_union}. For the latter method, the non-conformal patches are further enclosed with two conformal layers, as introduced in \cref{subsec:ibcmexpl} and  illustrated in \cref{fig:straight_HSM_ibcm}. 
The results in \cref{fig:bmag_y01_union_IBCM} show a comparison of the evaluation of the magnetic flux density over the line $y=0.1$ for both methods with the reference solution. The \emph{union NC} numerical solution is evaluated with a polynomial degree of $p=5$ and each patch is discretized with uniform element size $h=0.01$, resulting in a total number of 13,288 DOFs. For the \emph{union CL} solution, we use the same elements size and polynomial degree, resulting in a slightly higher total number of 16,638 DOFs. 
Here, it can clearly be observed that both methods align perfectly with the multi-patch IGA evaluation. It is evident that both methods mitigate the fictitious oscillations over the interfaces at a high polynomial degree, even with a smaller number of $\text{DOFs}$ compared with the reference solution, which has $17,545$ degrees of freedom.

\begin{figure}[t!]
    \centering
    \begin{subfigure}[b]{0.42\linewidth}
        \centering
	    \resizebox{!}{62.5mm}{\input{figures/StriaghtHSM_H1s_error}}
	    \caption{$H^{1}_s$-seminorm error over $h$}
        \label{fig:straigh_HSM_H1s_error}
    \end{subfigure}
    \begin{subfigure}[b]{0.57\linewidth}
        \centering
	    \resizebox{!}{62.5mm}{\input{figures/StriaghtHSM_L2_error}}
	    \caption{$L^2$-norm error over $h$}
        \label{fig:straigh_HSM_L2_error}
    \end{subfigure}
    \caption{Comparison of convergence behaviors of the \emph{union with non-conformal patches} and the \emph{union with conformal layers} methods for the horseshoe magnet problem} \label{fig:straigh_HSM_error}
\end{figure}
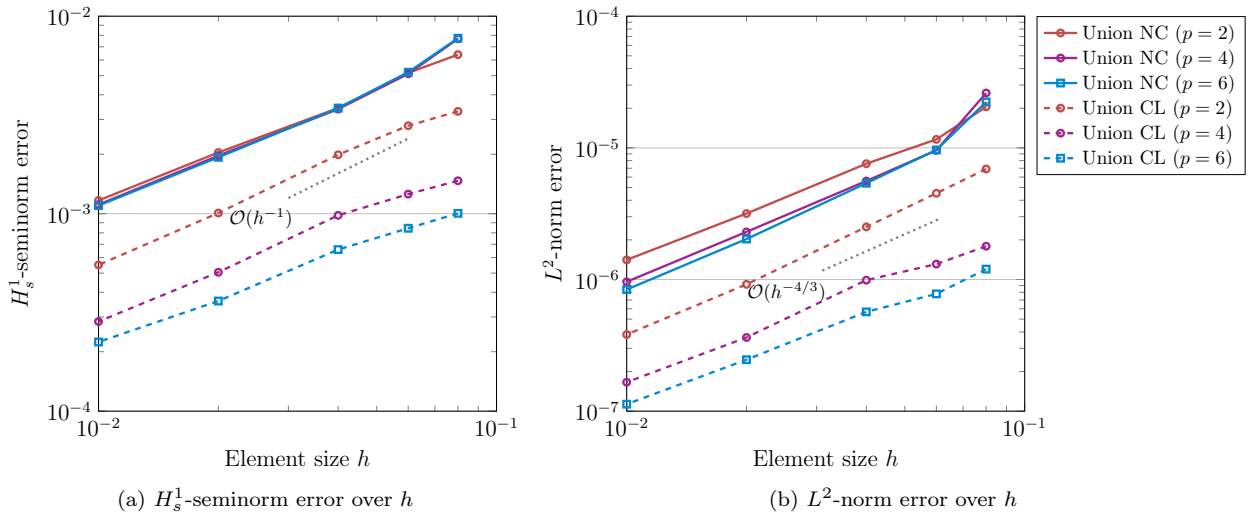

To further evaluate the performance of both approaches, we perform an $h$-ref analysis study for both methods for different polynomial orders $p=\left\{2,4,6\right\}$. The stabilization parameter is set to $\beta \nu_{max} p/h$, where $\beta=10^4$, $\nu_{\max}$ denotes the maximum reluctivity (corresponding to air in the present configuration), $p$ is the polynomial order, and $h$ corresponds to the element size, as every patch is discretized with uniform element size $h$. The $L^2$-norm and $H^1_s$-seminorm error measures as defined in \cref{eq:errors} are considered in comparison to the conformal multi-patch IGA reference solution, as the horseshoe magnet problem has no exact solution. For the calculations of the error norms, a $(p+10)\times (p+10)$ Gauss-Legendre quadrature rule is used within each element. 
The results shown in \cref{fig:straigh_HSM_error} indicate that, for the \emph{union NC} method, mesh refinement through knot insertion improves accuracy in both the $L^2$-norm and the $H_{s}^{1}$-seminorm errors. In contrast, increasing the polynomial order does not lead to a noticeable improvement in accuracy for this approach.
For the \emph{union CL} method, however, the error decreases with increasing polynomial order, indicating a more effective exploitation of higher-order basis functions. Nevertheless, neither method achieves the optimal theoretical convergence rate. This behavior can be attributed to geometric discontinuities at the rectangles' corners, which reduce the regularity of the solution. Consequently, the convergence rate is limited by the reduced solution regularity rather than by the order of the basis functions.

\begin{figure}[tp!]
    \centering
    \begin{subfigure}[b]{0.49\linewidth}
    \centering
    \includegraphics[width = 0.99\linewidth]{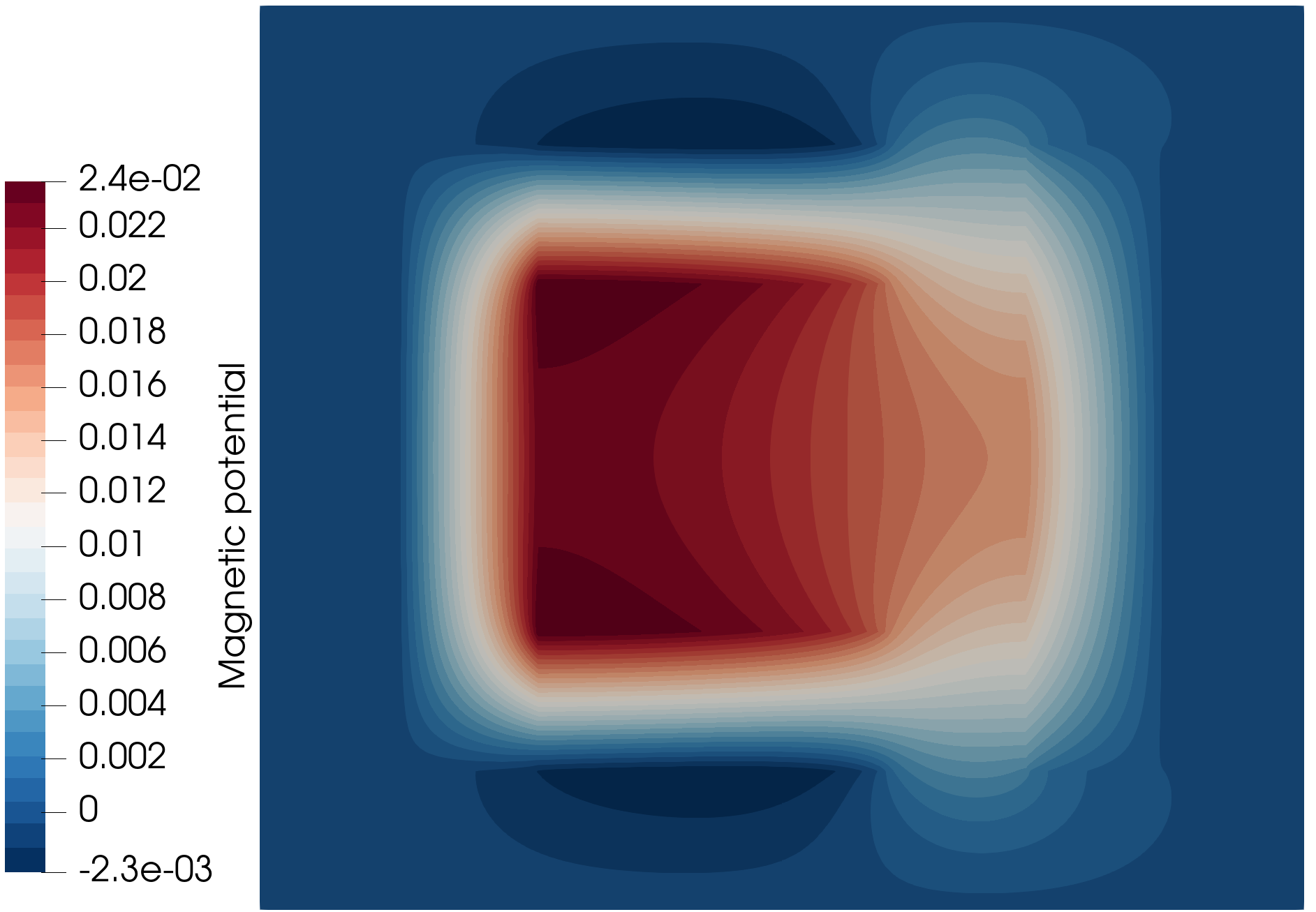}
    \caption{$A_z$ for reference multi-patch IGA}
    \label{fig:ex_pot}
    \end{subfigure}
    \hfill
    \begin{subfigure}[b]{0.49\linewidth}
    \centering
    \includegraphics[width = 0.99\linewidth]{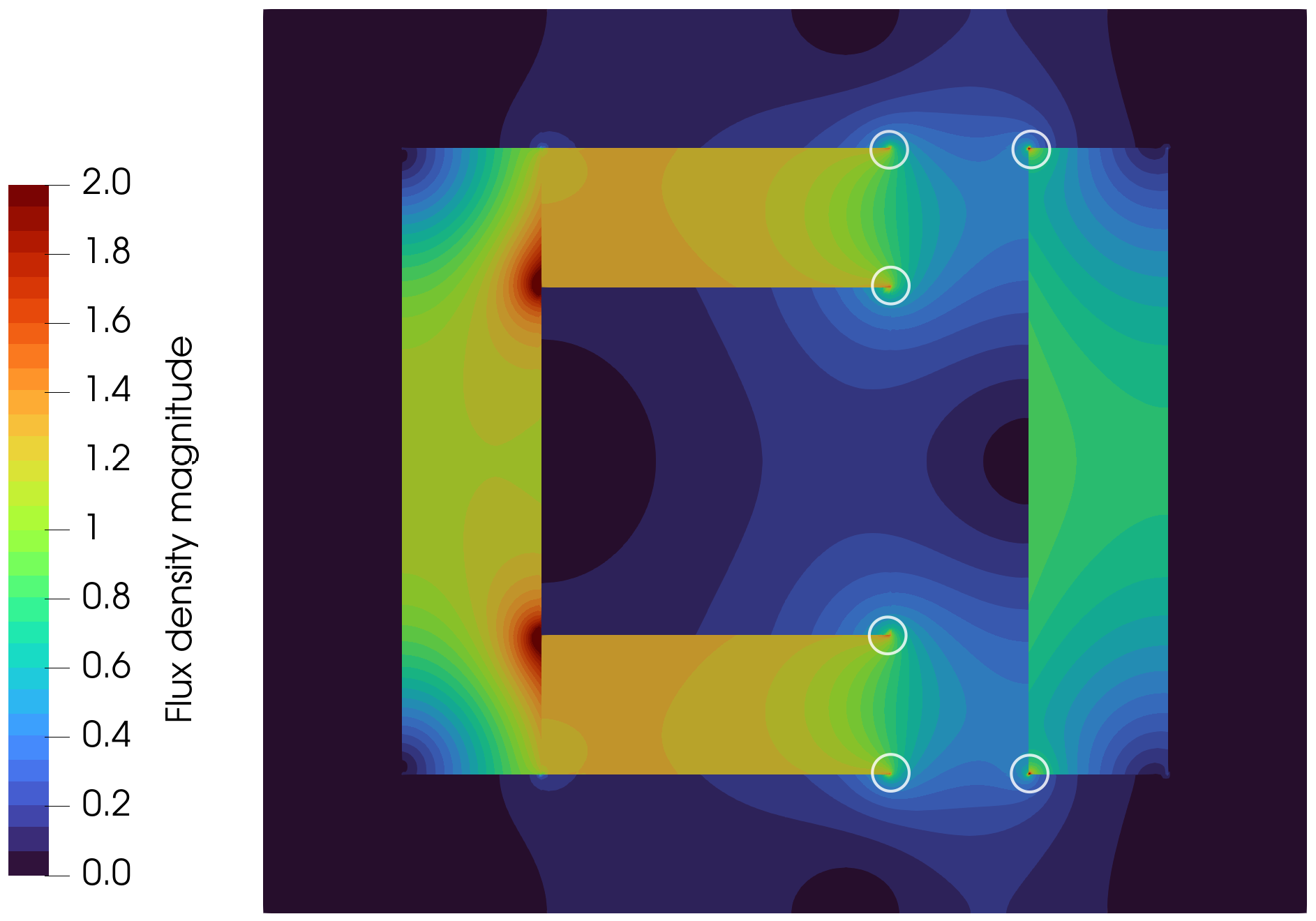}
    \caption{$\|\tbB\|$ for reference multi-patch IGA}
    \label{fig:ex_flux}
    \end{subfigure}
    \\[5mm]
    \begin{subfigure}[b]{0.49\linewidth}
    \centering
    \includegraphics[width = 0.99\linewidth]{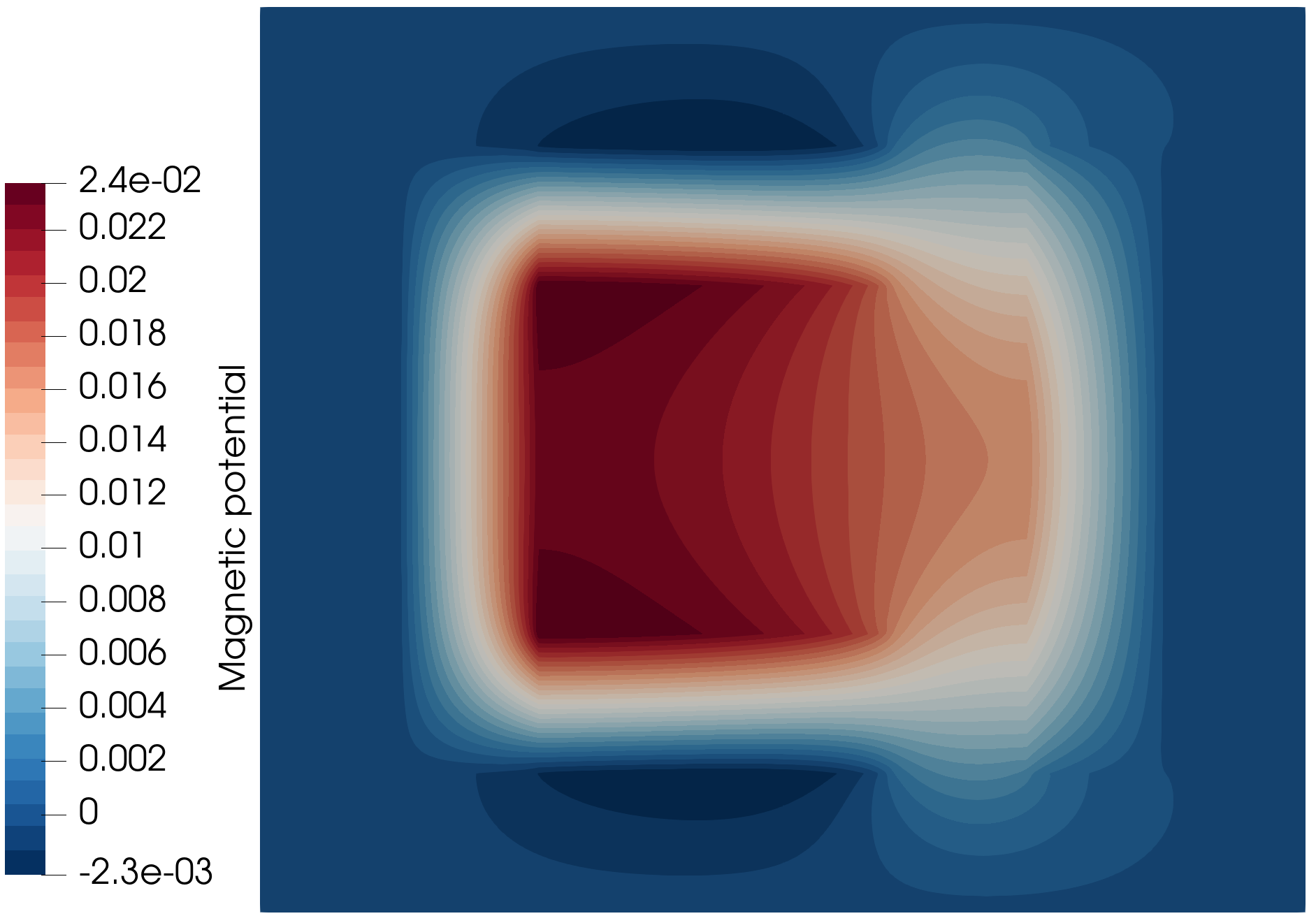}
    \caption{$A_z$ for \emph{union with non-conformal patches} ($h=0.01$)}
    \label{fig:union_pot}
    \end{subfigure}
    \hfill
    \begin{subfigure}[b]{0.495\linewidth}
    \centering
    \includegraphics[width = 0.985\linewidth]{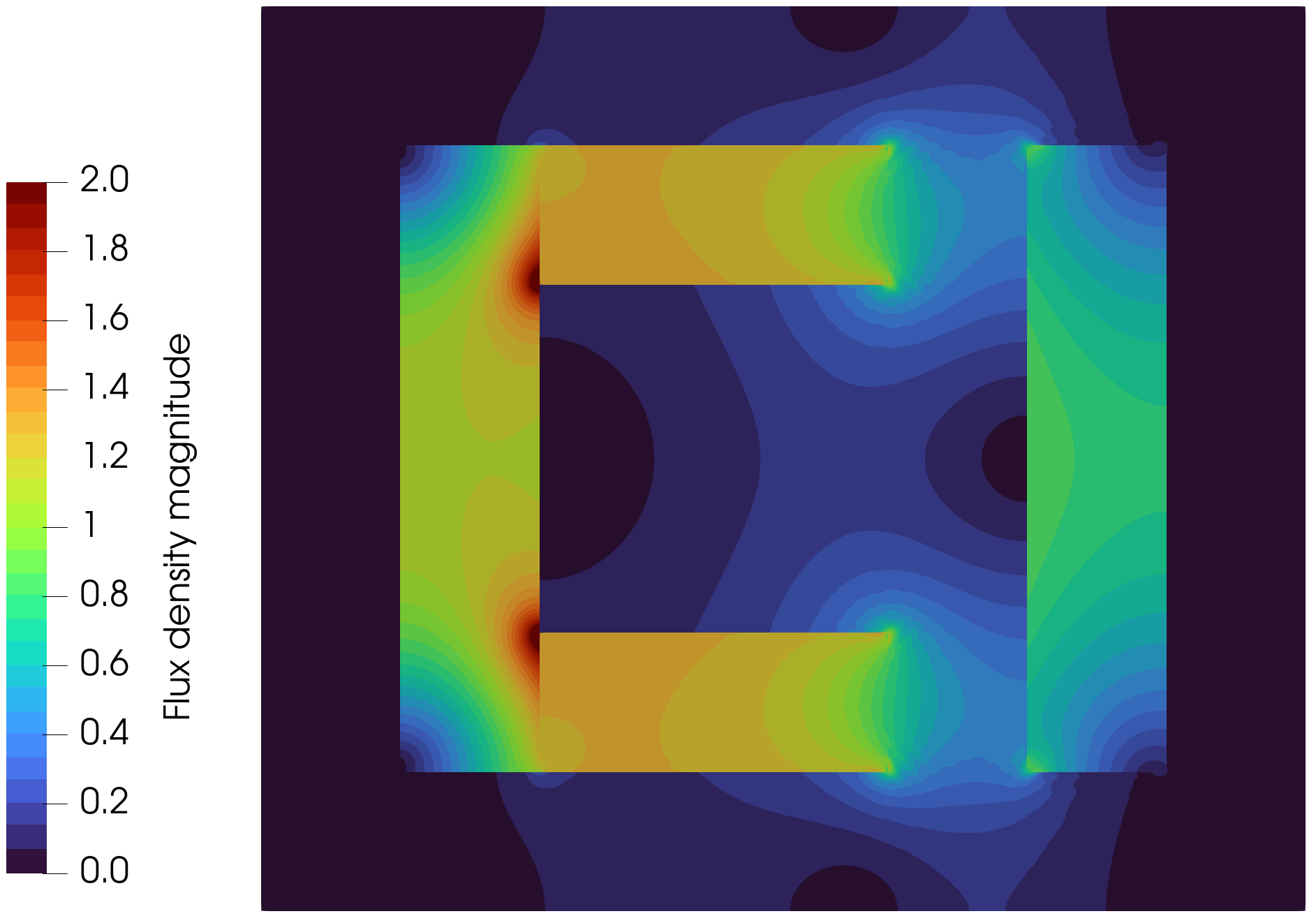}
    \caption{$\|\tbB\|$ for \emph{union with non-conformal patches} ($h=0.01$)}
    \label{fig:union_flux}
    \end{subfigure}
    \\[5mm]
    \begin{subfigure}[b]{0.49\linewidth}
    \centering
    \includegraphics[width = 0.99\linewidth]{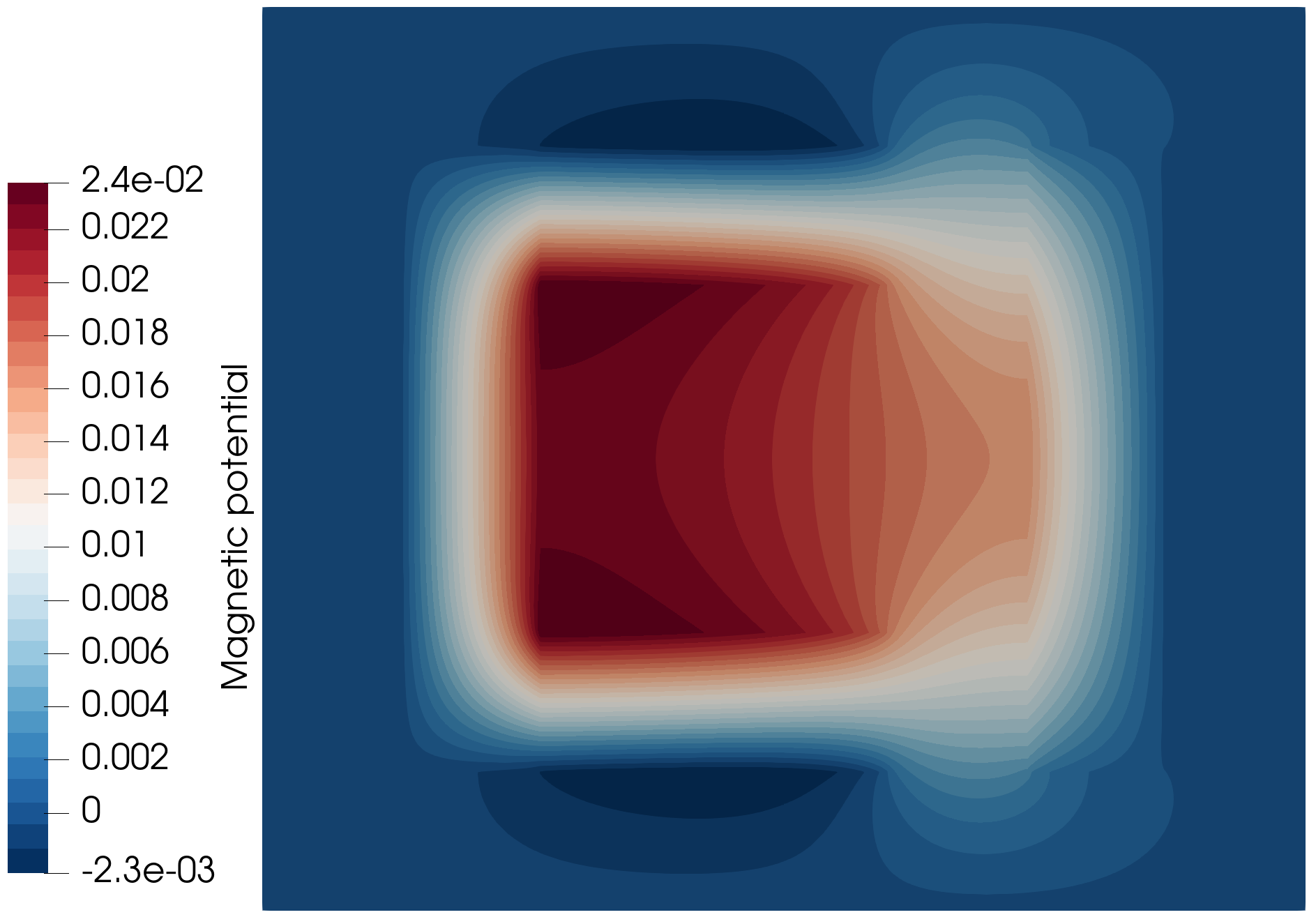}
    \caption{$A_z$ for \emph{union with conformal layers} ($h=0.01$)}
    \label{fig:ibcm_pot}
    \end{subfigure}
    \hfill
    \begin{subfigure}[b]{0.49\linewidth}
    \centering
    \includegraphics[width = 0.99\linewidth]{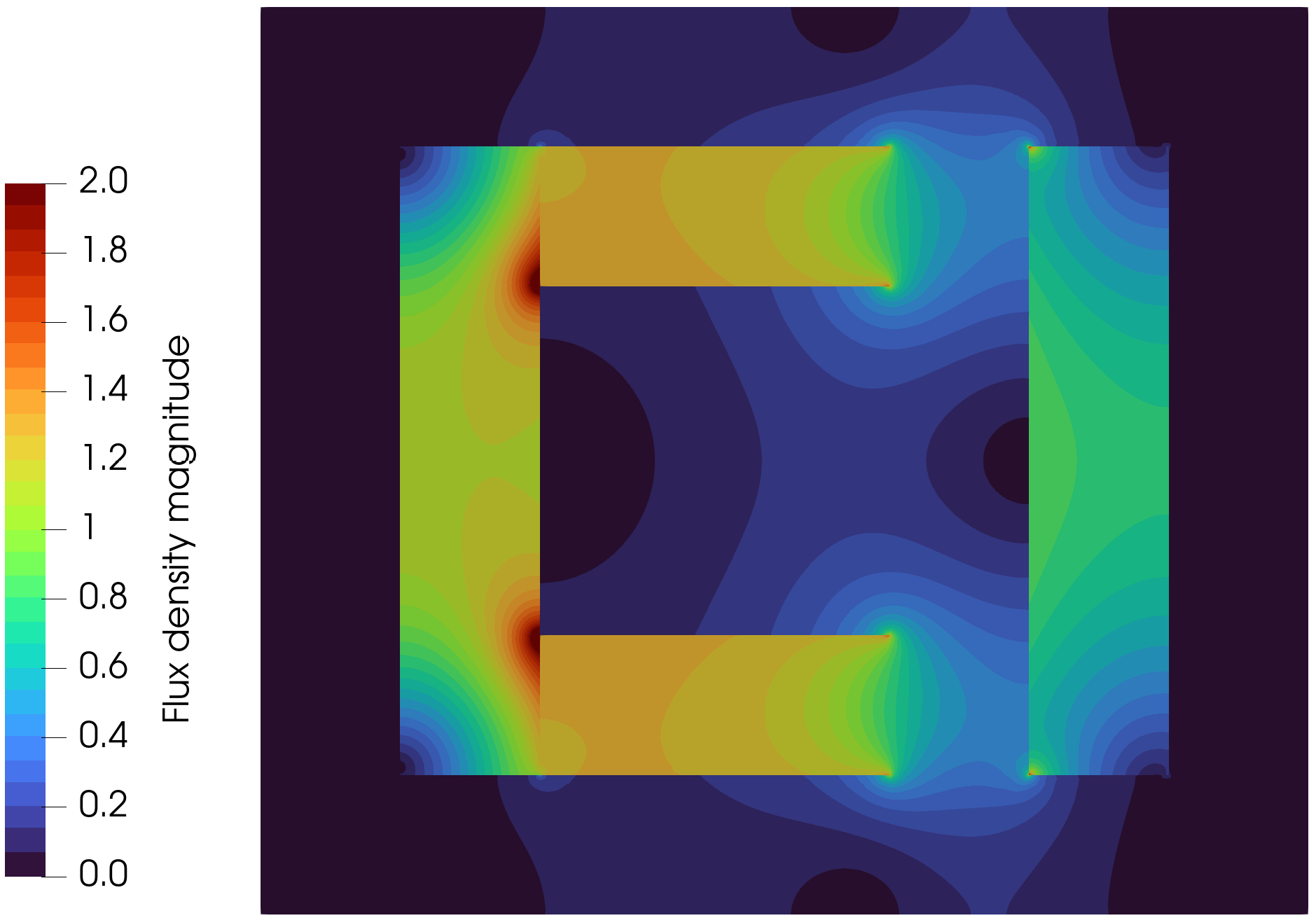}
    \caption{$\|\tbB\|$ for \emph{union with conformal layers} ($h=0.01$)}
    \label{fig:ibcm_flux}
    \end{subfigure}
    \caption{Contours of scalar potential (left) and magnetic flux density (right) for horseshoe magnet problem}\label{fig:straighHSM_comp_contours}
\end{figure}

Finally, \cref{fig:straighHSM_comp_contours} shows the contours of the scalar potential field $A_z$ and the magnitude of the magnetic flux density $\|\tbB\|$ for the conformal multi-patch IGA reference solution, as well as the union methods.
The \emph{union CL} demonstrate an improved capability to capture singular field behavior compared to the \emph{union NC} method. In particular, the singularities occurring at the corners of the left boundary of the right iron rod are clearly resolved in \cref{fig:ex_flux,fig:ibcm_flux}. In contrast, these localized singular features are not adequately captured by the \emph{union NC} method, as observed in \cref{fig:union_flux}.

\subsection{Permanent magnet assembly for a magnetocaloric cooling device}
\label{subsec:PMAssem_MCC}

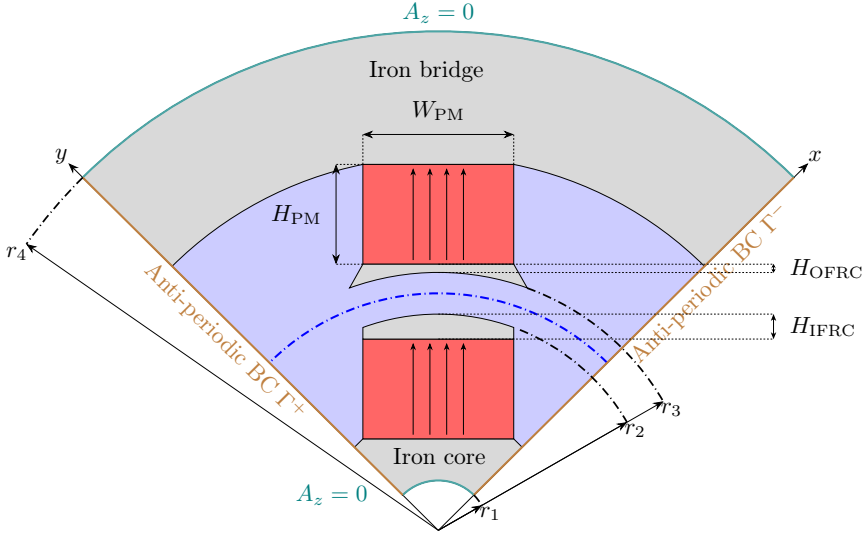
\begin{figure}[t!]
    \centering
    \resizebox{0.7\linewidth}{!}{\input{figures/PMA_MCC_schem}}
    \caption{Schematic representation of the permanent magnet assembly. The permanent magnets are shown in red, the iron components in gray, and the air regions in light blue. Each magnet a magnetization direction with $\theta = 45^{o}$ measured from the $x$-axis
    } \label{fig:PMA_MCC_schem}
\end{figure}

As a final benchmark, we apply the union methods to an industry-related problem: the simulation of a permanent magnet assembly of a magnetocaloric cooling device. Magnetocaloric systems can adopt different design configurations; in this study, we consider a co-rotary configuration consisting of an inner rotor and an outer rotor separated by an air gap (see \cref{fig:PMA_MCC_schem}). The air gap typically hosts the active magnetic regenerator units, where heat exchange occurs between the working fluid (e.g., water) and the magnetocaloric material \cite{kitanovski2020a}. 

In the present work, however, the air gap is assumed to contain only air \cite{wiesheu2023}. Consequently, assessing the cooling performance of such a device requires an accurate evaluation of the magnetic field along a fixed radius within the air gap \cite{bjork2011a, wiesheu2023}. In this study, the analysis is limited to the accurate evaluation of the magnetic field distribution, whereas a subsequent thermodynamic performance analysis lies beyond the scope.

In the following, we briefly describe the main geometric dimensions of the design. The permanent magnet assembly shown in \cref{fig:PMA_MCC_schem} consists of an inner rotor with an inner radius of $r_1=30$ mm and an outer radius of $r_2=130$ mm, as well as an outer rotor with an inner radius of $r_3=155$ mm and an outer radius of $r_4=300$ mm. Both permanent magnets have a dimension of $60\text{ mm}\times60$ mm. The flux collector in the outer rotor has a  height of $H_{\text{ORFC}} = 5$ mm, while it is $H_{\text{IRFC}} = 15$ mm for the inner rotor. 
Dirichlet boundary conditions $A_z = 0$ are imposed strongly by elimination on the inner radius of the inner rotor and on the outer radius of the outer rotor, which physically implies that no magnetic flux crosses these boundaries. Additionally, an anti-periodic boundary conditions are imposed on the sides $\Gamma^+$ and $ \Gamma^-$. 
The material properties of the permanent magnets, iron, and air are the same as those used in the horseshoe magnet problem, see \cref{subsec:horseshoe_magnet}. For comparison, we solve the problem using conformal multi-patch IGA with 21 conformal patches, each being discretized with $30\times30$ knot spans and polynomial order of $p=2$, resulting in a total number of 20,492 DOFs. We evaluate the magnetic flux density over the mean radius between both rotors at $r_a = 142.5$ mm, as highlighted by the blue dashed curve in \cref{fig:PMA_MCC_schem}.

\begin{figure}[t!]
    \centering
    \begin{subfigure}[b]{0.42\linewidth}
    \centering
    \includegraphics[width = 0.97\linewidth]{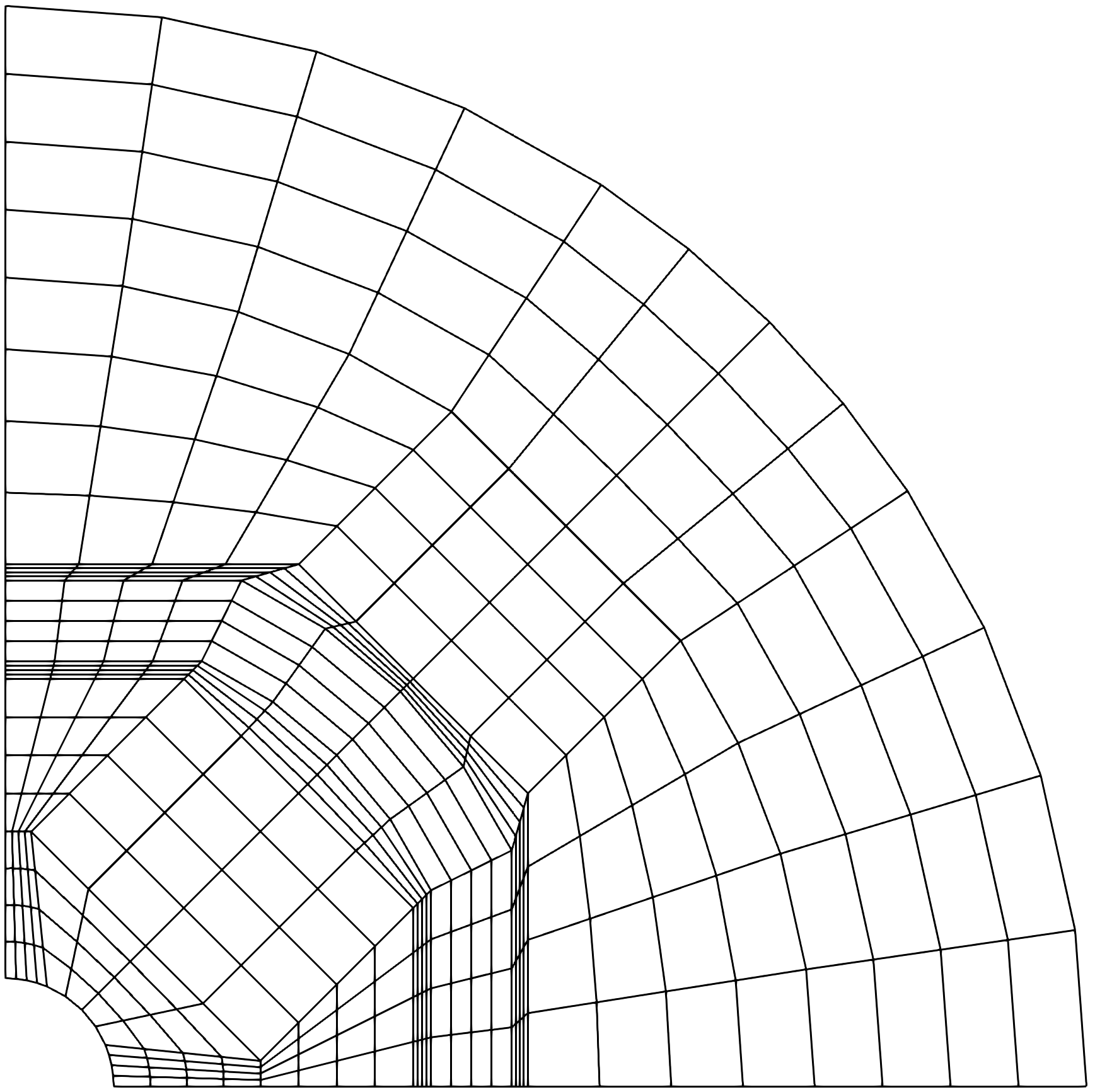}
    \caption{Multi-patch IGA}
    \label{fig:PMA_grid_confIGA}
    \end{subfigure}
    ~~~~~~~~~
    \begin{subfigure}[b]{0.42\linewidth}
    \centering
    \includegraphics[width = 0.97\linewidth]{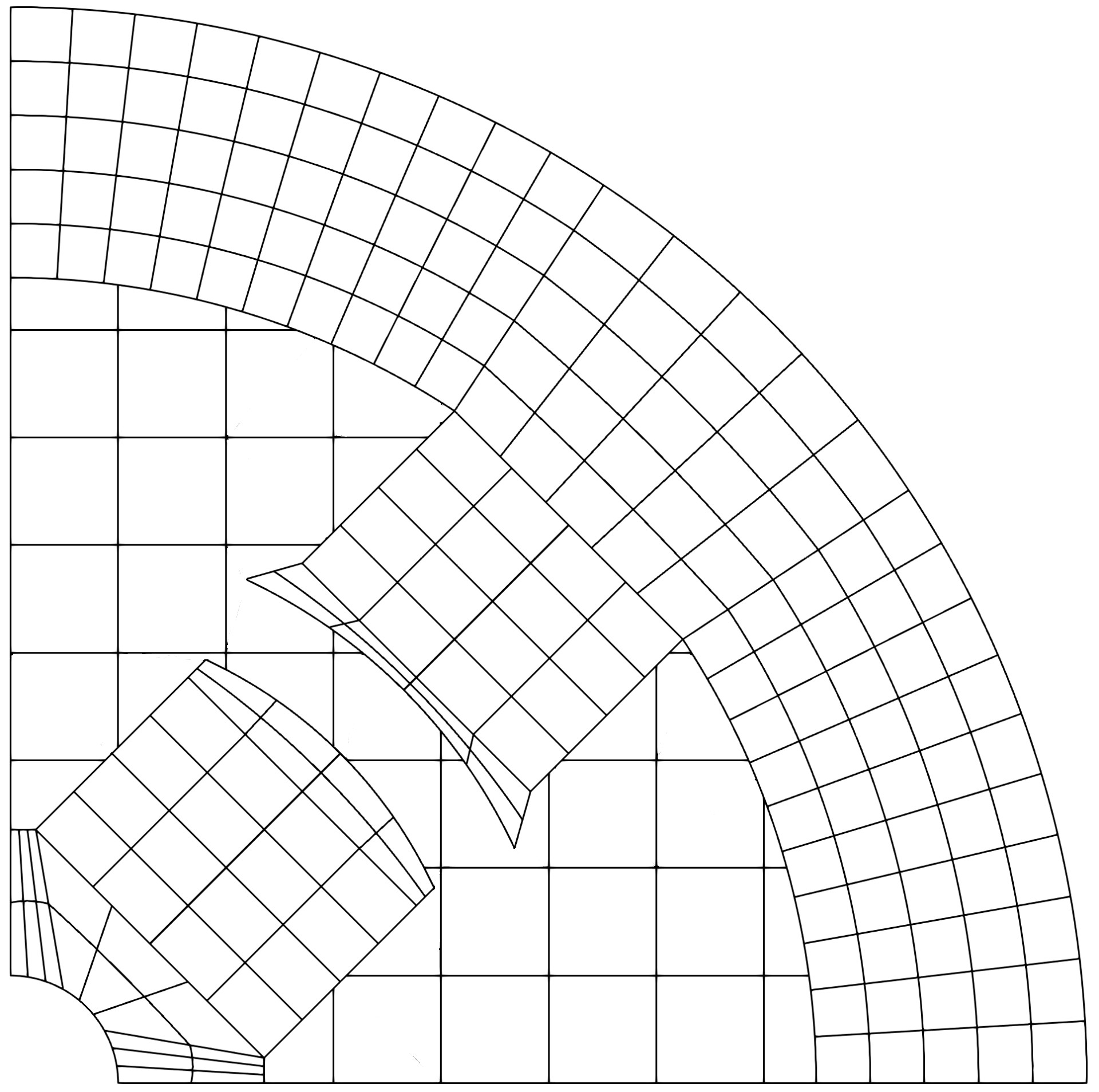}
    \caption{Union with non-conformal patches}
    \label{fig:PMA_grid_union}
    \end{subfigure}\\
    \begin{subfigure}[b]{0.42\linewidth}
    \centering
    \includegraphics[width = 0.97\linewidth]{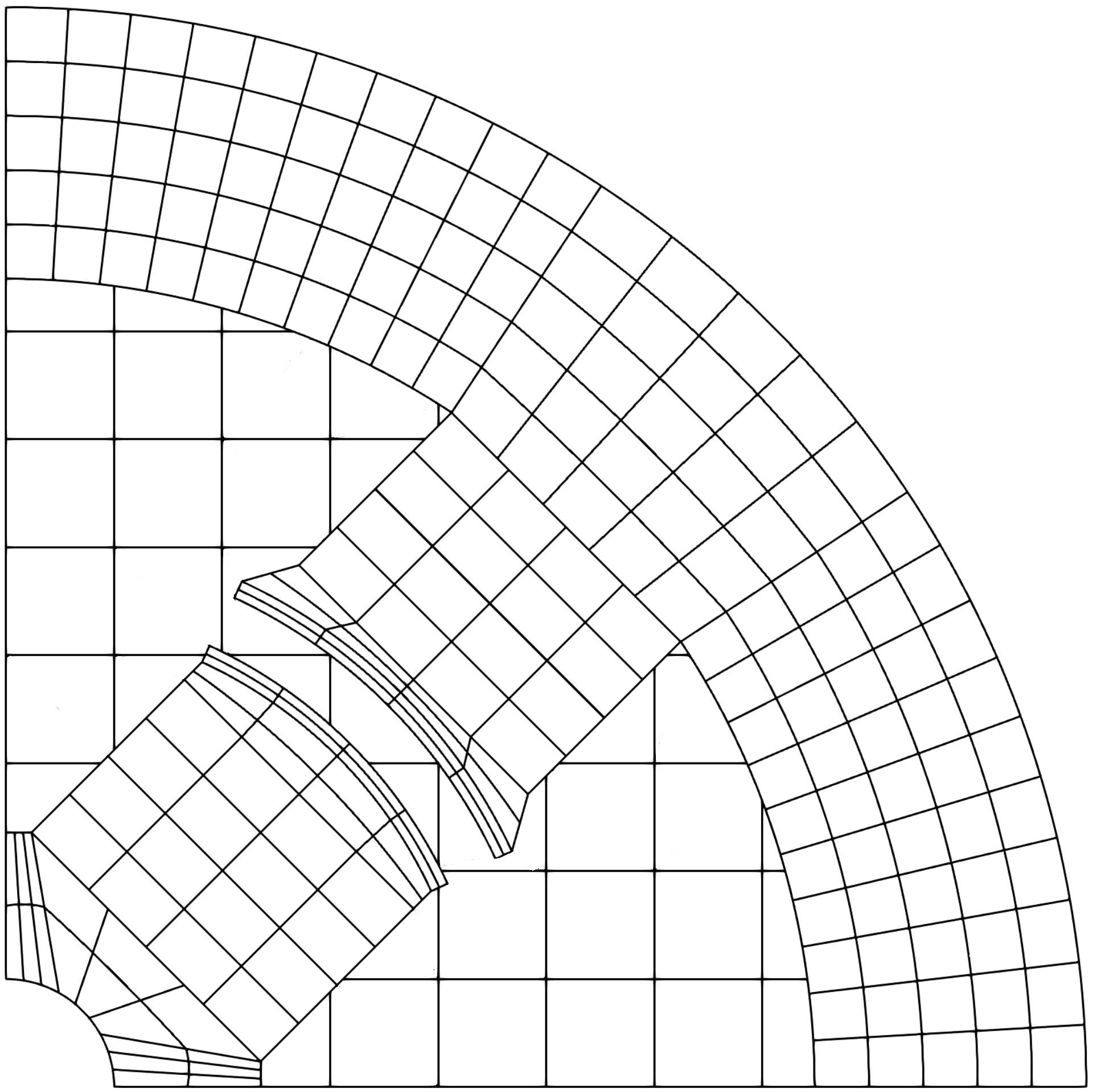}
    \caption{Union with conformal layers (version 1)}
    \label{fig:PMA_grid_ibcm1}
    \end{subfigure}
    ~~~~~~~~~
    \begin{subfigure}[b]{0.42\linewidth}
    \centering
    \includegraphics[width = 0.97\linewidth]{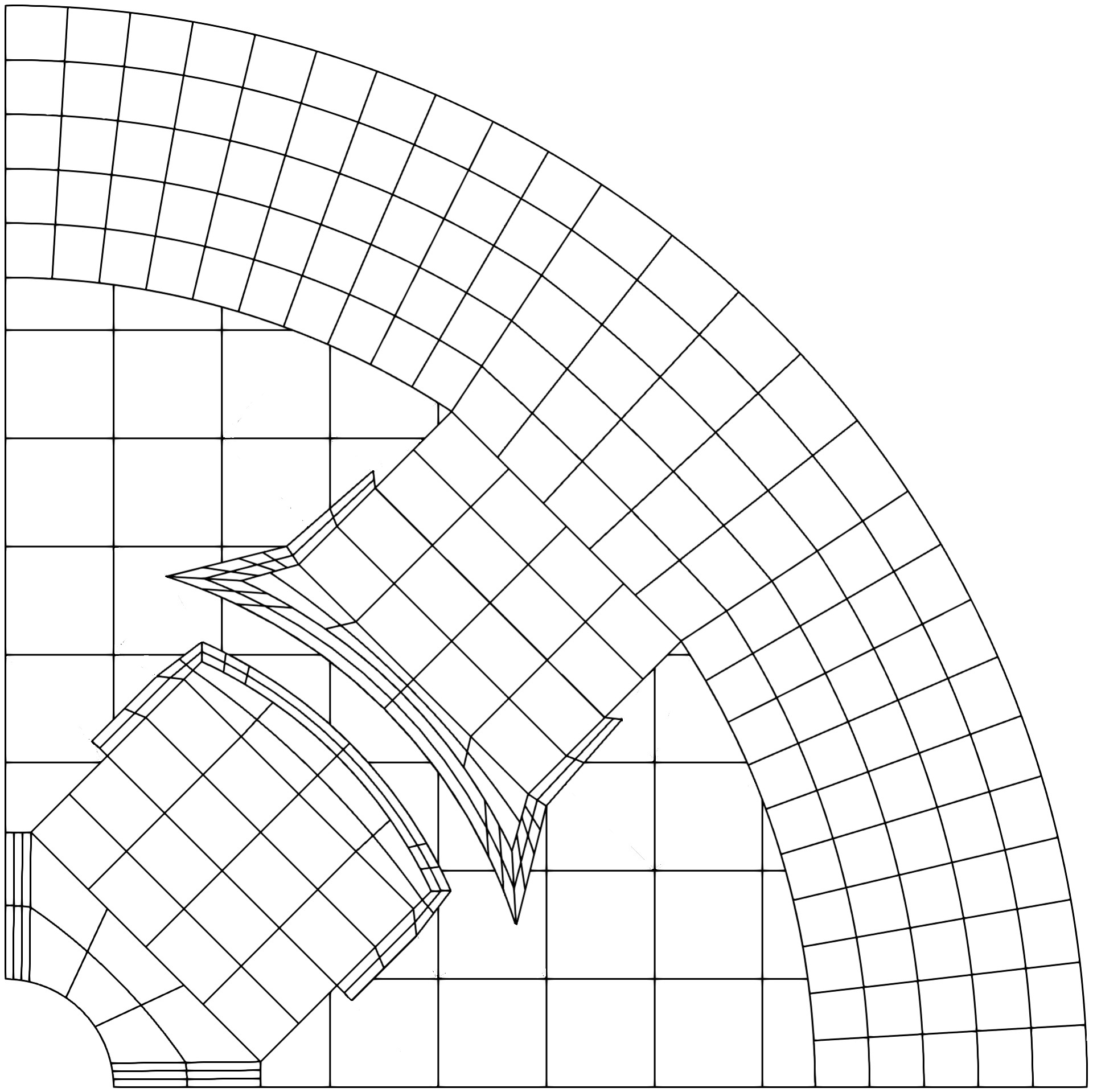}
    \caption{Union with conformal layers (version 2)}
    \label{fig:PMA_grid_ibcm2}
    \end{subfigure}
    \caption{Illustration of the different geometric discretizations of the permanent magnet assembly}\label{fig:PMA_grid}
\end{figure}

For illustration purposes, examples of the spatial discretization for the considered approaches are shown in \cref{fig:PMA_grid}. In \cref{fig:PMA_grid_confIGA}, the geometry is parametrized using 21 conformal patches, each discretized with $4\times4$ knot spans. Constructing such a parametrization is a cumbersome process, as the corner points of each patch must be carefully identified and conformity between adjacent patches must be ensured.
In contrast, for the \emph{union with non-conformal patches} approach shown in \cref{fig:PMA_grid_union}, the geometry can be parametrized by unifying six non-conformal patches over a trimmed background patch, which significantly reduces the parametrization effort.
For the \emph{union with conformal layers}, two configurations are considered for constructing the conformal layers. In the first configuration, see \cref{fig:PMA_grid_ibcm1}, the conformal layer is constructed only along the outer boundaries of the flux collectors in the inner and outer rotors. In the second configuration, see \cref{fig:PMA_grid_ibcm2}, the conformal layers wrap around the flux collectors and part of the permanent magnets in both the inner and outer rotors. The total number of patches for both configurations is nine  (eight non-conformal patches unified with a trimmed background patch).

For the \emph{union with non-conformal patches} approach, the number of elements used for the discretization of each patch is shown in \cref{tab:PMA_kntspans_grid}. The total number of degrees of freedom is $23,869$. For the \emph{union with conformal layers} configurations, the total number of degrees of freedom is $25,621$ for version 1 and $24,477$ for version 2. The conformal layers in version 2 are constructed in such a way to allow the solution field to be resolved more accurately in the vicinity of the singularities. Consequently, a coarser background mesh is used in this configuration, resulting in fewer degrees of freedom compared to version 1, see \cref{tab:PMA_kntspans_grid}. 
For all three methods, basis functions with a polynomial degree of $p=2$ are used, and the stabilization parameter for Nitsche's method was set to $\beta \nu p / h = 5\cdot 10^{11}$.

\begin{table}[t!]
    \centering
    \begin{tabular}{cccc} 
    \hline
    Patch & Union NC & Union CL (v1) & Union CL (v2) \\
    \hline
    Iron core & $20\times30$ & $20\times30$ & $20\times30$ \\ 
    PM1 & $40\times40$ & $40\times40$ & $40\times40$ \\ 
    Iron IRFC & $20\times40$ & $20\times40$ & $20\times40$ \\
    Iron ORFC & $20\times40$ & $20\times40$ & $20\times40$\\
    PM2 & $40\times40$ & $40\times40$ & $40\times40$ \\
    Iron bridge & $50\times50$ & $50\times50$ & $50\times50$\\
    Boundary layers& -- & $20\times40$ & $10\times80$\\
    Air (Background) & $100\times100$ & $100\times100$ & $60\times60$\\
    \hline
    \end{tabular}
    \caption{Space discretization for each approach in terms of knot spans}
    \label{tab:PMA_kntspans_grid}
\end{table}

\begin{figure}[t!]
    \centering
    \resizebox{0.9\linewidth}{!}{\input{figures/PMA_gapfield_comparison}}
    \caption{Evaluation of flux density magnitude in the air gap of the permanent magnet assembly at $r_a=142.5$ mm \label{fig:PMA_MCC_gapfieldcomp}}
\end{figure}
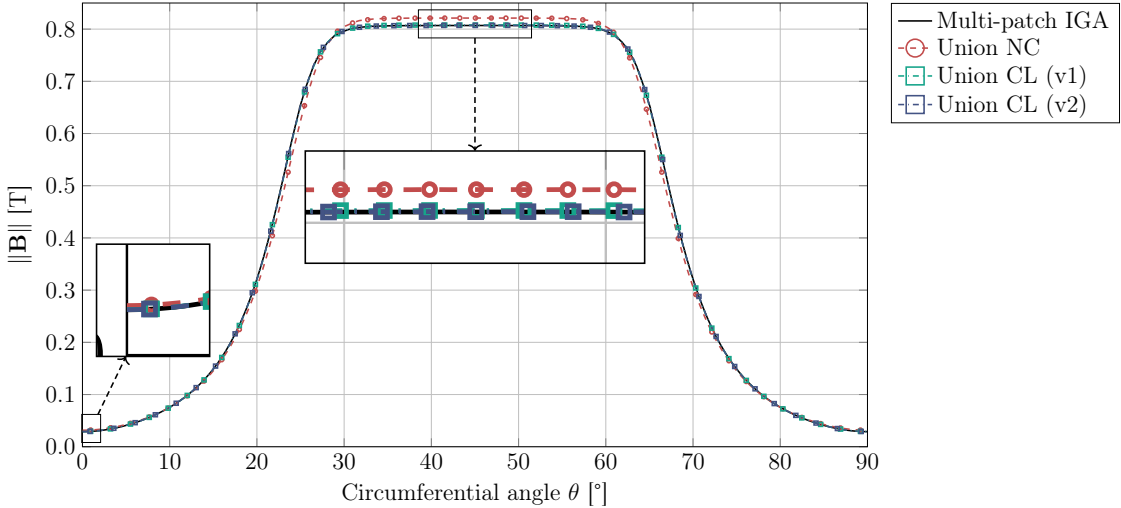

The evaluation of the magnitudes of the flux field $\|\tbB\|$ at $r_a=142.5$ mm shown in \cref{fig:PMA_MCC_gapfieldcomp} indicates that the \emph{union with non-conformal patches} approach yields the lowest accuracy, with the largest discrepancies occurring in the high field region between the two flux collectors (inner and outer rotors), where the flux density is overestimated. Along the circumferential angle $\theta$, the field is generally underestimated, except near both ends of the profile, where the solution again exhibits overestimation. Overall, these trends suggest that the union NC approach requires a finer discretization of the trimmed background patch to achieve a more accurate field evaluation.

For version 1 of the \emph{union with conformal layers}, the predicted field closely matches the multi-patch IGA reference solution, with only a very slight overestimation in the angular interval $\theta \in [35,55]$. This deviation can be reduced by additional refinement of the conformal boundary layers. Finally, the version 2 configuration exhibits an almost perfect agreement with the multi-patch IGA solution, despite employing fewer degrees of freedom than version 1. 

\begin{figure*}[t!]
    \centering
    \begin{subfigure}[b]{0.48\linewidth}
    \centering
    \includegraphics[width = 0.98\linewidth]{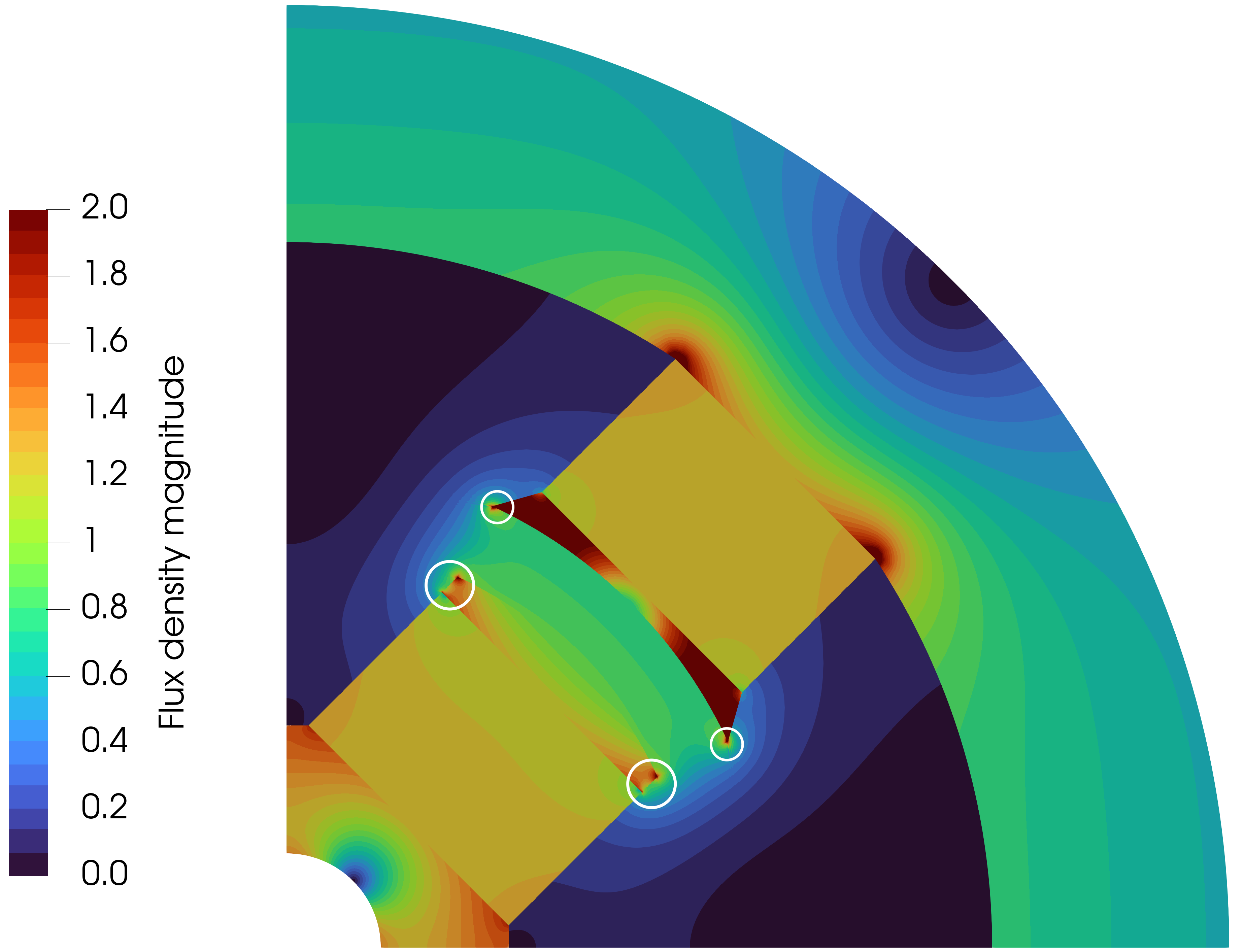}
    \caption{Conformal multi-patch IGA reference solution}
    \label{fig:PMA_flux_confIGA}
    \end{subfigure}
    \hfill
    \begin{subfigure}[b]{0.48\linewidth}
    \centering
    \includegraphics[width = 0.98\linewidth]{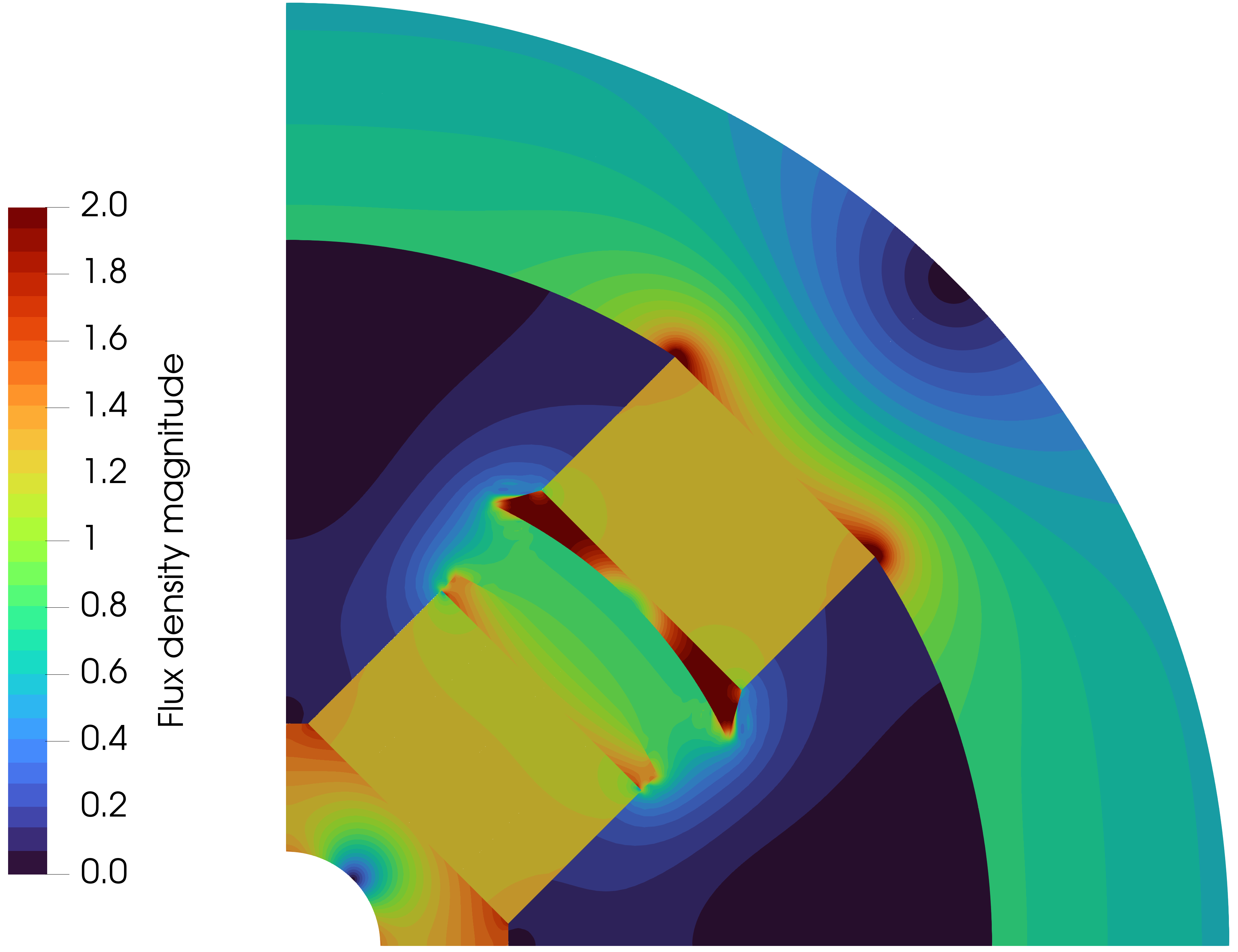}
    \caption{Union with non-conformal patches}
    \label{fig:PMA_flux_union}
    \end{subfigure}
    \\[5mm]
    \begin{subfigure}[b]{0.48\linewidth}
    \centering
    \includegraphics[width = 0.98\linewidth]{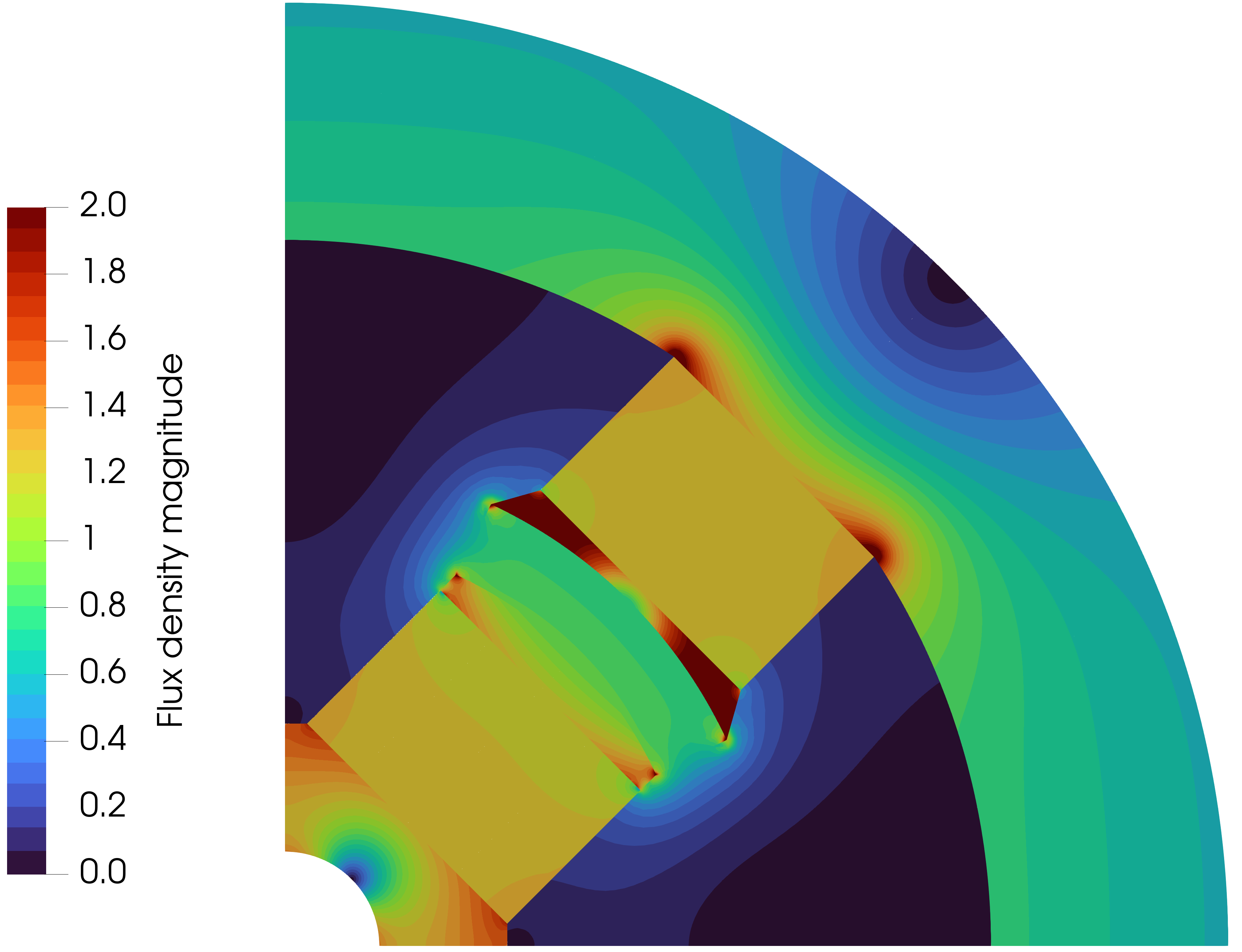}
    \caption{Union with conformal layers (version 1)}
    \label{fig:PMA_flux_ibcm1}
    \end{subfigure}
    \hfill
    \begin{subfigure}[b]{0.48\linewidth}
    \centering
    \includegraphics[width = 0.98\linewidth]{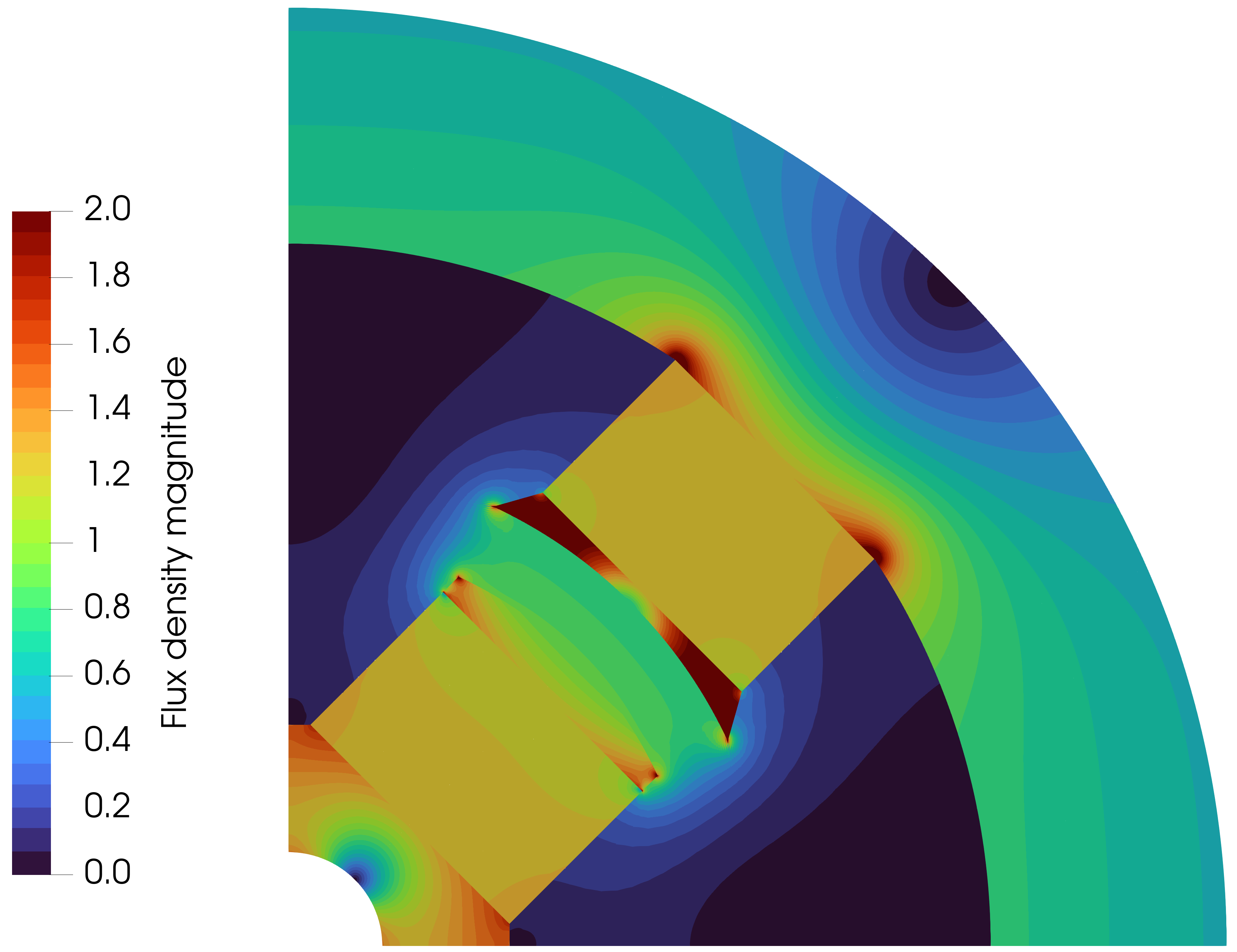}
    \caption{Union with conformal layers (version 2)}
    \label{fig:PMA_flux_ibcm2}
    \end{subfigure}
    \caption{Contours of magnetic flux density of the permanent magnet assembly for different  methods} \label{fig:PMA_flux}
\end{figure*}

To finalize the comparison, the contours of the magnetic flux density magnitude $\|\tbB\|$ are shown in \cref{fig:PMA_flux}. For the union NC approach, the contours generally agree with those of the reference solution. However, small but noticeable discrepancies are observed within the air gap and in the vicinity of the flux collectors. As discussed earlier, these discrepancies could be mitigated by further refinement of the background patch. In contrast, both union CL configurations show excellent agreement with the reference contours throughout the domain.

\subsection{Discussion}

Overall, the three examples presented pose different challenges to the three immersed IGA approaches adopted in this work (\emph{fully immersed}, \emph{union with non-conformal patches}, and \emph{union with conformal layers}).

The \textbf{coaxial cable} problem introduces a discontinuous source current density across different subdomains. 
Here, the \emph{fully immersed} approach is able to accurately predict the magnetic potential and the flux density within each subdomain. However, it cannot capture  the flux density profile across subdomain interfaces exactly. This is due to the inability to represent the discontinuities in the solution gradient through smooth basis functions across the interfaces. 
On the contrary, the \emph{Union NC} approach accurately resolves the flux density over subdomain interfaces. In addition, the approach exhibits optimal convergence rates in the $h$-refinement analysis for different polynomial orders, unlike the \emph{fully immersed} approach.

The \textbf{horseshoe magnet} problem introduces a different challenge compared to the coaxial cable problem: discontinuous material parameters across several subdomains. 
The \emph{fully immersed} approach with $k$-refinement was able to represent the scalar potential and the magnetic flux density within each material well. However, as the magnetic flux density, i.e., a gradient field, is approximated with highly continuous basis function, Gibbs-type phenomena are observed at material interfaces. Although this can be reduced when $p$-refinement is employed, the Gibbs-type phenomena can still be observed within the first few knot spans around the interfaces. 
For the \emph{Union NC} approach, these oscillations were suppressed, as each material is represented by an independent patch and thus independent basis functions. However, the method requires refinement, mainly near singularities, to accurately capture them. This is possible with the implementation of the \emph{Union CL} method, where critical areas are wrapped with conformal layers to accurately capture local features and accurately resolve the magnetic flux density. This can be seen in particular at the corners of the iron rods, where singularities occur, indicated by spikes in the high flux density. Overall, the \emph{Union CL} has shown to have less error in both $L^2$-norm and $H^1_s$-seminorm by approximately one order of magnitude.

The \textbf{permanent magnet assembly} problem features similar challenges as the horseshoe magnet problem, however, the geometry is more complex. Here, the industry-related target was to accurately evaluate the magnetic field within the air gap. This requires an accurate representation of the geometry and ensuring that no cross-talk occurs between different materials/subdomains. The \emph{Union NC} and \emph{Union CL} approaches were examined against a reference solution obtained by conventional conformal multi-patch IGA. The \emph{Union NC} was found to overestimate the field at the high-field region, and slightly underestimate the field near the low-field region. However, with two different boundary conformal configurations, the \emph{Union CL} was able to accurately evaluate the gap field.

%% file: figures/coaxial_schem_quarter.tex
\begin{tikzpicture}
\fill [DarkOrchid!40] (0,0) -- (0:1) arc (0:90:1) -- cycle;

\fill [SeaGreen!60] (0,0) -- (0:2/3) arc (0:90:2/3) -- cycle;

\fill [Goldenrod!60] (0,0) -- (0:1/3) arc (0:90:1/3) -- cycle;
\node[scale=0.36] at (0.1,0.2) {\footnotesize $\Omega_1$};
\node[scale=0.36] at (0.36,0.36) {\footnotesize $\Omega_2$};
\node[scale=0.36] at (0.60,0.60) {\footnotesize $\Omega_3$};

\node[scale=0.5] at (0.2,0.1) {\footnotesize $\odot$};

\node[scale=0.5] at (0.2,0.75) {\footnotesize $\otimes$};

\draw[line width=0.05mm,-{Stealth[length=2.5pt, width=1.5pt]}] (0,0) -- (0:1/3);
\node[scale=0.35] at (0.3,-0.08) {\footnotesize $r_1$};

\draw[line width=0.05mm,-{Stealth[length=2.5pt, width=1.5pt]}] (0,0) -- (90:2/3);
\node[scale=0.35] at (-0.08,2/3-0.05) {\footnotesize $r_2$};

\draw[line width=0.05mm,-{Stealth[length=2.5pt, width=1.5pt]}] (0,0) -- (90:1);
\node[scale=0.35] at (-0.08,1-0.05) {\footnotesize $r_3$};

\draw[teal!70, thin]  (0:1) arc (0:90:1);
\node[scale = 0.35] at (0.83,0.80) {\footnotesize \textcolor{teal}{$A_z = 0$}};

\end{tikzpicture}

%% file: figures/coaxial_trim_a.tex
\begin{tikzpicture}
\def\R{1}    
\def\step{0.25} 

\fill[gray!5] (0,0) rectangle (1,1);
\draw[step=0.25, gray!50, dash pattern=on 0.5pt off 0.5pt, very thin]  (0,0) grid (1,1);

\fill[Goldenrod!30] (0,0) -- (0:0.333) arc (0:90:0.333) -- cycle;
\draw[very thin] (0,0) -- (0:0.333) arc (0:90:0.333) -- cycle;

\def\n{1} 

\newcommand{\annularsector}[3]{
    \fill[#3] (0:#1) arc (0:90:#1) -- (90:#2) arc (90:0:#2) -- cycle;
    
    \foreach \i in {0,...,\n} {
        \pgfmathsetmacro{\r}{#1 + (\i/\n)*(#2-#1)}
        \draw[line width=0.1pt] (0:\r) arc (0:90:\r); 
        \draw[line width=0.1pt] (\i*90/\n: #1) -- (\i*90/\n: #2); 
    }
}

\annularsector{0.333}{0.667}{SeaGreen!30} 
\annularsector{0.667}{1}{Orchid!30}       

\begin{scope}
    \clip (0,0) -- (0:\R) arc (0:90:\R) -- cycle;
    
    \draw[step=\step, gray!50, very thin] (0,0) grid (\R,\R);
\end{scope}

\node[scale = 0.35] at (0.13,0.13) {\footnotesize $\Omega_1^\square$};
\node[scale = 0.35] at (0.36,0.36) {\footnotesize $\Omega_2^\square$};
\node[scale = 0.35] at (0.60,0.60) {\footnotesize $\Omega_3^\square$};
\node[scale = 0.35] at (0.9,0.92) {\color{gray}\footnotesize $\Omega^\square$};
\draw[teal!70, thin]  (0:1) arc (0:90:1);
\node[scale = 0.35] at (0.83,0.80) {\footnotesize \textcolor{teal}{$A_z = 0$}};
\draw[{Stealth[length=2pt, width=1.5pt]}-{Stealth[length=2.5pt, width=1.5pt]},very thin] (0,-0.08)--(1,-0.08) node[midway,sloped,below,rotate=0, scale=0.2] {$r_3$};
\draw[dashed, very thin] (0,-0.12)--(0,-0.02);
\draw[dashed, very thin] (1,-0.12)--(1,-0.02);
\end{tikzpicture}

%% file: figures/coaxial_union_schem_a.tex
\begin{tikzpicture}

\def\R{1}    
\def\step{0.25} 
\fill[Goldenrod!30] (0,0) -- (0:0.333) arc (0:90:0.333) -- cycle;

\begin{scope}
    \clip (0,0) -- (0:\R) arc (0:90:\R) -- cycle;
    
    \draw[step=\step, gray!50, line width=0.1pt] (0,0) grid (\R,\R);
\end{scope}

\def\n{2} 

\newcommand{\annularsector}[3]{
    \fill[#3] (0:#1) arc (0:90:#1) -- (90:#2) arc (90:0:#2) -- cycle;
    
    \foreach \i in {0,...,\n} {
        \pgfmathsetmacro{\r}{#1 + (\i/\n)*(#2-#1)}
        \draw[very thin,gray!50] (0:\r) arc (0:90:\r); 
        \draw[very thin=0.1pt,gray!50] (\i*90/\n: #1) -- (\i*90/\n: #2); 
    }
}

\annularsector{0.333}{0.667}{SeaGreen!30} 
\annularsector{0.667}{1}{Orchid!30}       

\draw[line width=0.1pt] (0,0) -- (90:\R);
\draw[line width=0.1pt] (0,0) -- (0:\R);
\draw[line width=0.1pt] (0:0.333) arc (0:90:0.333);
\draw[line width=0.1pt] (0:0.667) arc (0:90:0.667);
\draw[line width=0.1pt] (0:1) arc (0:90:1);

\draw[red, thin]  (0:2/3) arc (0:90:2/3);
\draw[red, thin]  (0:1/3) arc (0:90:1/3);

\node[scale = 0.35] at (0.13,0.13) {\footnotesize{$\Omega_{1}^{\square}$}};
\node[scale = 0.35] at (0.36,0.36) {\footnotesize{$\Omega_{2}^p$}};
\node[scale = 0.35] at (0.60,0.60) {\footnotesize{$\Omega_{3}^p$}};
\draw[teal!70, thin]  (0:1) arc (0:90:1);
\node[scale = 0.35] at (0.83,0.80) {\footnotesize \textcolor{teal}{$A_z = 0$}};
\draw[{Stealth[length=2pt, width=1.5pt]}-{Stealth[length=2.5pt, width=1.5pt]},very thin] (0,-0.08)--(1,-0.08) node[midway,sloped,below,rotate=0, scale=0.2] {$r_3$};
\draw[dashed, very thin] (0,-0.12)--(0,-0.02);
\draw[dashed, very thin] (1,-0.12)--(1,-0.02);
\end{tikzpicture}


%% file: figures/coaxial_H1snorm_href.tex
\begin{tikzpicture} 

    \pgfplotsset{table/search path={figures}}
    
	\begin{loglogaxis}[ height=25em, 
						width=25em,
						grid=major,
						xmin=1/33,
						xmax=1.1, 
                        xticklabel style={font=\large},
    yticklabel style={font=\large},
						xlabel={\large Element size $h$},
						ymin=1e-15,
						ymax=5e-4, 
						ylabel={\large $H^{1}_s$-seminorm error},
						legend style={
							cells={anchor=west},
							legend pos=outer north east, 
						}
					  ] 
	
 \addplot[color=CPSgreen, dashed,very thick,mark=triangle, mark size=2] table[x=helem,y=errh1s_p1] {data/coaxial_errors_trim.txt};
        
        \addplot[color=CPSred, dashed,very thick,mark=o,mark options={solid}, mark size=1.7] table[x=helem,y=errh1s_p2] {data/coaxial_errors_trim.txt};

		\addplot[color=TUDa-10b, dashed,very thick,mark=*,mark options={solid}, mark size=1.7] table[x=helem,y=errh1s_p4] {data/coaxial_errors_trim.txt};
        
        \addplot[color=TUDa-2b, dashed,very thick,mark=square,mark options={solid}, mark size=1.7] table[x=helem,y=errh1s_p6] {data/coaxial_errors_trim.txt};

		\addplot[color=CPSgreen, solid,very thick,mark=triangle, mark size=2] table[x=helem,y=errh1s_p1] {data/coaxial_errors.txt};
        
        \addplot[color=CPSred, solid,very thick,mark=o,mark options={solid}, mark size=1.7] table[x=helem,y=errh1s_p2] {data/coaxial_errors.txt};

		\addplot[color=TUDa-10b, solid,very thick,mark=*,mark options={solid}, mark size=1.7] table[x=helem,y=errh1s_p4] {data/coaxial_errors.txt};
        
        \addplot[color=TUDa-2b, solid,very thick,mark=square,mark options={solid}, mark size=1.7] table[x=helem,y=errh1s_p6] {data/coaxial_errors.txt};

        \addplot[color=gray, dotted, very thick, domain=0.065:0.4] {0.00001*x^1};
		\node at (axis cs:0.065,6e-7) [anchor=east] {\textcolor{black}{$\mathcal{O}(h^{-1})$}};
        \addplot[color=gray, dotted, very thick, domain=0.065:0.4] {0.000003*x^2};
		\node at (axis cs:0.065,5e-9) [anchor=east] {\textcolor{black}{$\mathcal{O}(h^{-2})$}};
		\addplot[color=gray, dashed, very thick, domain=0.065:0.4] {.0000005*x^4};
		\node at (axis cs:0.065,1.4e-12) [anchor=east] {\textcolor{black}{$\mathcal{O}(h^{-4})$}};
		\addplot[color=gray, dashdotted, very thick, domain=0.065:0.4] {.0000002*x^6};
		\node at (axis cs:0.065,1e-14) [anchor=east] {\textcolor{black}{$\mathcal{O}(h^{-6})$}};
	
	\end{loglogaxis} 

\end{tikzpicture}

%% file: figures/coaxial_L2norm_href.tex
\begin{tikzpicture} 

    \pgfplotsset{table/search path={figures}}
    
	\begin{loglogaxis}[ height=25em, 
						width=25em,
						grid=major,
						xmin=1/33,
						xmax=1.1, 
                        xticklabel style={font=\large},
    yticklabel style={font=\large},
						xlabel={\large Element size $h$},
						ymin=1e-17,
						ymax=5e-4, 
                        ylabel={\large $L^2$-norm error},
						legend style={
							cells={anchor=west},
							legend pos=outer north east, 
						}
					  ] 
        \addplot[color=CPSgreen, dashed,very thick,mark=triangle, mark size=2] table[x=helem,y=errl2_p1] {data/coaxial_errors_trim.txt};
 		\addlegendentry{Immersed ($p=1$)};
 		\addplot[color=CPSred, dashed,very thick,mark=o,mark options={solid}, mark size=1.7] table[x=helem,y=errl2_p2] {data/coaxial_errors_trim.txt};
 		\addlegendentry{Immersed ($p=2$)};

		\addplot[color=TUDa-10b, dashed,very thick,mark=*,mark options={solid}, mark size=1.7] table[x=helem,y=errl2_p4] {data/coaxial_errors_trim.txt};
 		\addlegendentry{Immersed ($p=4$)};
 		\addplot[color=TUDa-2b, dashed,very thick,mark=square,mark options={solid}, mark size=1.7] table[x=helem,y=errl2_p6] {data/coaxial_errors_trim.txt};
 		\addlegendentry{Immersed ($p=6$)};
		
		\addplot[color=CPSgreen, solid,very thick,mark=triangle, mark size=2] table[x=helem,y=errl2_p1] {data/coaxial_errors.txt};
 		\addlegendentry{Union NC ($p=1$)};
 		\addplot[color=CPSred, solid,very thick,mark=o,mark options={solid}, mark size=1.7] table[x=helem,y=errl2_p2] {data/coaxial_errors.txt};
 		\addlegendentry{Union NC ($p=2$)};

		\addplot[color=TUDa-10b, solid,very thick,mark=*,mark options={solid}, mark size=1.7] table[x=helem,y=errl2_p4] {data/coaxial_errors.txt};
 		\addlegendentry{Union NC ($p=4$)};
 		\addplot[color=TUDa-2b, solid,very thick,mark=square,mark options={solid}, mark size=1.7] table[x=helem,y=errl2_p6] {data/coaxial_errors.txt};
 		\addlegendentry{Union NC ($p=6$)};

        \addplot[color=gray, dotted, very thick, domain=0.065:0.4] {0.000003*x^2};
		\node at (axis cs:0.065,5e-9) [anchor=east] {\textcolor{black}{$\mathcal{O}(h^{-2})$}};

        \addplot[color=gray, dotted, very thick, domain=0.065:0.4] {0.0000002*x^3};
		\node at (axis cs:0.065,1e-11) [anchor=east] {\textcolor{black}{$\mathcal{O}(h^{-3})$}};
		\addplot[color=gray, dashed, very thick, domain=0.065:0.4] {.000000015*x^5};
		\node at (axis cs:0.065,3e-15) [anchor=east] {\textcolor{black}{$\mathcal{O}(h^{-5})$}};
		\addplot[color=gray, dashdotted, very thick, domain=0.065:0.4] {.00000002*x^7};
		\node at (axis cs:0.065,3e-17) [anchor=east] {\textcolor{black}{$\mathcal{O}(h^{-7})$}};
	
	\end{loglogaxis} 

\end{tikzpicture}

%% file: figures/coaxial_H1snorm_sqrtdofs.tex
\begin{tikzpicture} 

    \pgfplotsset{table/search path={figures}}
    
	\begin{loglogaxis}[ height=25em, 
						width=25em,
						grid=major,
						log basis x=2,
						xmin=2,
                        xticklabel style={font=\large},
    yticklabel style={font=\large},
						xmax=64, 
						xlabel={\large $\sqrt{\mathrm{DOFs}}$},
						ymin=1e-15,
						ymax=5e-4, 
						ylabel={\large $H^{1}_s$-seminorm error},
						legend style={
							cells={anchor=west},
							legend pos=outer north east, 
						}
					  ]

		\addplot[color=CPSgreen, solid,very thick,mark=triangle, mark size=2] table[x=sqrtndofs1,y=errh1s_p1] {data/coaxial_errors.txt};
        
        \addplot[color=CPSred, solid,very thick,mark=o,mark options={solid}, mark size=1.7] table[x=sqrtndofs2,y=errh1s_p2] {data/coaxial_errors.txt};

		\addplot[color=TUDa-10b, solid,very thick,mark=*,mark options={solid}, mark size=1.7] table[x=sqrtndofs4,y=errh1s_p4] {data/coaxial_errors.txt};
        
        \addplot[color=TUDa-2b, solid,very thick,mark=square,mark options={solid}, mark size=1.7] table[x=sqrtndofs6,y=errh1s_p6] {data/coaxial_errors.txt};
        \addplot[color=CPSgreen, dashed,very thick,mark=triangle, mark size=2] table[x=sqrtndofs1,y=errh1s_p1] {data/coaxial_errors_trim.txt};
        
        \addplot[color=CPSred, dashed,very thick,mark=o,mark options={solid}, mark size=1.7] table[x=sqrtndofs2,y=errh1s_p2] {data/coaxial_errors_trim.txt};

		\addplot[color=TUDa-10b, dashed,very thick,mark=*,mark options={solid}, mark size=1.7] table[x=sqrtndofs4,y=errh1s_p4] {data/coaxial_errors_trim.txt};
        
        \addplot[color=TUDa-2b, dashed,very thick,mark=square,mark options={solid}, mark size=1.7] table[x=sqrtndofs6,y=errh1s_p6] {data/coaxial_errors_trim.txt};

	
	\end{loglogaxis} 

\end{tikzpicture}

%% file: figures/coaxial_L2norm_sqrtdofs.tex
\begin{tikzpicture} 

    \pgfplotsset{table/search path={figures}}
    
	\begin{loglogaxis}[ height=25em, 
						width=25em,
						grid=major,
						log basis x=2,
                        xticklabel style={font=\large},
                        yticklabel style={font=\large},
						xmin=2,
						xmax=64, 
						xlabel={\large $\sqrt{\mathrm{DOFs}}$},
						ymin=1e-17,
						ymax=5e-4, 
                        ylabel={\large $L^2$-norm error},
						legend style={
							cells={anchor=west},
							legend pos=outer north east, 
						}
					  ]

        \addplot[color=CPSgreen, dashed,very thick,mark=triangle, mark size=2] table[x=sqrtndofs1,y=errl2_p1] {data/coaxial_errors_trim.txt};
 		\addlegendentry{Immersed ($p=1$)};
 		\addplot[color=CPSred, dashed,very thick,mark=o,mark options={solid}, mark size=1.7] table[x=sqrtndofs2,y=errl2_p2] {data/coaxial_errors_trim.txt};
 		\addlegendentry{Immersed ($p=2$)};

		\addplot[color=TUDa-10b, dashed,very thick,mark=*,mark options={solid}, mark size=1.7] table[x=sqrtndofs4,y=errl2_p4] {data/coaxial_errors_trim.txt};
 		\addlegendentry{Immersed ($p=4$)};
 		\addplot[color=TUDa-2b, dashed,very thick,mark=square,mark options={solid}, mark size=1.7] table[x=sqrtndofs6,y=errl2_p6] {data/coaxial_errors_trim.txt};
 		\addlegendentry{Immersed ($p=6$)};

		\addplot[color=CPSgreen, solid,very thick,mark=triangle, mark size=2] table[x=sqrtndofs1,y=errl2_p1] {data/coaxial_errors.txt};
 		\addlegendentry{Union NC ($p=1$)};
 		\addplot[color=CPSred, solid,very thick,mark=o,mark options={solid}, mark size=1.7] table[x=sqrtndofs2,y=errl2_p2] {data/coaxial_errors.txt};
 		\addlegendentry{Union NC ($p=2$)};

		\addplot[color=TUDa-10b, solid,very thick,mark=*,mark options={solid}, mark size=1.7] table[x=sqrtndofs4,y=errl2_p4] {data/coaxial_errors.txt};
 		\addlegendentry{Union NC ($p=4$)};
 		\addplot[color=TUDa-2b, solid,very thick,mark=square,mark options={solid}, mark size=1.7] table[x=sqrtndofs6,y=errl2_p6] {data/coaxial_errors.txt};
 		\addlegendentry{Union NC ($p=6$)};

        \addplot[color=gray, dotted, very thick, domain=0.065:0.4] {0.00002*x^2};
		\node at (axis cs:0.065,5e-9) [anchor=east] {\textcolor{black}{$\mathcal{O}(\ell^{-2})$}};

        \addplot[color=gray, dotted, very thick, domain=0.065:0.4] {0.0000002*x^3};
		\node at (axis cs:0.065,1e-11) [anchor=east] {\textcolor{black}{$\mathcal{O}(\ell^{-3})$}};
		\addplot[color=gray, dashed, very thick, domain=0.065:0.4] {.000000015*x^5};
		\node at (axis cs:0.065,5e-15) [anchor=east] {\textcolor{black}{$\mathcal{O}(\ell^{-5})$}};
		\addplot[color=gray, dashdotted, very thick, domain=0.065:0.4] {.00000002*x^7};
		\node at (axis cs:0.065,3e-17) [anchor=east] {\textcolor{black}{$\mathcal{O}(\ell^{-7})$}};
	
	\end{loglogaxis} 

\end{tikzpicture}

%% file: figures/coaxial_exact_sol_comp.tex
\begin{tikzpicture}
\begin{axis}[
        height=17em, 
		width=30em, 
		grid=both,xmin=0,
		xmax=1, 
		xlabel={\large $r$},
		ymin=0,
        xticklabel style={font=\large},
    yticklabel style={font=\large},
		ymax=3e-4, 
		y tick label style={
                /pgf/number format/.cd,
                    fixed,
                    precision=1,
                    zerofill,
                /tikz/.cd,
            },
		ylabel={\large $A_z$},
		legend style={
			cells={anchor=west},
			legend pos=north east, 
		}
        ]
        				
        \addplot[color=black, thick, mark options={solid}, mark size=1.7,mark phase=8,legend image post style={mark size=4.5}] table[x=r,y=Az] {data/exact_sol_along_xaxis.txt};
 		\addlegendentry{Exact};

        \addplot[color=CPSlightblue, solid, thick, mark=x,mark options={solid},mark repeat=20, mark size=0.5,mark phase=8,legend image post style={mark size=4.5}] table[x=x,y=solution] {data/p2_64x64_sol_along_xaxis_trim.txt};
 		\addlegendentry{Fully immersed};

 		\addplot[color=CPSred, dashed, thick, mark=o,mark options={solid},mark repeat=20, mark size=0.5,mark phase=8,legend image post style={mark size=4.5}] table[x=x,y=solution] {data/p2_64x64_sol_along_xaxis.txt};
 		\addlegendentry{Union NC};

\end{axis}

\end{tikzpicture}

%% file: figures/coaxial_exact_grad_comp.tex
\begin{tikzpicture}[spy using outlines={circle, magnification=6, size=2cm, connect spies}]
\begin{axis}[
        height=17em, 
		width=30em, 
		grid=both,xmin=0,
		xmax=1, 
		xlabel={\large $r$},
		ymin=0,
		ymax=6e-4, 
        xticklabel style={font=\large},
    yticklabel style={font=\large},
		y tick label style={
                /pgf/number format/.cd,
                    fixed,
                    precision=1,
                    zerofill,
                /tikz/.cd,
            },
		ylabel={\large $B_{\theta}$}, 
		legend style={
			cells={anchor=west},
			legend pos=north east, 
		}
        ]
        				
 		\addplot[color=black, solid, thick, mark options={solid}, mark repeat=15,mark size=1,legend image post style={mark size=4.5}] table[x=r,y=btheta] {data/exact_sol_along_xaxis.txt};
 		\addlegendentry{Exact};

        \addplot[color=CPSlightblue, solid, thick, mark=x,mark options={solid},mark repeat=20, mark size=0.5,mark phase=8,legend image post style={mark size=4.5}] table[x=x,y=gradmag] {data/p2_64x64_sol_along_xaxis_trim.txt};
 		\addlegendentry{Fully immersed};
        
 		\addplot[color=CPSred, dash dot, thick, mark=o,mark options={solid},mark repeat=20, mark size=0.5,mark phase=8,legend image post style={mark size=4.5}] table[x=x,y=gradmag] {data/p2_64x64_sol_along_xaxis.txt};
 		\addlegendentry{Union NC};

        \path (axis cs:0.333,5.9e-4) coordinate (onA);
        \path (axis cs:0.155,1.7e-4) coordinate (atA);
        \spy [black] on (onA) in node [right,fill=white] at (atA);
        \path (axis cs:0.666,3e-4) coordinate (onB);
        \path (axis cs:0.42,1.7e-4) coordinate (atB);
        \spy [black] on (onB) in node [right,fill=white] at (atB);
\end{axis}
\end{tikzpicture}

%% file: figures/StriaghtHSM_schem_a.tex
\begin{tikzpicture}
  \fill[blue!20,draw=black, line width=0.1pt] (0,0) rectangle (15mm,13mm);

  \fill[gray!30,draw=black, line width=0.1pt] (2mm,2mm) rectangle (4mm,11mm);
  \fill[red!60,draw=black, line width=0.1pt] (4mm,2mm) rectangle (9mm,4mm);
  \fill[red!60,draw=black, line width=0.1pt]  (4mm,9mm) rectangle (9mm,11mm);
  \fill[gray!30,draw=black, line width=0.1pt] (11mm,2mm) rectangle (13mm,11mm);

    \draw[{Stealth[length=2pt, width=1.5pt]}-{Stealth[length=2.5pt, width=1.5pt]},very thin] (0mm,-0.8mm)--(15mm,-0.8mm) node[midway,sloped,below,rotate=0, scale=0.2] {$W$};
    \draw[dashed, very thin] (0,-1.2mm)--(0,-0.2mm);
    \draw[dashed, very thin] (15mm,-1.2mm)--(15mm,-0.2mm);

    \draw[{Stealth[length=2pt, width=1.5pt]}-{Stealth[length=2.5pt, width=1.5pt]},very thin] (-0.8mm,0mm)--(-0.8mm,13mm) node[midway,sloped,above,rotate=0, scale=0.2] {$H$};
    \draw[dashed, very thin] (-0.2mm,0.0mm)--(-1.2mm,0.0mm);
    \draw[dashed, very thin] (-0.2mm,13mm)--(-1.2mm,13mm);

    \draw[{Stealth[length=1pt, width=0.5pt]}-{Stealth[length=1pt, width=.5pt]},very thin] (11mm,1.6mm)--(13mm,1.6mm) node[midway,sloped,below,rotate=0, scale=0.2] {$W_{\text{iron}}$};
    \draw[dashed, very thin] (13mm,1.8mm)--(13mm,1.4mm);
    \draw[dashed, very thin] (11mm,1.8mm)--(11mm,1.4mm);

    \draw[{Stealth[length=1pt, width=0.5pt]}-{Stealth[length=1pt, width=.5pt]},very thin] (13.4mm,2mm)--(13.4mm,11mm) node[midway,sloped,below,rotate=0, scale=0.2] {$H_{\text{iron}}$};
    \draw[dashed, very thin] (13.2mm,2mm)--(13.6mm,2mm);
    \draw[dashed, very thin] (13.2mm,11mm)--(13.6mm,11mm);

    \draw[{Stealth[length=1pt, width=0.5pt]}-{Stealth[length=1pt, width=.5pt]},very thin] (9.4mm,9mm)--(9.4mm,11mm) node[midway,sloped,below,rotate=0, scale=0.2] {$H_{\text{PM}}$};
    \draw[dashed, very thin] (9.2mm,9mm)--(9.6mm,9mm);
    \draw[dashed, very thin] (9.2mm,11mm)--(9.6mm,11mm);

    \draw[{Stealth[length=1pt, width=0.5pt]}-{Stealth[length=1pt, width=.5pt]},very thin] (4mm,8.6mm)--(9mm,8.6mm) node[midway,sloped,below,rotate=0, scale=0.2] {$W_{\text{PM}}$};
    \draw[dashed, very thin] (9mm,8.8mm)--(9mm,8.4mm);

    \draw[-{Stealth[length=1pt, width=1pt]},very thin] (8mm,9.9mm) --(5mm,9.9mm) node[midway,sloped,above,rotate=0, scale=0.2] {$\tbB_r$};

    \draw[{Stealth[length=1pt, width=1pt]}-,very thin] (8mm,3mm)--(5mm,3mm) node[midway,sloped,below,rotate=0, scale=0.2] {$\tbB_r$};

    \draw[dash pattern=on 1pt off 1pt, very thin,blue] (-0.1mm,10mm)--(15.1mm,10mm) node[blue, right, scale=0.2] {$y=0.1$};

    \draw[teal!70,very thin] (0,0) -- (15mm,0) -- (15mm,13mm) -- (0,13mm) -- cycle;
    \node [teal, scale = 0.2] at (16.3mm,7mm) {$A_z = 0$};

\end{tikzpicture}

%% file: figures/StriaghtHSM_schem_mpiga.tex
\begin{tikzpicture}
\fill[blue!20,draw=black, line width=0.1pt] (0,0) rectangle (15mm,13mm);

\fill[gray!30,draw=black, line width=0.1pt] (2mm,2mm) rectangle (4mm,11mm);
\fill[red!60,draw=black, line width=0.1pt] (4mm,2mm) rectangle (9mm,4mm);
\fill[red!60,draw=black, line width=0.1pt]  (4mm,9mm) rectangle (9mm,11mm);
\fill[gray!30,draw=black, line width=0.1pt] (11mm,2mm) rectangle (13mm,11mm);

\draw[black, line width=0.1pt] (0,2mm) -- (15mm,2mm);
\draw[black, line width=0.1pt] (0,4mm) -- (15mm,4mm);
\draw[black, line width=0.1pt] (0,9mm) -- (15mm,9mm);
\draw[black, line width=0.1pt] (0,11mm) -- (15mm,11mm);

\draw[black, line width=0.1pt] (2mm,0) -- (2mm,13mm);
\draw[black, line width=0.1pt] (4mm,0) -- (4mm,13mm);
\draw[black, line width=0.1pt] (9mm,0) -- (9mm,13mm);
\draw[black, line width=0.1pt] (11mm,0) -- (11mm,13mm);
\draw[black, line width=0.1pt] (13mm,0) -- (13mm,13mm);


\end{tikzpicture}


%% file: figures/StriaghtHSM_schem_b.tex
\begin{tikzpicture}
  \fill[blue!20,draw=black, line width=0.1pt] (0,0) rectangle (15mm,13mm);

  \fill[gray!30,draw=black, line width=0.1pt] (2mm,2mm) rectangle (4mm,11mm);
  \fill[red!60,draw=black, line width=0.1pt]  (4mm,2mm) rectangle (9mm,4mm);
  \fill[red!60,draw=black, line width=0.1pt]  (4mm,9mm) rectangle (9mm,11mm);
  \fill[gray!30,draw=black, line width=0.1pt] (11mm,2mm) rectangle (13mm,11mm);

  \draw[xstep=1.5mm, ystep=1.3mm, gray!60, line width=0.1pt]
    (0.001,0.001) grid (14.98mm,12.99mm);
\end{tikzpicture}

%% file: figures/StriaghtHSM_schem_c.tex
\begin{tikzpicture}
  \fill[blue!20,draw=black, line width=0.1pt] (0,0) rectangle (15mm,13mm);
  \draw[xstep=1.5mm, ystep=1.3mm, gray!60, line width=0.1pt]
    (0.001,0.001) grid (14.98mm,12.98mm);

  \fill[gray!30,draw=black, line width=0.1pt] (2mm,2mm) rectangle (4mm,11mm);
    \draw[xstep=0.5mm, ystep=1mm, gray!60, line width=0.1pt]
    (2mm,2mm) grid (3.99mm,10.99mm);
    
  \fill[red!60,draw=black, line width=0.1pt]  (4mm,2mm) rectangle (9mm,4mm);
    \draw[xstep=1mm, ystep=0.5mm, gray!60, line width=0.1pt]
    (4mm,2mm) grid (8.99mm,3.99mm);
    
  \fill[red!60,draw=black, line width=0.1pt]  (4mm,9mm) rectangle (9mm,11mm);
    \draw[xstep=1mm, ystep=0.5mm, gray!60, line width=0.1pt]
    (4mm,9mm) grid (8.99mm,10.99mm);
    
  \fill[gray!30,draw=black, line width=0.1pt] (11mm,2mm) rectangle (13mm,11mm);
    \draw[xstep=0.5mm, ystep=1mm, gray!60, line width=0.1pt]
    (11mm,2mm) grid (12.99mm,10.99mm);

\end{tikzpicture}

%% file: figures/StriaghtHSM_schem_d.tex
\begin{tikzpicture}
\def\done{0.25mm}   
\def\dtwo{0.50mm}   

\fill[blue!20,draw=black, line width=0.1pt] (0,0) rectangle (15mm,13mm);
\draw[xstep=1.5mm, ystep=1.3mm, gray!60, line width=0.1pt]
  (0.001mm,0.001mm) grid (14.98mm,12.98mm);


\fill[blue!35, even odd rule]
  (2mm-\dtwo,2mm-\dtwo) --
  (2mm-\dtwo,11mm+\dtwo) --
  (9mm+\dtwo,11mm+\dtwo) --
  (9mm+\dtwo,9mm-\dtwo) --
  (4mm+\dtwo,9mm-\dtwo) --
  (4mm+\dtwo,4mm+\dtwo) --
  (9mm+\dtwo,4mm+\dtwo) --
  (9mm+\dtwo,2mm-\dtwo) -- cycle
  (2mm-\done,2mm-\done) --
  (2mm-\done,11mm+\done) --
  (9mm+\done,11mm+\done) --
  (9mm+\done,9mm-\done) --
  (4mm+\done,9mm-\done) --
  (4mm+\done,4mm+\done) --
  (9mm+\done,4mm+\done) --
  (9mm+\done,2mm-\done) -- cycle;

\fill[blue!35, even odd rule]
  (2mm-\done,2mm-\done) --
  (2mm-\done,11mm+\done) --
  (9mm+\done,11mm+\done) --
  (9mm+\done,9mm-\done) --
  (4mm+\done,9mm-\done) --
  (4mm+\done,4mm+\done) --
  (9mm+\done,4mm+\done) --
  (9mm+\done,2mm-\done) -- cycle
  (2mm,2mm) --
  (2mm,11mm) --
  (9mm,11mm) --
  (9mm,9mm) --
  (4mm,9mm) --
  (4mm,4mm) --
  (9mm,4mm) --
  (9mm,2mm) -- cycle;

\fill[blue!35, even odd rule]
  (11mm-\dtwo,2mm-\dtwo) rectangle (13mm+\dtwo,11mm+\dtwo)
  (11mm-\done,2mm-\done) rectangle (13mm+\done,11mm+\done);

\fill[blue!35, even odd rule]
  (11mm-\done,2mm-\done) rectangle (13mm+\done,11mm+\done)
  (11mm,2mm) rectangle (13mm,11mm);

\fill[gray!30,draw=black, line width=0.1pt] (2mm,2mm) rectangle (4mm,11mm);
\draw[xstep=0.5mm, ystep=1mm, gray!60, line width=0.1pt]
  (2mm,2mm) grid (3.99mm,10.99mm);

\fill[red!60,draw=black, line width=0.1pt] (4mm,9mm) rectangle (9mm,11mm);
\draw[xstep=1mm, ystep=0.5mm, gray!60, line width=0.1pt]
  (4mm,9mm) grid (8.99mm,10.99mm);

\fill[red!60,draw=black, line width=0.1pt] (4mm,2mm) rectangle (9mm,4mm);
\draw[xstep=1mm, ystep=0.5mm, gray!60, line width=0.1pt]
  (4mm,2mm) grid (8.99mm,3.99mm);

\fill[gray!30,draw=black, line width=0.1pt] (11mm,2mm) rectangle (13mm,11mm);
\draw[xstep=0.5mm, ystep=1mm, gray!60, line width=0.1pt]
  (11mm,2mm) grid (12.99mm,10.99mm);

\draw[line width=0.1pt]
  (2mm-\done,2mm-\done) --
  (2mm-\done,11mm+\done) --
  (9mm+\done,11mm+\done) --
  (9mm+\done,9mm-\done) --
  (4mm+\done,9mm-\done) --
  (4mm+\done,4mm+\done) --
  (9mm+\done,4mm+\done) --
  (9mm+\done,2mm-\done) -- cycle;

\draw[line width=0.1pt]
  (2mm-\dtwo,2mm-\dtwo) --
  (2mm-\dtwo,11mm+\dtwo) --
  (9mm+\dtwo,11mm+\dtwo) --
  (9mm+\dtwo,9mm-\dtwo) --
  (4mm+\dtwo,9mm-\dtwo) --
  (4mm+\dtwo,4mm+\dtwo) --
  (9mm+\dtwo,4mm+\dtwo) --
  (9mm+\dtwo,2mm-\dtwo) -- cycle;

\draw[line width=0.1pt] (2mm,2mm) -- (2mm-\done,2mm-\done);
\draw[line width=0.1pt] (2mm,11mm) -- (2mm-\done,11mm+\done);
\draw[line width=0.1pt] (9mm,11mm) -- (9mm+\done,11mm+\done);
\draw[line width=0.1pt] (4mm,9mm) -- (4mm+\done,9mm-\done);
\draw[line width=0.1pt] (4mm,4mm) -- (4mm+\done,4mm+\done);
\draw[line width=0.1pt] (9mm,2mm) -- (9mm+\done,2mm-\done);

\draw[line width=0.1pt] (9mm,9mm) -- (9mm+\done,9mm-\done);
\draw[line width=0.1pt] (9mm,4mm) -- (9mm+\done,4mm+\done);

\draw[line width=0.1pt] (2mm-\done,2mm-\done) -- (2mm-\dtwo,2mm-\dtwo);
\draw[line width=0.1pt] (2mm-\done,11mm+\done) -- (2mm-\dtwo,11mm+\dtwo);
\draw[line width=0.1pt] (9mm+\done,11mm+\done) -- (9mm+\dtwo,11mm+\dtwo);
\draw[line width=0.1pt] (4mm+\done,9mm-\done) -- (4mm+\dtwo,9mm-\dtwo);
\draw[line width=0.1pt] (4mm+\done,4mm+\done) -- (4mm+\dtwo,4mm+\dtwo);
\draw[line width=0.1pt] (9mm+\done,2mm-\done) -- (9mm+\dtwo,2mm-\dtwo);

\draw[line width=0.1pt] (9mm+\done,9mm-\done) -- (9mm+\dtwo,9mm-\dtwo);
\draw[line width=0.1pt] (9mm+\done,4mm+\done) -- (9mm+\dtwo,4mm+\dtwo);

\draw[line width=0.1pt]
  (11mm-\done,2mm-\done) rectangle (13mm+\done,11mm+\done);

\draw[line width=0.1pt]
  (11mm-\dtwo,2mm-\dtwo) rectangle (13mm+\dtwo,11mm+\dtwo);

\draw[line width=0.1pt] (11mm,2mm) -- (11mm-\done,2mm-\done);
\draw[line width=0.1pt] (11mm,11mm) -- (11mm-\done,11mm+\done);
\draw[line width=0.1pt] (13mm,11mm) -- (13mm+\done,11mm+\done);
\draw[line width=0.1pt] (13mm,2mm) -- (13mm+\done,2mm-\done);

\draw[line width=0.1pt] (11mm-\done,2mm-\done) -- (11mm-\dtwo,2mm-\dtwo);
\draw[line width=0.1pt] (11mm-\done,11mm+\done) -- (11mm-\dtwo,11mm+\dtwo);
\draw[line width=0.1pt] (13mm+\done,11mm+\done) -- (13mm+\dtwo,11mm+\dtwo);
\draw[line width=0.1pt] (13mm+\done,2mm-\done) -- (13mm+\dtwo,2mm-\dtwo);

\end{tikzpicture}

%% file: figures/y01-h127_trim.tex
\begin{tikzpicture}[
spy using outlines={circle,magnification =3.5,size=4cm,connect spies,
spy connection path = {
\draw[->,thick,densely dashed](tikzspyonnode) -- (tikzspyinnode);
}}]
    \begin{axis}[axis lines=left,
    width =\linewidth,
    height = 0.6\linewidth,
    xmin=0,
    xmax=0.15, 
    xticklabel style={font=\Large},
    yticklabel style={font=\Large}, 
    xlabel={\Large $x$},
    xtick={0,0.02,0.04,0.09,0.11,0.13,0.15},
    xticklabel style={/pgf/number format/fixed, /pgf/number format/precision=2},
    ylabel={\Large $\|\mathbf{B}\|$ [T]},
    ymin=0,
    ymax=1.7,
    legend columns=1, 
    legend style={column sep=1ex,
        font=\large},
    legend cell align=left,
    legend pos=outer north east,
    ] 
        \addplot[color=black, solid,thick, no markers] table[col sep=comma,x=x,y=y] {data/IGA_MagFluxMagnit_at_y=0.1_p=5_h=19.txt};
        \addlegendentry{Multi-patch IGA};
        

        \addplot[color=CPSgreen,solid,thick, mark=*,mark size=0.8,mark options={solid},mark repeat=8, mark phase=8,legend image post style={mark size=4.5}]table[col sep=comma,x=x,y=p1] {data/IBM_MagFluxMagnit_at_y=0.1_h=127_k.txt};
        \addlegendentry{$k/p$-ref. ($p=1$)};
        \addplot[color=CPSlightblue,solid,thick, mark=square,mark size=0.8,mark options={solid},mark repeat=8, mark phase=8,legend image post style={mark size=4.5}]table[col sep=comma,x=x,y=p2] {data/IBM_MagFluxMagnit_at_y=0.1_h=127_k.txt};
        \addlegendentry{$k$-ref. ($p=2$)};
        \addplot[color=CPSorange,solid,thick, mark=triangle,mark size=0.8,mark options={solid},mark repeat=8, mark phase=8,legend image post style={mark size=4.5}]table[col sep=comma,x=x,y=p5] {data/IBM_MagFluxMagnit_at_y=0.1_h=127_k.txt};
        \addlegendentry{$k$-ref. ($p=5$)};


        \addplot[color=CPSred,solid,thick, mark=o,mark size=0.8,mark options={solid},mark repeat=8,mark phase=8,legend image post style={mark size=4.5}]table[col sep=comma,x=x,y=y] {data/straight_B_at_y=0.1_p=5_128x128_p-ver.txt};
        \addlegendentry{$p$-ref. ($p=5$)};

        \addplot[thin, samples=2, densely dotted ,black] coordinates {(0.02,0)(0.02,3)};
        \addplot[thin, samples=2, densely dotted ,black] coordinates {(0.04,0)(0.04,3)};
        \addplot[thin, samples=2, densely dotted ,black] coordinates {(0.09,0)(0.09,3)};
        \addplot[thin, samples=2, densely dotted ,black] coordinates {(0.11,0)(0.11,3)};
        \addplot[thin, samples=2, densely dotted ,black] coordinates {(0.13,0)(0.13,3)};
        \node at (axis cs:0.01, 1.65)[font=\Large]{Air};
        \node at (axis cs:0.03, 1.65)[font=\Large]{Iron}; 
        \node at (axis cs:0.065, 1.65)[font=\Large]{Magnet}; 
        \node at (axis cs:0.1, 1.65)[font=\Large]{Air}; 
        \node at (axis cs:0.12, 1.65)[font=\Large]{Iron};
        \node at (axis cs:0.14, 1.65)[font=\Large]{Air};

        \path (axis cs:0.022,0.55) coordinate (onA);
        \path (axis cs:0.038,1.33) coordinate (onB);
        \path (axis cs:0.1134,0.65) coordinate (onC);
        \path (axis cs:0.1284,0.35) coordinate (onD);
        \path (axis cs:-0.037,0.45) coordinate (atA);
        \path (axis cs:0.065,0.6) coordinate (atB);
        \path (axis cs:0.132,1.15) coordinate (atC);
        \path (axis cs:0.17,0.5) coordinate (atD);
    \end{axis}
    \spy[size=4cm] on (onA) in node[fill=white] at (atA);
    \spy[size=4cm] on (onB) in node[fill=white] at (atB);
    \spy[size=4cm] on (onC) in node[fill=white] at (atC);
    \spy[size=4cm] on (onD) in node[fill=white] at (atD);

\end{tikzpicture}

%% file: figures/y01_union.tex
\begin{tikzpicture}[
spy using outlines={circle,magnification =3.5,size=4cm,connect spies,
spy connection path = {
\draw[->,thick,densely dashed](tikzspyonnode) -- (tikzspyinnode);
}}]
    \begin{axis}[axis lines=left,
    width =\linewidth,
    height = 0.6\linewidth,
    xmin=0,
    xmax=0.15, 
            xticklabel style={font=\Large},
            yticklabel style={font=\Large}, 
    xlabel={\Large $x$},
    xtick={0,0.02,0.04,0.09,0.11,0.13,0.15},
    xticklabel style={/pgf/number format/fixed, /pgf/number format/precision=2},
    ylabel={\Large $\|\mathbf{B}\|$ [T]},
    ymin=0,
    ymax=1.7,
    legend columns=1, 
    legend style={column sep=1ex,
        font=\large},
    legend cell align=left,
    legend pos=outer north east,font =\
    ] 
        \addplot[color=black, solid,thick, no markers] table[col sep=space,x=x,y=flux] {data/ex_20x20_p5.txt};
        \addlegendentry{Multi-patch IGA};


        \addplot[color=CPSred,solid,thick, mark=*,mark size=0.8,mark options={solid},mark repeat=8, mark phase=8,legend image post style={mark size=4.5}]table[col sep=space,x=x,y=flux] {data/union_10000_p5_0.01_y01.txt};
        \addlegendentry{Union NC ($p=5$)};

        \addplot[color=CPSlightblue,solid,thick, mark=square,mark size=0.8,mark options={solid},mark repeat=8, mark phase=8,legend image post style={mark size=4.5}]table[col sep=space,x=x,y=flux] {data/ibcm_10000_p5_0.01_y001.txt};
        \addlegendentry{Union CL ($p=5$)};
        
        \addplot[thin, samples=2, densely dotted ,black] coordinates {(0.02,0)(0.02,3)};
        \addplot[thin, samples=2, densely dotted ,black] coordinates {(0.04,0)(0.04,3)};
        \addplot[thin, samples=2, densely dotted ,black] coordinates {(0.09,0)(0.09,3)};
        \addplot[thin, samples=2, densely dotted ,black] coordinates {(0.11,0)(0.11,3)};
        \addplot[thin, samples=2, densely dotted ,black] coordinates {(0.13,0)(0.13,3)};
        \node at (axis cs:0.01, 1.65)[font=\Large]{Air};
        \node at (axis cs:0.03, 1.65)[font=\Large]{Iron}; 
        \node at (axis cs:0.065, 1.65)[font=\Large]{Magnet}; 
        \node at (axis cs:0.1, 1.65)[font=\Large]{Air}; 
        \node at (axis cs:0.12, 1.65)[font=\Large]{Iron};
        \node at (axis cs:0.14, 1.65)[font=\Large]{Air};

        \path (axis cs:0.022,0.55) coordinate (onA);
        \path (axis cs:0.038,1.33) coordinate (onB);
        \path (axis cs:0.09,0.75) coordinate (onC);
        \path (axis cs:0.1284,0.35) coordinate (onD);
        \path (axis cs:-0.037,0.45) coordinate (atA);
        \path (axis cs:0.065,0.6) coordinate (atB);
        \path (axis cs:0.132,1.15) coordinate (atC);
        \path (axis cs:0.17,0.5) coordinate (atD);
    \end{axis}
    \spy[size=4cm] on (onA) in node[fill=white] at (atA);
    \spy[size=4cm] on (onB) in node[fill=white] at (atB);
    \spy[size=4cm] on (onC) in node[fill=white] at (atC);
    \spy[size=4cm] on (onD) in node[fill=white] at (atD);

\end{tikzpicture}

%% file: figures/StriaghtHSM_H1s_error.tex
\begin{tikzpicture} 

    \pgfplotsset{table/search path={figures}}
    
	\begin{loglogaxis}[ height=25em, 
						width=25em,
						grid=major,
						xmin=0.01,
						xmax=0.1, 
						xlabel={\large Element size $h$},
						ymin=1e-4,
						ymax=1e-2, 
                        xticklabel style={font=\large},
            yticklabel style={font=\large}, 
						ylabel={\large $H^{1}_s$-seminorm error},
						legend style={
							cells={anchor=west},
							legend pos=outer north east, 
						}
					  ]

        
 		\addplot[color=CPSred, solid,very thick,mark=o,mark options={solid}, mark size=1.7] table[x=helem,y=H1s_p2] {data/ncmp_straighthorseshoe.txt};

		\addplot[color=TUDa-10b, solid,very thick,mark=o,mark options={solid}, mark size=1.7] table[x=helem,y=H1s_p4] {data/ncmp_straighthorseshoe.txt};
        
 		\addplot[color=TUDa-2b, solid,very thick,mark=square,mark options={solid}, mark size=1.7] table[x=helem,y=H1s_p6] {data/ncmp_straighthorseshoe.txt};

        \addplot[color=CPSred, dashed,very thick,mark=o,mark options={solid}, mark size=1.7] table[x=helem,y=H1s_p2_IBCM_100] {data/IBCM_straighthorseshoe_3elem_100penal.txt};

		\addplot[color=TUDa-10b, dashed,very thick,mark=o,mark options={solid}, mark size=1.7] table[x=helem,y=H1s_p4_IBCM_100] {data/IBCM_straighthorseshoe_3elem_100penal.txt};
        
 		\addplot[color=TUDa-2b, dashed,very thick,mark=square,mark options={solid}, mark size=1.7] table[x=helem,y=H1s_p6_IBCM_100] {data/IBCM_straighthorseshoe_3elem_100penal.txt};

        \addplot[color=gray, dotted, very thick, domain=0.03:0.06] {0.04*x^1};
		\node at (axis cs:0.032,9.5e-4) [anchor=east] {\textcolor{black}{$\mathcal{O}(h^{-1})$}};
	
	\end{loglogaxis} 

\end{tikzpicture}

%% file: figures/StriaghtHSM_L2_error.tex
\begin{tikzpicture} 

    \pgfplotsset{table/search path={figures}}
    
	\begin{loglogaxis}[ height=25em, 
						width=25em,
						grid=major,
                        xticklabel style={font=\large},
            yticklabel style={font=\large}, 
						xmin=0.01,
						xmax=0.1, 
						xlabel={\large Element size $h$},
						ymin=1e-7,
						ymax=1e-4, 
                        ylabel={\large $L^2$-norm error},
						legend style={
							cells={anchor=west},
							legend pos=outer north east, 
						}
					  ]

 		\addplot[color=CPSred, solid,very thick,mark=o,mark options={solid}, mark size=1.7] table[x=helem,y=L2_p2] {data/ncmp_straighthorseshoe.txt};
 		\addlegendentry{Union NC ($p=2$)};

		\addplot[color=TUDa-10b, solid,very thick,mark=o,mark options={solid}, mark size=1.7] table[x=helem,y=L2_p4] {data/ncmp_straighthorseshoe.txt};
 		\addlegendentry{Union NC ($p=4$)};
        
 		\addplot[color=TUDa-2b, solid,very thick,mark=square,mark options={solid}, mark size=1.7] table[x=helem,y=L2_p6] {data/ncmp_straighthorseshoe.txt};
 		\addlegendentry{Union NC ($p=6$)};

        \addplot[color=CPSred, dashed,very thick,mark=o,mark options={solid}, mark size=1.7] table[x=helem,y=L2_p2_IBCM_100] {data/IBCM_straighthorseshoe_3elem_100penal.txt};
 		\addlegendentry{Union CL ($p=2$)};

		\addplot[color=TUDa-10b, dashed,very thick,mark=o,mark options={solid}, mark size=1.7] table[x=helem,y=L2_p4_IBCM_100] {data/IBCM_straighthorseshoe_3elem_100penal.txt};
 		\addlegendentry{Union CL ($p=4$)};
        
 		\addplot[color=TUDa-2b, dashed,very thick,mark=square,mark options={solid}, mark size=1.7] table[x=helem,y=L2_p6_IBCM_100] {data/IBCM_straighthorseshoe_3elem_100penal.txt};
 		\addlegendentry{Union CL ($p=6$)};
        \addplot[color=gray, dotted, very thick, domain=0.031:0.061] 
        {0.00012*x^(4/3)};
		\node at (axis cs:0.033,8e-7) [anchor=east] {\textcolor{black}{$\mathcal{O}(h^{-4/3})$}};

	
	\end{loglogaxis} 

\end{tikzpicture}

%% file: figures/PMA_MCC_schem.tex
\begin{tikzpicture}[rotate=45]

\def\startangle{0}
\def\endangle{90}

\newcommand{\FourArrowsInSquare}[4]{%
  \coordinate (sqC) at ($ #1 !0.5! #3 $);

  \begin{scope}
    \clip #1 -- #2 -- #3 -- #4 -- cycle;

    \begin{scope}[shift={(sqC)}, rotate=45]
      \def\L{55mm}      
      \def\S{10mm}      

      \foreach \k in {-1.5,-0.5,0.5,1.5}{%
        \draw[black, line width=1.2pt, -{Stealth[length=4mm,width=3mm]}]
          (-0.5*\L, \k*\S) -- (0.5*\L, \k*\S);
      }
    \end{scope}
  \end{scope}
}

\draw[black, line width=1.2pt, -{Stealth[length=6mm,width=4.5mm]}]
          (0,0) -- (312mm, 0) node[scale=4] at (318mm, 0mm) {$x$};
          
\draw[black, line width=1.2pt, -{Stealth[length=6mm,width=4.5mm]}]
          (0,0) -- (0mm, 312mm) node[scale=4] at (0mm, 318mm) {$y$};
\filldraw[fill=blue!20, draw=black, very thin]
  (\startangle:300mm)
  arc[start angle=\startangle, end angle=\endangle, radius=300mm]
  -- (\endangle:30mm)
  arc[start angle=\endangle, end angle=\startangle, radius=30mm]
  -- cycle;


\filldraw[fill=gray!30, draw=black, very thin]
  (0:300mm)
  arc[start angle=0, end angle=90, radius=300mm]
  -- (90:224.5551mm)
  arc[start angle=90, end angle=56.56012838, radius=224.5551mm]
  -- (33.43987162:224.5551mm) 
  arc[start angle=33.43987162, end angle=0, radius=224.5551mm]
  -- cycle;

\draw [fill=red!60, draw=black, very thin] (187.3833mm,123.7437mm) -- (144.957mm,81.3174mm) -- (81.3174mm,144.957mm) -- (123.7437mm,187.383mm) -- cycle;

\FourArrowsInSquare
  {(187.3833mm,123.7437mm)}
  {(144.957mm,81.3174mm)}
  {(81.3174mm,144.957mm)}
  {(123.7437mm,187.383mm)}
  
\draw[fill=gray!30, draw=black, very thin] (0:30mm) arc (0:90:30mm) -- (0,70.7106mm) -- (7.07106mm,70.7106mm) -- (70.7106mm,7.07106mm)
-- (70.7106mm,0) -- cycle;

\draw [fill=red!60, draw=black, very thin] (7.07106mm,70.7106mm) -- (70.7106mm,7.07106mm) -- (113.1372mm,49.4976mm) -- (49.4976mm,113.1372mm) -- cycle;

\FourArrowsInSquare
  {(7.07106mm,70.7106mm)}
  {(70.7106mm,7.07106mm)}
  {(113.1372mm,49.4976mm)}
  {(49.4976mm,113.1372mm)}

\draw[fill=gray!30, draw=black, very thin] (49.4976mm,113.1372mm) -- (113.1372mm,49.4976mm) -- (118.0608mm,54.4212mm) arc [start angle=24.74776888, end angle=65.25223112, radius=130mm] -- cycle;

\draw[fill=gray!30, draw=black, very thin] (144.957mm,81.3174mm) -- (81.3174mm,144.957mm) -- (65.6069mm,140.4777mm) arc [start angle=65, end angle=25, radius=155mm] -- cycle;

\draw[teal!70, line width=3.5pt] (0:30mm) arc (0:90:30mm);
\draw[teal!70, line width=3.5pt] (0:300mm) arc (0:90:300mm);
\node[scale = 4] at (-30mm,60mm) {\textcolor{teal}{$A_z = 0$}};
\node[scale = 4] at (220mm,220mm) {\textcolor{teal}{$A_z = 0$}};
\node[scale = 4] at (190mm,200mm) {\text{Iron bridge}};
\node[scale = 4] at (32mm,31mm) {\text{Iron core}};

\draw[blue,line width=3.5pt,dash pattern=on 15pt off 6pt on 2pt off 6pt] (0:142.5mm) arc (0:90:142.5mm);
\draw[brown, line width=3.5pt] (30mm,0) -- (300mm,0);
\node[scale = 4,rotate=45] at (220mm,-10mm) {\textcolor{brown}{\text{Anti-periodic BC} $\Gamma^{-}$}};
\draw[brown, line width=3.5pt] (0,30mm) -- (0,300mm);
\node[scale = 4,rotate=315] at (-10mm,170mm) {\textcolor{brown}{\text{Anti-periodic BC} $\Gamma^{+}$}};

\draw[black,line width=3pt,dash pattern=on 15pt off 6pt on 2pt off 6pt] (90:300mm) arc (90:100:300mm);

\draw[black, line width=1.2pt, -{Stealth[length=6mm,width=4.5mm]}]
          (0,0) -- (-52.0944mm, 295.44232mm) node[scale=4] at (-60.0944mm, 295.44232mm) {$r_4$};

\draw[black,line width=3pt,dash pattern=on 15pt off 6pt on 2pt off 6pt] (0:30mm) arc (0:-15:30mm);
\draw[black, line width=1.2pt, -{Stealth[length=6mm,width=4.5mm]}]
          (0,0) -- (28.9777mm, -7.76457mm) node[scale=4] at (28.9777mm, -14.8094mm) {$r_1$};

\draw[black,line width=3pt,dash pattern=on 15pt off 6pt on 2pt off 6pt] (25:155mm) arc (25:-15:155mm);
\draw[black, line width=1.2pt, -{Stealth[length=6mm,width=4.5mm]}]
          (0,0) -- (149.7185mm,-40.11695mm ) node[scale=4] at (149.7185mm, -47.11695mm) {$r_3$};
          
\draw[black,line width=3pt,dash pattern=on 15pt off 6pt on 2pt off 6pt] (23:130mm) arc (23:-15:130mm);
\draw[black, line width=1.2pt, -{Stealth[length=6mm,width=4.5mm]}]
          (0,0) -- (125.57035mm, -33.64647mm) node[scale=4] at (125.57035mm,-40.54647mm) {$r_2$};


\coordinate (TR) at (187.3833mm,123.7437mm);   
\coordinate (TL) at (123.7437mm,187.383mm);    
\coordinate (BL) at (81.3174mm,144.957mm);     

\coordinate (TLw) at ($(TL)+(45:18mm)$);
\coordinate (TRw) at ($(TR)+(45:18mm)$);

\draw[dashed, line width=1.2pt] (TL) -- (TLw);
\draw[dashed, line width=1.2pt] (TR) -- (TRw);

\draw[line width=1.4pt, {Stealth[length=6mm,width=4.5mm]}-{Stealth[length=6mm,width=4.5mm]}]
  (TLw) -- (TRw)
  node[midway, above=3mm, scale=4] {$W_{\text{PM}}$};

\coordinate (TLh) at ($(TL)+(135:16mm)$);
\coordinate (BLh) at ($(BL)+(135:16mm)$);

\draw[dashed, line width=1.2pt] (TL) -- (TLh);
\draw[dashed, line width=1.2pt] (BL) -- (BLh);

\draw[line width=1.4pt, {Stealth[length=6mm,width=4.5mm]}-{Stealth[length=6mm,width=4.5mm]}]
  (BLh) -- (TLh)
  node[midway, left=3mm, scale=4] {$H_{\text{PM}}$};


\coordinate (TopGap)    at (45:155mm);   
\coordinate (MidGap)    at (45:142.5mm); 
\coordinate (BottomGap) at (45:130mm);   
\coordinate (TopMagR) at (45:115mm);   
\coordinate (StartMagS) at (45:160mm);   

\coordinate (TopDim)    at ($(TopGap)+(-45:200mm)$);
\coordinate (BottomDim) at ($(BottomGap)+(-45:200mm)$);
\coordinate (TopMagRDim) at ($(TopMagR)+(-45:200mm)$);
\coordinate (StartMagSDim) at ($(StartMagS)+(-45:200mm)$);

\draw[dashed,line width=1.2pt] (TopGap) -- (TopDim);
\draw[dashed,line width=1.2pt] (TopMagR) -- (TopMagRDim);
\draw[dashed,line width=1.2pt] (StartMagS) -- (StartMagSDim);
\draw[dashed,line width=1.2pt] (BottomGap) -- (BottomDim);

\draw[line width=1.4pt,{Stealth[length=2mm,width=2mm]}-{Stealth[length=2mm,width=2mm]}]
(StartMagSDim) -- (TopDim)
node[midway, right=4mm, scale=4] {$H_{\text{OFRC}}$};

\draw[line width=1.4pt,
{Stealth[length=4mm,width=3mm]}-{Stealth[length=4mm,width=3mm]}]
(BottomDim) -- (TopMagRDim)
node[midway, right=4mm, scale=4] {$H_{\text{IFRC}}$};
\end{tikzpicture}

%% file: figures/PMA_gapfield_comparison.tex
\begin{tikzpicture}[
spy using outlines={
rectangle,
magnification=3,
connect spies,
spy connection path={
\draw[->,thick,densely dashed] (tikzspyonnode) -- (tikzspyinnode);
}
}]
\begin{axis}[
        height=27em, 
		width=44em, 
		grid=both,xmin=0,
		xmax=90, 
		xlabel={\large Circumferential angle $\theta$ [\textdegree]},
		ymin=0,
		ymax=0.85, 
		y tick label style={
                /pgf/number format/.cd,
                    fixed,
                    precision=1,
                    zerofill,
                /tikz/.cd,
            },
            xticklabel style={font=\large},
            yticklabel style={font=\large}, 
		ylabel={\large $\|\tbB\|$ [T]},
		legend style={
			cells={anchor=west},
			legend pos=outer north east,
        font=\large
		},
  clip=false
        ]
		
 		\addplot[color=black, solid, thick,mark options={solid}, mark size=1.7, mark phase=8,legend image post style={mark size=4.5}] table[x=theta_conf,y=fluxconf] {data/PMA_gapfield_confIGA.txt};
 		\addlegendentry{Multi-patch IGA};

 		\addplot[color=CPSred, dashed, thick,mark=o,mark options={solid},mark repeat=20, mark size=1, mark phase=8,legend image post style={mark size=4.5}] table[x=theta_union,y=fluxunion] {data/PMA_gapfield_union.txt};
 		\addlegendentry{Union NC};

        \addplot[color=CPSgreen, dash dot, thick,mark=square,mark options={solid},mark repeat=20, mark size=1, mark phase=8,legend image post style={mark size=4.5}] table[x=theta_ibcm1,y=fluxibcm1] {data/PMA_gapfield_ibcm1.txt};
 		\addlegendentry{Union CL (v1)};

        \addplot[color=CPSdarkblue, dash dot, thick,mark=square,mark options={solid},mark repeat=20, mark size=1, mark phase=8,legend image post style={mark size=4.5}] table[x=theta_ibcm2,y=fluxibcm2] {data/PMA_gapfield_ibcm2.txt};
 		\addlegendentry{Union CL (v2)};

        \coordinate (spypoint) at (axis cs:45,0.81);
        \coordinate (spypoint1) at (axis cs:1,0.035);    
            \spy[
              every spy on node/.append style={width=6cm, minimum height=1.5cm}, 
              every spy in node/.append style={minimum width=6cm,  minimum height=2cm}    
            ] on (spypoint)
              in node [fill=white] at (rel axis cs:0.5,0.54);

              \spy[
              every spy on node/.append style={minimum width=1cm, minimum height=1.5cm}, 
              every spy in node/.append style={minimum width=2cm,  minimum height=2cm}    
            ] on (spypoint1)
              in node [fill=white] at (rel axis cs:0.09,0.33);
\end{axis}

\end{tikzpicture}

%% file: sections/conclusions.tex
\section{Conclusions and future work}
\label{sec:conclusion}

In this work, we developed an isogeometric immersed boundary-conformal framework for two-dimensional magnetostatic problems, in which three different methods are employed, namely: \emph{fully immersed},  \emph{union with non-conformal patches}, and \emph{union with conformal layers}. In all three approaches, trimmed elements in the immersed background patch are  efficiently handled by adopting a boundary-conformal quadrature method, leading to accurate integration with a low number of quadrature points. Coupling non-conformal patches is enabled by employing Nitsche's method while ensuring stability by considering the flux from the non-trimmed patches, which alleviates the need for additional stabilization methods.

The results demonstrate that the proposed framework achieves accuracy comparable to conformal multi-patch IGA while significantly reducing the geometry preprocessing effort. We investigated the application of the aforementioned methods for three different problems, each presenting distinct challenges. The coaxial cable problem is intended to represent a discontinuous current field between several regions. For the horseshoe magnet problem, discontinuous material parameters, i.e., permeability, introduce difficulties in treating Gibbs-type phenomena. In the last example, the geometry represents a cross-section of an industrial application, where accurate evaluation of the field in the air gap plays a crucial role in predicting the cooling power of the device.

The proposed immersed IGA framework with non-conformal patches and conformal layers offers several advantages compared to conventional frameworks, e.g., multi-patch IGA:
\begin{itemize}

\item By employing the union approaches, multiple subdomains can be described directly by immersing their B-rep -- which can be extracted from CAD -- and conformal layers can then be generated automatically without human intervention. This is not the case for standard multi-patch IGA, where conforming patches must be constructed manually.
\item The geometric preprocessing effort and the number of patches required are significantly reduced, e.g., from 30 to 5 in the horseshoe magnet problem and from 21 to 7 for the permanent magnet assembly.
\item The method circumvents the requirement for conformity between independent patches, unlike multi-patch IGA, thereby reducing the effort associated with generating a conformal multi-patch mesh. This also applies to weakly coupled multi-patch IGA based on Nitsche's method \cite{nguyen2014}. However, such approaches are typically limited to regions that are topologically equivalent to a square, e.g., making it difficult to represent the air region in \cref{subsec:PMAssem_MCC} using a single patch. 
\item The framework removes the need for Lagrange multiplier methods or harmonic mortaring methods to couple different subdomains, resulting in a system with fewer unknowns.
\item The union concept (used in the \emph{Union NC} and \emph{Union CL} methods) allows for strong imposition of Dirichlet boundary conditions on domain boundaries, unlike the \emph{fully immersed} fictitious domain procedure, where this is only possible when the physical domain boundaries coincide with the extended domain boundaries.
\item The \emph{Union CL} enables accurate capture of solution singularities, resulting in improved accuracy.
\item The \emph{Union CL} demonstrates its effectiveness in mitigating spurious oscillations across material interfaces due to the construction of a conformal layer within the trimmed field, which leads to smoother transitions across interfaces.
\item Both the \emph{Union NC} and \emph{Union CL} provide accurate evaluation of the field within critical regions (e.g., the air gap in \cref{subsec:PMAssem_MCC}).
\end{itemize}

Overall, the \emph{union with non-conformal patches} method provides accurate results for the coaxial cable problem. However, for the horseshoe magnet problem and the permanent magnet assembly problem, the method requires refinement near critical regions to mitigate over- or underestimation of the field. In contrast, the \emph{union with conformal layers} provides highly accurate results for all considered problems, demonstrating its accuracy, efficacy, and robustness for two-dimensional magnetostatic simulations. 

Future work will investigate the effect of local refinement strategies, such as Truncated Hierarchical B-splines \cite{giannelli2012,schmidt2023,grendas2026}, particularly near critical regions, which may reduce the number of conformal layers required. Furthermore, the method will be extended to three-dimensional magnetostatic problems and applied to additional electric machine configurations \cite{gangl2026}.